\documentclass[12pt]{article} 
\usepackage{amsmath}
\usepackage{amssymb,amsfonts,euscript}
\usepackage{hyperref}
\renewcommand{\baselinestretch}{1.2}
\newcommand\nn{\nonumber}

\newcommand{\D}{{\cal D}}

\newcommand{\cL}{\cal{L}}

\newcommand{\Zint}{\mathbb{Z}}

\newcommand{\bJ}{\bar{J}}
\newcommand{\bz}{\bar{z}}

\def\a{\alpha}
\def\be{\beta}
\def\l{\lambda}
\def\m{\mu}
\def\n{\nu}
\def\r{\rho}

\def\de{\delta}
\def\k{\kappa}

\def\hB{\hat{B}}

\def\hg{\hat{g}}

\def\hb{\hat{B}}

\def\part{\partial}

\def\thickone{{\rm 1\mskip-4.5mu l}}

\newcommand{\s}{\sigma}
\newcommand{\srange}{\sigma=0,...,N_c-1}

\newcommand{\ws}{\omega (h_\s)}

\newcommand{\hc}{$\hat{J}_{\gst}$}

\newcommand{\tp}{{2\pi i}}

\newcommand{\Ord}{{\cal{O}}}

\newcommand{\sgb}{{\mbox{\scriptsize{\gb}}}}

\newcommand{\gfraks}{{\mbox{\scriptsize{\mbox{${\mathfrak g}$}}}}}
\def\gb            {\mbox{$\hat{\mathfrak g}$}}

\def\sm#1      {\mbox{\scriptsize $#1$}}

\def\d             {\mbox{$\mathbb D$}}

\def\srac#1#2{\smal{\frac{#1}{#2}}}
\def\trac#1#2{\scrs{\frac{#1}{#2}}}
\def\foot#1{\mbox{\footnotesize $#1$}}
\def\scrs#1{\mbox{\scriptsize $#1$}}
\def\tyny#1{\mbox{\tiny $#1$}}
\def\smal#1{\mbox{\small $#1$}}

\def\Big#1{\mbox{\Large $#1$}}
\def\BIG#1{\mbox{\Huge $#1$}}

\def\hjb{\hat{\bar{J}}}

\def\hjbb{ \hat{\bar{J}}^\sharp }

\def\hjbb{ \hat{\bar{J}}^\sharp }

\def\hc{^\dagger}

\def\one{{\mathchoice {\rm 1\mskip-4mu l} {\rm 1\mskip-4mu} {\rm
1\mskip-4.5mu l}
{\rm 1\mskip-5mu l}}}
\def\d{\delta}

\def\nnsrs{n+\srac{n(s)}{\r(\s)}}

\def\nrm{{n(r)\m}}

\def\nrn{{n(r)\n}}
\def\mnrn{{-n(r),\n}}
\def\nsn{{n(s)\n}}
\def\ntd{{n(t)\d}}

\def\nue{{n(u)\epsilon}}
\def\nvk{{n(v)\kappa}}
\def\nwl{{n(w)\lambda}}

\def\mnnrnsrs{{m+n+\srac{n(r)+n(s)}{\r(\s)}}}

\def\mnrrs{{m+\srac{n(r)}{\r(\s)}}}

\def\nrrs{{\srac{n(r)}{\r(\s)}}}
\def\nrrsf{{\frac{n(r)}{\r(\s)}}}
\def\nsrs{{\srac{n(s)}{\r(\s)}}}
\def\scf{{\cal F}}
\def\sG{{\cal G}}

\def\gfrak{\mbox{$\mathfrak g$}}

\def\hj{\hat{J}}
\def\nrn{{n(r)\n}}
\def\nsn{{n(s)\n}}

\def\schisig{{\foot{\chi(\s)}}}

\def\ntd{{n(t)\delta}}
\def\hc{^\dagger}
\def\st{{\cal T}}
\def\0b{\ }
\def\pl{\partial}

\def\Nrm{{N(r)\m}}

\def\Nsn{{N(s)\n}}
\def\Ntd{{N(t)\d}}

\def\srange{\s=0,\ldots,N_c-1}
\def\sm{{\cal M}}

\def\ho{{\hat{\Omega}}}
\def\hY{{\hat{Y}}}

\def\sxh{{\hat{\smal{\EuScript{X}}}}}

\def\he{{\hat{e}}}
\def\hei{{ \hat{e}^{-1}}}
\def\heb{{\hat{\bar{e}\hspace{.02in}}\hspace{-.02in}}}
\def\hebi{{ \hat{\bar{e}}^{-1}}}
\def\hx{{\hat{x}}}

\def\hp{\hat{p}}
\def\sh{{\cal H}}
\def\hG{\hat{G}}
\def\hpl{\hat{\pl}}
\def\k{\kappa}
\def\l{\lambda}

\def\su{{ \mathfrak{su} }}
\def\so{{ \mathfrak{so} }}

\def\bigspc{{ \quad \quad \quad \quad}}
\def\gscfwt{{ \hat{\Delta}_0 (\s)}}
\def\ep{{ \epsilon}}
\def\pom{{\tilde{\omega}}}
\def\hD{{\hat{\Delta}}}
\def\hPs{{\hat{\Psi}}}
\def\hps{{\hat{\psi}}}
\def\hL{{\hat{\Lambda}}}
\def\hU{{\hat{U}}}
\def\Nrpm{{ N'(r')\m'}}
\def\Nspn{{ N'(s')\n'}}

\def\bnrrs{{ \frac{\bar{n}(r)}{\r(\s)}}}
\def\sbnrrs{{ \srac{\bar{n}(r)}{\r(\s)}}}
\def\Dg{{ \D_{\sgb(\s)}}}
\def\bkspc{{ \!\!\!\!}}
\def\lpl{{ \overleftarrow{\pl}}}
\def\bpl{{ \bar{\pl}}}
\def\bT{{ \bar{T}}}

\makeatletter
\renewcommand{\@makefnmark}{\mbox{$^{\ddagger\@thefnmark}$}}
\renewcommand{\subsection}{\@startsection
{subsection}{2}{0pt
}{-\baselineskip}{0.5\baselineskip}
{\normalfont\normalsize\bf}}
\renewcommand{\section}{\@startsection
{section}{2}{0pt
}{-\baselineskip}{0.5\baselineskip}
{\bf\large}}

\makeatother
\numberwithin{equation}{section}
\numberwithin{table}{section}

\newcounter{myfigctr}
\def\myfig#1{\refstepcounter{myfigctr}%
 \label{#1}%
}

\newcommand{\publititle}[8]
{ 
  \vspace*{-3cm}
  \begin{flushright} #1 \\ {\tt #2} \end{flushright}
  \vfill
  \begin{center}{\Large
    \bfseries #3}\end{center}
  \vskip 8mm
  \begin{center}{\large #4}\end{center}
  \begin{center}{\normalsize #5}\end{center}
  \vskip 8mm
  \nopagebreak
  \noindent #6
  \vfill
  \begin{flushleft} #7
  \end{flushleft}
  \hrule width 6.7cm \vskip.1mm
  {\small #8}
  \thispagestyle{empty}
  \clearpage
}

\begin{document}

\publititle{ ${}$ \\ UCB-PTH-04/03 \\ LBNL-54534 \\ hep-th/0402108}{}{A
Basic Class of Twisted Open WZW Strings
}{M.B.Halpern$^{1a}$ and C. Helfgott$^{1b}$}
{$^1$ {\em Department of Physics, University of California and \\
Theoretical Physics Group,  Lawrence Berkeley National
Laboratory \\
University of California, Berkeley, California 94720, USA}
\\[2mm]} {Recently, Giusto and Halpern reported the open-string description
of a certain basic class of untwisted open WZW strings, including their
associated non-commutative geometry and open-string KZ equations. In this
paper, we combine this development with results from the theory
of current-algebraic orbifolds to find the open-string description of a
corresponding basic class of {\it twisted} open WZW strings,
which begin and end on different WZW branes. The basic class of twisted
open WZW strings is in 1-to-1 correspondence with the twisted sectors of all
closed-string WZW orbifolds, and moreover, the basic class can be
decomposed into a large
collection of open-string WZW orbifolds. At the classical level, these
open-string orbifolds exhibit new {\it twisted non-commutative geometries},
and we also find the relevant {\it twisted open-string KZ equations} which
describe these orbifolds at the quantum level. In a related development,
we also formulate the closed-string description (in terms of twisted
boundary states) of the {\it general} twisted open WZW string.
} {$^a${\tt halpern@physics.berkeley.edu} \\ $^b${\tt
helfgott@socrates.berkeley.edu}
}

\clearpage

\renewcommand{\baselinestretch}{.4}\rm
{\footnotesize
\tableofcontents
}
\renewcommand{\baselinestretch}{1.2}\rm

\pagebreak

\section{Introduction}

In recent years, the orbifold program [1-11] has in large part completed
the local description
of {\it closed-string orbifolds}, including:

\noindent $\bullet$ the twisted current algebras and stress tensors of
all sectors of the general current-algebraic orbifold [1-5],

\noindent $\bullet$ the twisted affine primary fields, twisted operator
algebras and twisted KZ equations \cite{Big,Big',Perm,so2n} of
all WZW orbifolds,

\noindent $\bullet$ the action formulation [6-8,10] of all WZW
and coset orbifolds, in terms of group
orbifold elements with definite monodromy,

\noindent $\bullet$ the action formulation and twisted Einstein equations
\cite{Geom} of a large class of sigma-model orbifolds,

\noindent $\bullet$  free-bosonic avatars \cite{Big',Perm,Geom} of these
constructions and the explicit form of their
twisted vertex operators.

\noindent A short review of the program can be found in Ref.~\cite{so2n}.  Recent progress
at the level of characters
has also been reported in Refs.~[12,1,13-15].

Subsequently, the techniques of the orbifold program were also applied to
construct a new class of so-called {\it orientation orbifolds}
\cite{Orient1,Orient2},
which arise by twisting world-sheet orientation-reversing automorphisms in
closed-string WZW, coset and sigma models. Like conventional orientifolds
[18-21], the orientation orbifolds contain both closed- and
open-string sectors,but the orientation-orbifold sectors are
characterized by fractional moding, including twisted Virasoro operators
\cite{Chr,DV2,Perm} in the open-string sectors.

The open-string sectors of the WZW orientation orbifolds are certainly not
the most general twisted open WZW strings, but because orientation-reversing
automorphisms are quite special, it is not immediately clear how to
generalize the construction of Refs.~\cite{Orient1,Orient2}.

In pursuit of more general twisted open WZW strings, we have therefore
reexamined the ``{\it open-string description}'' of untwisted open WZW
strings\footnote{ In the closed-string description of open strings, one
uses the closed-string currents $J ,\bJ$ to define boundary states, while
in the open-string description of open strings, one works directly with a
single set of current modes $J$.}
given in Ref.~\cite{Giusto}, including the non-commutative geometry and the
open-string KZ equations of these strings. We find that this construction
can
be straightforwardly combined with the theory of closed-string
current-algebraic orbifolds to give a large set of new twisted open WZW
strings, which we
call the {\it basic class}.

The basic class is a particular class of twisted open WZW strings, which

\noindent $\bullet$ is entirely disjoint from the open-string sectors of the
 WZW orientation orbifolds,

\noindent $\bullet$ is in 1-to-1 correspondence with the twisted sectors of
 all closed-string WZW orbifolds,

\noindent $\bullet$ can be decomposed into a large collection of
open-string WZW orbifolds
\begin{gather}
\frac{A_g^{open}(H)}{H} ,\quad H \subset Aut(g)
\end{gather}
$\,\,\,$ where $A_g^{open}(H)$ is any Giusto-Halpern open string with a
symmetry $H$. $\bigspc$ \linebreak Examples of simple open-string
orbifolds have been
discussed e.~g.~in Refs.~\cite{DM,Angel}.

An overview of our construction is given in Subsec.~$2.1$, and the
necessary background material from the orbifold program is reviewed in
Subsec.~$2.2$ and Apps.~A,B.
Central results for the basic class include the generalized WZW branes in
Subsec.~$3.5$, the new twisted non-commutative geometry in Subsec.~$3.6$
and the twisted open-string KZ equations in Subsec.~$4.5$. Explicit
non-abelian examples are given in Subsecs.~$3.5$, $4.5$ and App.~D, and
free-bosonic
analogues are worked out in Subsec.~$3.7$. In a parallel development,
App.~A formulates the {\it closed-string description} (in terms of twisted
boundary states)
of the {\it general} twisted open WZW string.

Taken together, the general twisted boundary state equation and the
open-string sectors
of the WZW orientation orbifolds give important clues for further
generalization of our main development (see the final Discussion in
Sec.~5).
Based on these observations, we will return elsewhere to construct the
open-string description of the general twisted open WZW string -- which
includes
both the orientation orbifolds and the basic class as special cases.

\section{Preliminaries}

\subsection{The Open-String WZW Orbifolds $A_g^{open}(H)/H$}

In Ref.~\cite{Giusto} a procedure was given to construct an open WZW string
$A_g^{open}$ from the left-mover sector of the closed-string WZW model
$A_g$ on affine $g$.  In this paper we combine this procedure with the
local theory of current-algebraic orbifolds [1-11] to construct a
corresponding set of
$N_c$ {\it twisted} open WZW strings, one from each of the left-mover
sectors $\s =0\ldots N_c -1$ of the general closed-string WZW orbifold $A_g
(H)/H$ (see
Fig.~\ref{fig:Giu-reln}).

\vskip 10pt
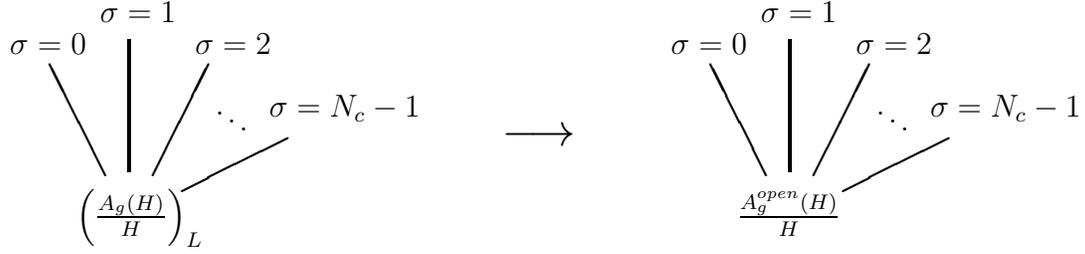
\begin{picture}(380,230)

\thicklines

\put(46,142){\line(-1,2){21}}
\put(55,143){\line(0,1){50}}
\put(64,142){\line(1,2){21}}
\put(75,136){\line(2,1){40}}
\put(35,123){$\left( \frac{A_g(H)}{H} \right)_L $}

\put(10,188){$\s=0$}
\put(44,200){$\s =1$}
\put(80,188){$\s =2$}
\put(87,158){$\ddots$}
\put(108,164){$\s =N_c -1$}

\put(196,153){$\Big{\longrightarrow}$}

\put(296,142){\line(-1,2){21}}
\put(305,143){\line(0,1){50}}
\put(314,142){\line(1,2){21}}
\put(325,136){\line(2,1){40}}
\put(285,123){$\frac{A_g^{open}(H)}{H} $}

\put(260,188){$\s=0$}
\put(294,200){$\s =1$}
\put(330,188){$\s =2$}
\put(337,158){$\ddots$}
\put(358,164){$\s =N_c -1$}

\put(15,83){$\left( \frac{A_g(H)}{H} \right)_L $: The left-mover data of
any closed-string WZW orbifold.}
\put(15,58){$\frac{A_g^{open}(H)}{H} $: The corresponding open-string WZW
orbifold.}
\put(15,38){$N_c$: The number of conjugacy classes of symmetry group $H
\!\subset \!Aut(g)$ and }
\put(36,25){the number of sectors $\s$ of $A_g(H)/H$ or $A_g^{open}(H)/H$.}

\put(5,2){Fig.\,\ref{fig:Giu-reln}: Construction of twisted open WZW
strings from closed-string WZW orbifold}

\end{picture}
\myfig{fig:Giu-reln}
\vskip 10pt

Although our construction follows the path shown in
Fig.~\ref{fig:Giu-reln}, it is also possible to consider these new
constructions as
{\it open-string orbifolds} $A_g^{open} (H)/H$ of any $H$-symmetric
untwisted open WZW string $A_g^{open}(H)$ (see Fig.~\ref{fig:open-orb}).

\vskip 10pt
\begin{picture}(380,130)

\thicklines

\put(85,70){$A_g^{open}(H)$}
\put(150,70){$\Big{\longrightarrow}$}
\put(200,70){$\frac{A_g^{open}(H)}{H} $}

\put(241,79){\line(1,1){35}}
\put(243,75){\line(2,1){43}}
\put(244,71){\line(1,0){50}}
\put(242,67){\line(2,-1){43}}

\put(279,115){$\s =0$}
\put(290,95){$\s =1$}
\put(299,69){$\s =2$}
\put(280,55){$\vdots$}
\put(289,42){$\s =N_c -1$}

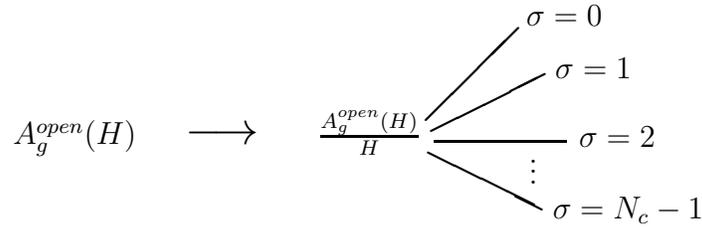
\put(5,15){Fig.\,\ref{fig:open-orb}: The open-string WZW orbifold
$A_g^{open}(H)/H$ associated to the untwisted }
\put(35,2){open WZW string $A_g^{open}(H)$ with a symmetry $H$.}

\end{picture}
\myfig{fig:open-orb}
\vskip 10pt

We infer that $A_g^{open}(H)/H$ is an open-string WZW orbifold for the
following reasons:

$\bullet$ Each sector $\srange$ of $A_g^{open} (H)/H$ is a twisted open
string, in particular, the $\s =0$ sector is the untwisted open string
$A_g^{open}(H)$ of Ref.~\cite{Giusto} with a symmetry H.

$\bullet$ As in closed-string orbifold theory, each sector $\s$ of
$A_g^{open}(H)/H$ is labelled by a conjugacy class of $H$. Moreover, sector
$\s$
contains the appropriate twisted current algebra, obtained by twisting
affine $g$ by an element $h_\s \in H$ in this conjugacy class.

$\bullet$ At the classical level, the target space of each sector of
$A_g^{open}(H)/H$ is the appropriate sector of the group orbifold $g/H$,
with the corresponding group orbifold elements [6-8], and with
generalized WZW branes at each end of the open string.

\noindent Because of modifications needed to apply the principle of local
isomorphisms \cite{Dual,More,Big} to equal-time formulations \cite{Geom} and
open strings \cite{Orient1}, a direct realization of the path shown in
Fig.~\ref{fig:open-orb} will not be discussed in this paper.

In what follows we will therefore refer to $A_g^{open}(H)/H$ as an
open-string WZW orbifold, while the {\it basic class} of twisted open
strings will
denote the set of all sectors of all $A_g^{open}(H)/H$. Presumably, these
open-string WZW orbifolds are associated via non-planar processes
to closed-string WZW orbifolds, but we will not study this issue here.

\subsection{The Left-Mover Data of the Closed-String Orbifold $A_g(H)/H$}

Ref.~\cite{Giusto} constructed a basic class of untwisted open WZW strings from
the untwisted affine-Sugawara construction [26-30]
on affine $g$ \cite{Kac,Moody,BH}, and it is clear from this development that an
open-string conformal field theory can be constructed from any single
chiral current-algebraic stress tensor [26,27,33-38]

For the twisted construction of this paper, we therefore begin with the
left-mover\footnote{If the algebra of the twisted right-mover currents
$\hjb$ of $A_g (H)/H$ is not {\it rectifiable} \cite{Big,Big', Perm,so2n} into a
copy of the twisted left-mover current algebra, then in principle there
could
be another set of twisted open strings based on $\hjb$. However, as
reviewed in App.~B, all basic orbifold types are known to be rectifiable.}
twisted
affine-Sugawara construction \cite{Dual,More,Big} of sector $\s$ of the
closed-string WZW orbifold $A_g(H)/H$:
\vspace{-0.1in}
\begin{subequations}
\label{Eq2.1}
\begin{gather}
L_\s(m) = {\cL}_{\sgb(\s)}^{\nrm;\mnrn}(\s) \sum_{p \in \Zint} :\!\hj_\nrm
(p\!+\!\nrrs) \hj_\mnrn (m\!-\!p\! -\!\nrrs) \!: \label{Eq 2.1a}
\end{gather}
\begin{align}
&\!\!\![\hj_\nrm (m\!+\!\nrrs) ,\hj_\nsn (n\!+\!\nsrs)] \!=\!
i\scf_{\nrm;\nsn}{}^{\!\!\!\!\!n(r)+n(s),\de} (\s)
\hj_{n(r)+n(s),\de}(m\!+\!n\!
   +\!\srac{n(r)+n(s)}{\r(\s)}) \quad \nn \\
& \bigspc \bigspc \quad \quad +(m\!+\!\nrrs)
\de_{m+n+\frac{n(r)+n(s)}{\r(\s)},0} \sG_{\nrm;\mnrn}(\s) \label{Eq 2.1b}
\end{align}
\begin{gather}
[L_\s (m) ,\hj_\nrm (n\!+\!\nrrs)] = -(n\!+\!\nrrs) \hj_\nrm
(m\!+\!n\!+\!\nrrs) \label{Eq 2.1c} \\
[L_\s (m) ,L_\s (n)] =(m-n) L_\s (m+n) +\de_{m+n ,0} \frac{\hat{c}}{12} m
(m^2 -1) \label{Eq 2.1d} \\
\hat{c} =2{\cL}_{\sgb(\s)}^{\nrm;\mnrn}(\s) \sG_{\nrm;\mnrn}(\s) =2L_g^{ab}
G_{ab} =c_g ,\quad \s =0\ldots N_c -1 \,.
\end{gather}
\end{subequations}
The normal ordering $:\,\cdot \,:$ in Eq.~\eqref{Eq 2.1a} is the mode form
of operator-product normal ordering in the orbifold \cite{TVME,Dual,More,Big},
and the alternate mode normal-ordered form $:\,\cdot \,:_M$ of the twisted
affine-Sugawara construction is given in Eq.~\eqref{EqA.3}.

The numerical coefficients in Eq.~\eqref{Eq2.1} are called twisted tensors
or {\it duality transformations}. For sector $\s$ of $A_g(H)/H$, the
standard duality transformations have the following explicit forms:
\begin{subequations}
\label{Eq2.2}
\begin{align}
\sG_{\nrm;\nsn} (\s) &\equiv \schisig_\nrm \schisig_\nsn U(\s)_\nrm{}^a
U(\s)_\nsn{}^b G_{ab} \nn \\
&= \sG_{\nsn;\nrm} (\s)= \de_{n(r)+n(s) ,0\, \text{mod } \r(\s)}
\sG_{\nrm;\mnrn} (\s) \label{Eq 2.2a} \\
\scf_{\nrm;\nsn}{}^{\!\!\ntd} &(\s) \equiv \schisig_\nrm \schisig_\nsn
\schisig^{-1}_\ntd U(\s)_\nrm{}^a U(\s)_\nsn{}^b f_{ab}{}^c U\hc
(\s)_c{}^\ntd \nn \\
&\!\!\!\!= -\scf_{\nsn;\nrm}{}^{\!\!\!\ntd} (\s) =\!\de_{n(r)+n(s)-n(t),0\,
\text{mod } \r(\s)}
   \scf_{\nrm;\nsn}{}^{\!\!\!n(r)+n(s),\de} (\s) \label{Eq 2.2b} \\
{\cL}_{\sgb (\s)}^{\nrm;\nsn} (\s) &= \schisig^{-1}_\nrm \schisig^{-1}_\nsn
L_g^{ab} U\hc (\s)_a{}^\nrm U\hc (\s)_b{}^\nsn \nn \\
&={\cL}_{\sgb (\s)}^{\nsn;\nrm} (\s) =\de_{n(r)+n(s),0\,\text{mod }\r(\s)}
{\cL}_{\sgb (\s)}^{\nrm;\mnrn} (\s) \label{Eq 2.2c}
\end{align}
\begin{gather}
\st_\nrm (T,\s) \equiv \schisig_\nrm U(\s)_\nrm{}^a U(T,\s) T_a U\hc (T,\s)
\label{Eq 2.2d} \\
e^{\tp \nrrs} \st_\nrm (T,\s) = E (T,\s) \st_\nrm (T,\s) E(T,\s)^\ast \,.
\label{Eq 2.2e} \\
[\st_\nrm (T,\s) ,\st_\nsn (T,\s)] =i\scf_{\nrm;\nsn}{}^{n(r)+n(s),\de}(\s)
\st_{n(r)+n(s),\de}(T,\s)\label{Eq 2.2f} \\
\widehat{Tr} (\sm (\st,\s) \st_\nrm (T,\s) \st_\nsn (T,\s))
=\sG_{\nrm;\nsn}(\s) \label{Eq 2.2g}
\end{gather}
\begin{gather}
g=\oplus_I \gfrak^I ,\quad G_{ab} =\oplus_I k_I \eta_{ab}^I ,\quad
f_{ab}{}^c =\oplus_I f_{ab}^I {}^c ,\quad L_g^{ab} =\oplus_I
\frac{\eta^{ab}_I}
   {2k_I +Q_I} ,\quad T_a =\oplus_I T_a^I \,.
\end{gather}
\end{subequations}
Here the quantities $G, f,L_g$ and $T$ are the generalized Killing metric,
structure constants, inverse inertia tensor and representation matrices of
the untwisted theory on $g$. The twisted tensors $\sG ,\scf ,{\cL}$ and
$\st$ are the duality transformations (discrete Fourier transforms) of the
corresponding untwisted quantities. Similarly, the twisted data matrix
$\sm$ \cite{Big} is the duality transformation of the data matrix, which records
the level $k_I$ of each affine $\gfrak^I$ and the Dynkin index of each rep
$T^I$.

In Eq.~\eqref{Eq2.2}, the normalization constants $\schisig$ are
essentially arbitrary and the unitary matrices (Fourier elements) $U(\s)$
and $U(T,\s)$ solve the {\it $H$-eigenvalue problems} \cite{Dual,More,Big,Big',so2n} of orbifold theory:
\begin{subequations}
\label{Eq2.3}
\begin{gather}
\ws_a{}^b U\hc (\s)_b{}^\nrm = U\hc (\s)_a{}^\nrm E_{n(r)}(\s) ,\quad
E_{n(r)} (\s) = e^{-\tp \nrrs} \label{Eq 2.4a} \\
W(h_\s;T)_\a{}^\be U\hc (T,\s)_\be{}^\Nrm =U\hc (T,\s)_\a{}^\Nrm E_{N(r)}
(T,\s) ,\quad E_{N(r)}(T,\s) =e^{-\tp \frac{N(r)}{R(\s)}} \\
E(T,\s)_\Nrm{}^\Nsn = \de_\m^\n \de_{N(r) +N(s),0\,\text{mod }R(\s)}
E_{N(r)} (T,\s) \\
\s=0\ldots N_c -1 \,.
\end{gather}
\end{subequations}
Here the matrices $\ws$ and $W(h_\s;T)$ are the actions (in the
underlying untwisted theory) of the automorphism $h_\s \in H$ in the
adjoint rep
and rep $T$ respectively, and $E_{n(r)}(\s),$ $E_{N(r)}(T,\s)$ are the
eigenvalues of $\ws ,W(h_\s;T)$. All these quantities are periodic $n(r)
\rightarrow
n(r)\pm \r(\s),\, N(r) \rightarrow N(r)\pm R(\s)$ in any spectral index,
with period equal to the order $\r(\s)$ or $R(\s)$ of the corresponding
automorphic
action $\omega$ or $W$. We denote the pullbacks to the fundamental ranges
by the usual symbols $\bar{n}(r), \bar{N}(r)$.

In the untwisted sector $\s=0$, we have
\begin{subequations}
\label{Eq2.4}
\begin{gather}
\ws =W(h_\s ;T) =\one ,\quad U(\s) =U(T,\s) =\one ,\quad \schisig =1 \\
\sG \rightarrow G ,\quad \scf \rightarrow f ,\quad {\cL} \rightarrow L_g
,\quad \st \rightarrow T
\end{gather}
\end{subequations}
and the system \eqref{Eq2.1} and \eqref{Eq2.2} reduces to the left-mover
sector of the WZW model $A_g(H)$ with symmetry $H \subset Aut(g)$. Except
for the $H$-symmetry of the untwisted theory (which was modded out to
construct the twisted sectors), this is precisely the starting point of
Ref.~\cite{Giusto}.

The reader should therefore bear in mind that the $\s =0$ sectors of our
open-string WZW orbifolds below will agree with all the untwisted results
of
Ref.~\cite{Giusto} -- although one should $H$-symmetrize the correlators of the
untwisted string when it is included as an orbifold
sector in our construction.

For detailed information on particular classes of closed-string WZW
orbifolds, we direct the reader to the following references:

$\bullet$ the WZW permutation orbifolds \cite{Big,Big',Perm}

$\bullet$ the inner-automorphic WZW orbifolds \cite{Big, Perm}

$\bullet$ the (outer-automorphic) charge conjugation orbifold on $\su
(n\geq 3)$ \cite{Big'}

$\bullet$ the outer-automorphic WZW orbifolds on $\so (2n)$, including the
triality orbifolds \linebreak
\indent $\quad$ on $\so(8)$ \cite{so2n}.

\noindent Ref.~\cite{so2n} also contains a short review of the program.

\subsection{The Twisted Strip Currents and ``Locally Twisted WZW"}

Given the left-mover data of the previous subsection, we begin our
construction of the open-string orbifold $A_g^{open}(H)/H$ by defining the
twisted
left- and right-mover {\it open-string currents} on the strip
\begin{subequations}
\label{Eq2.5}
\begin{gather}
\hj_\nrm^{(\pm)} (\xi,t,\s) \equiv \sum_{m \in \Zint} \hj_\nrm (\mnrrs)
e^{-i(\mnrrs) (t \pm \xi)} ,\quad 0\leq \xi \leq\pi \label{Eq 2.5a} \\
\hj_\nrm^{(\pm )} (-\xi,t,\s) =\hj^{(\mp )} (\xi,t,\s) \\
\pl_\mp \hj^{(\pm)} (\xi,t,\s) =0 ,\quad \pl_\pm = \pl_t \pm \pl_\xi ,\quad
\srange  \label{Eq 2.5b}
\end{gather}
\end{subequations}
both of which are constructed from the same set \eqref{Eq 2.1b} of twisted
left-mover current modes. The strip currents satisfy the following {\it
boundary
conditions}
\begin{equation}
\hj_\nrm^{(+)} (0,t,\s) = \hj_\nrm^{(-)} (0,t,\s) ,\quad \hj_\nrm^{(+)}
(\pi,t,\s) = e^{-\tp \nrrs} \hj_\nrm^{(-)} (\pi,t,\s) \label{Eq2.6}
\end{equation}
which are the image on the strip of the monodromy \eqref{Eq B.3c} of the
corresponding cylinder current. With the mode expansion \eqref{Eq 2.5a} and
the mode algebra \eqref{Eq 2.1b}, we may compute the equal-time algebra of
the strip currents
\begin{subequations}
\label{Eq2.7}
\begin{align}
[ \hj_\nrm^{(+)} (\xi,t,\s) ,\hj_\nsn^{(+)} (\eta,t,\s) ] &= 2\pi i \Big{(}
\scf_{\nrm;\nsn}{}^{n(r)+n(s),\de} (\s)
\hj_{n(r)+n(s) ,\de}^{(+)} (\eta,t,\s) \quad \nn \\
& +\de_{n(r)+n(s) ,0\, \text{mod } \r(\s)} \sG_{\nrm;-\nrn} (\s) \pl_\xi
\Big{)} \de_{\nrrsf} (\xi -\eta)
\end{align}
\begin{align}
[ \hj_\nrm^{(+)} (\xi,t,\s) ,\hj_\nsn^{(-)} (\eta,t,\s) ] &= 2\pi i \Big{(}
\scf_{\nrm;\nsn}{}^{n(r)+n(s),\de} (\s)
\hj_{n(r)+n(s) ,\de}^{(-)} (\eta,t,\s) \quad \nn \\
& +\de_{n(r)+n(s) ,0\, \text{mod } \r(\s)} \sG_{\nrm;-\nrn} (\s) \pl_\xi
\Big{)} \de_\nrrsf (\xi +\eta) \label{Eq 2.7b}
\end{align}
\begin{align}
[ \hj_\nrm^{(-)} (\xi,t,\s) ,\hj_\nsn^{(-)} (\eta,t,\s) ] &= 2\pi i \Big{(}
\scf_{\nrm;\nsn}{}^{n(r)+n(s),\de} (\s)
\hj_{n(r)+n(s) ,\de}^{(-)} (\eta,t,\s) \quad \nn \\
& -\de_{n(r)+n(s) ,0\, \text{mod } \r(\s)} \sG_{\nrm;-\nrn} (\s) \pl_\xi
\Big{)} \de_{-\nrrsf} (\xi -\eta) \label{Eq 2.7c}
\end{align}
\end{subequations}
where $\sG (\s)$ and $\scf(\s)$ are the twisted tangent-space metric and
twisted structure constants given explicitly in Eq.~\eqref{Eq2.2}.

The {\it phase-modified Dirac delta functions} $\de_{n(r)/\r(\s)} (\xi \pm
\eta)$ in Eq.~\eqref{Eq2.7} are defined as follows:
\begin{subequations}
\label{Eq2.8}
\begin{gather}
\de_\nrrsf (\xi \pm \eta) \equiv e^{-i\frac{n(r)}{\r(\s)} (\xi \pm \eta)}
\de (\xi \pm \eta) =
  \frac{1}{2\pi} \sum_{m\in \Zint} e^{-i(\mnrrs )(\xi \pm \eta)}
=\de_{-\nrrsf}(-\xi \mp \eta) \\
\de_{\frac{n(r) \pm \r(\s)}{\r(\s)}} (\xi \pm \eta) = \de_\nrrsf (\xi \pm
\eta) ,\quad \de_{0} (\xi \pm \eta) = \de (\xi \pm \eta) \label{Eq 2.8b} \\
\de_\nrrsf (\xi \pm \eta +2\pi) = e^{-\tp \nrrs} \de_\nrrsf (\xi \pm \eta) \\
\de (\xi \pm \eta) \equiv \frac{1}{2\pi} \sum_{m\in \Zint} e^{-im(\xi \pm
\eta)} ,\quad \de (\xi \pm \eta +2\pi) =\de (\xi \pm \eta) \,.
\end{gather}
\end{subequations}
The quantity $\de_{n(r)/\r(\s)}(\xi \!-\!\eta)$ appeared previously in the
orbifold geometry of Ref.~\cite{Geom}, but the quantity $\de_{n(r)/\r(\s)}
(\xi \!+\!\eta)$ appears for the first time here. We note in particular
that $\de_{n(r)/\r(\s)} (\xi \!+\!\eta)$ has support only at the boundaries
of the
strip $\xi \!=\!\eta \!=\!0$ or $\pi$. The reader is referred to App.~C for
various useful identities involving these
phase-modified delta functions.

Following Ref.~\cite{Giusto}, we compare the twisted open-string equal-time
current algebra \eqref{Eq2.7} on the strip to the twisted closed-string
left- and right-mover current algebra
\eqref{EqB.4} on the cylinder under the map:
\begin{equation}
\text{(strip)  } \hj_\nrm^{(+)}(\xi,t) ,\,\,\hj_\nrm^{(-)} (\xi,t)
\,\,\stackrel{?}{\longleftrightarrow} \,\,\hj_\nrm(\xi,t) ,\,\,\hjb_\nrm
(\xi,t)
   \text{  (cylinder)} \,. \label{Eq2.10}
\end{equation}
We find that the two systems disagree by boundary terms in the
left/right-mover commutators \eqref{Eq B.4c} vs.~\eqref{Eq 2.7b}. More
importantly,
this comparison fails due to the form of the bulk terms
($\de_{n(r)/\r(\s)}(\xi-\eta)$ vs.~$\de_{-n(r)/\r(\s)}(\xi-\eta)$) in the
right/right-mover
commutators \eqref{Eq B.4b} vs.~\eqref{Eq 2.7c}. We find however a
successful comparison when we consider instead the {\it rectified}
right-mover
currents $\hjbb$ reviewed in App.~B:
\begin{equation}
\label{Eq2.11}
\text{(strip)  } \hj^{(+)}_\nrm (\xi,t) ,\,\, \hj^{(-)}_\nrm (\xi,t)
\,\,\longleftrightarrow \,\,\hj_\nrm (\xi,t) ,\,\, \hjbb_\nrm (\xi,t)
\text{  (cylinder)} \,.
\end{equation}
In this case, the strip current algebra \eqref{Eq2.7} is in fact isomorphic
in the bulk to the rectified cylinder current algebra
\eqref{EqB.6}, the two algebras differing only by terms $\de_{n(r)/\r(\s)}
(\xi \!+\!\eta)$ with support at the boundary.

In Ref.~\cite{Giusto}, the forms of various untwisted strip algebras were fixed
by the requirement that these algebras must be locally isomorphic in the
bulk to
the untwisted algebras of the corresponding closed WZW string. The reason
for this ``locally WZW'' requirement was that it guaranteed that all local
quantities and relations (such as the action density and equations of
motion) were isomorphic in the bulk to the corresponding relations in the
closed
WZW string, as expected intuitively \cite{AS}.

Following this intuition, Eq.~\eqref{Eq2.11} tells us that the ``locally
WZW'' requirement of Ref.~\cite{Giusto} must be generalized to a new ``locally
twisted WZW'' requirement for twisted open strings: All twisted strip
algebras must be isomorphic in the bulk to the {\it rectified} twisted
algebras
of the corresponding closed-string WZW orbifold sector. Drawing on the
rectified closed-string results collected in App.~B, and using this
requirement to fix certain
results below, we will find that all local quantities and relations in the
open-string WZW orbifold are indeed isomorphic in the bulk to the
corresponding relations in the closed-string WZW orbifold. We note in
particular that the locally twisted WZW condition reduces in untwisted
sector
$\s =0$ to the original locally WZW condition of Ref.~\cite{Giusto}.

\subsection{The Operator Stress Tensors on the Strip}

We turn now to the left- and right-mover stress tensors of sector $\s$ of
the open-string orbifold $A_g^{open}(H)/H$
\begin{subequations}
\label{Eq2.12}
\begin{gather}
\hat{T}_\s^{(\pm)} (\xi,t) \equiv \frac{1}{2\pi} {\cL}_{\sgb
(\s)}^{\nrm;\mnrn} (\s) :\!\hj_\nrm^{(\pm)}(\xi,t,\s)
   \hj_\mnrn^{(\pm)} (\xi,t,\s)\!: ,\quad 0\leq \xi \leq \pi \label{Eq
2.12a} \\
=\frac{1}{2\pi} \sum_{m\in \Zint} \!L_\s (m) e^{-im(t \pm \xi)} \bigspc
\bigspc \label{Eq 2.12b} \\
\hat{T}^{(\pm )}_\s (-\xi,t) =\hat{T}_\s^{(\mp )} (\xi,t) ,\quad \pl_\mp
\hat{T}_\s^{(\pm)} (\xi,t) =0 ,\quad \srange
\end{gather}
\end{subequations}
which are also constructed from the single left-mover stress tensor
$\hat{T}_\s (\xi) ,\,\,0 \!\leq \!\xi \!\leq \!2\pi$ of closed-string
orbifold sector $\s$.
The normal ordering $:\,\cdot \,:$ is the same as that shown in
Eq.~\eqref{Eq2.1}. With Eq.~\eqref{EqB.10}, we note that the forms of the
strip stress tensors
$\hat{T}^{(\pm)}$ are locally twisted WZW
\begin{equation}
\hj^{(+)} ,\hj^{(-)} ,\hat{T}_\s^{(+)} ,\hat{T}_\s^{(-)}
\longleftrightarrow \hj,\hjbb, \hat{T}_\s ,\hat{\bar{T}}_\s \quad
\label{Eq2.13}
\end{equation}
that is, isomorphic in form to the left- and right-mover stress tensors
$\hat{T} ,\hat{\bT}$ of the corresponding closed-string WZW orbifold sector.

The open-string stress tensors satisfy the following boundary conditions
\begin{equation}
\hat{T}_\s^{(+)} (\xi,t) = \hat{T}_\s^{(-)} (\xi,t) ,\,\,\,\, \text{at }
\xi=0,\pi \label{Eq2.14}
\end{equation}
which are obtained directly from Eq.~\eqref{Eq 2.12b}, or equivalently from
\eqref{Eq 2.12a} and the current boundary conditions \eqref{Eq2.6}.

The equal-time algebra of the open-string stress tensors
\begin{subequations}
\label{Eq2.15}
\begin{align}
&\!\![ \hat{T}_\s^{(+)} (\xi,t) ,\hat{T}_\s^{(\pm)} (\eta,t) ]=\!i \left(
(\hat{T}_\s^{(+)} (\xi,t) \!+\!\hat{T}_\s^{(\pm)} (\eta,t))
    -\frac{c_g}{24 \pi} (\pl^2_\xi +1)\right) \pl_\xi \de (\xi \mp \eta)  \\
&\!\![ \hat{T}_\s^{(-)} (\xi,t) ,\hat{T}_\s^{(\pm)} (\eta,t) ]= \!-i \left(
(\hat{T}_\s^{(-)} (\xi,t) \!+\!\hat{T}_\s^{(\pm)}
   (\eta,t)) -\frac{c_g}{24 \pi} (\pl^2_\xi +1) \right) \pl_\xi \de (\xi
\pm \eta)
\end{align}
\begin{gather}
[\hat{T}_\s^{(+)} (\xi,t) ,\hj_\nrm^{(\pm)} (\eta,t,\s)]= \mp i\pl_\eta
\left( \hj_\nrm^{(\pm)} (\eta,t,\s) \de (\xi \mp \eta) \right) \\
[\hat{T}_\s^{(-)} (\xi,t) ,\hj_\nrm^{(\pm)} (\eta,t,\s)]= \mp i\pl_\eta
\left( \hj_\nrm^{(\pm)} (\eta,t,\s) \de (\xi \pm \eta) \right)
\end{gather}
\end{subequations}
follows from the commutators in Eqs.~\eqref{Eq 2.1c}, \eqref{Eq 2.1d}. It
is easily checked that this strip algebra is also locally twisted WZW, that
is,
locally isomorphic in the bulk to the corresponding rectified closed-string
algebra \eqref{EqB.9}.

In each sector $\s$ of $A_g^{open}(H)/H$, the Hamiltonian is constructed
from the stress tensors by the usual integration:
\begin{subequations}
\label{Eq2.16}
\begin{align}
\hat{H}_\s &\!=\hat{H}_{\s}\hc \equiv L_\s(0)=\int_0^{\pi}\!\!d\xi
(\hat{T}_\s^{(+)}(\xi,t) +\hat{T}_\s^{(-)}(\xi,t)) = \bigspc \bigspc \quad
\quad \nn \\
&\!=\! \frac{1}{2\pi} \!\int_0^{\pi} \!\!\!d\xi {\cL}_{\sgb
(\s)}^{\nrm;\mnrn} (\s) \Big{(} :\! \hj_\nrm^{(+)} (\xi,t,\s)
\hj_\mnrn^{(+)}
   (\xi,t,\s) \nn \\
& \bigspc \bigspc \bigspc \bigspc +\! \hj_\nrm^{(-)} (\xi,t,\s)
\hj_\mnrn^{(-)} (\xi,t,\s) \!: \Big{)} \\
&\quad \quad \quad \bigspc \bigspc \srange \,.
\end{align}
\end{subequations}
Then we have the open-string equations of motion
\begin{subequations}
\label{Eq2.17}
\begin{gather}
\pl_t \hat{A} = i[\hat{H}_\s ,\hat{A}] \label{Eq 2.17a} \\
\pl_\mp \hj^{(\pm)} (\xi,t) = \pl_\mp \hat{T}_\s^{(\pm)} (\xi,t) =0
\label{Eq 2.17b}
\end{gather}
\end{subequations}
where the conservation laws in Eq.~\eqref{Eq 2.17b} follow from
Eqs.~\eqref{Eq 2.17a} and \eqref{Eq2.15}.

We may also define the bulk momentum operator
\begin{align}
\hat{P}_\s &\!=\hat{P}_{\s}\hc \!\equiv \!\int_0^{\pi} \!\!\!d\xi
(\hat{T}_\s^{(+)} (\xi,t) -\hat{T}_\s^{(-)} (\xi,t) ) \bigspc \nn \\
&\!= \!\frac{1}{2\pi} \!\int_0^{\pi} \!\!\!d\xi {\cL}_{\sgb
(\s)}^{\nrm;\mnrn} (\s) \Big{(} :\! \hj_\nrm^{(+)} (\xi,t,\s)
\hj_\mnrn^{(+)}
   (\xi,t,\s) \nn \\
& \bigspc \bigspc \bigspc \bigspc -\!\hj_\nrm^{(-)} (\xi,t,\s)
\hj_\mnrn^{(-)} (\xi,t,\s) \!: \Big{)} \label{Eq2.18}
\end{align}
in analogy to Ref.~\cite{Giusto}. As in the untwisted case, this quantity is not
conserved
\begin{equation}
\frac{d}{dt} \hat{P}_\s = 2 ( \hat{T}_\s^{(+)} (\pi) - \hat{T}_\s^{(+)}
(0)) \neq 0 \,. \label{Eq2.19}
\end{equation}
and it can be checked with Eq.~\eqref{Eq2.15} that $\hat{P}_\s$ generates
$-i\pl_\xi$ only in the bulk:
\begin{equation}
\label{Eq2.20}
i [ \hat{P}_\s ,\hj_\nrm^{(\pm)} (\xi,t,\s) ] = \pl_\xi \left\{
\begin{array}{ll}
\hj_\nrm^{(\pm)} (\xi,t,\s) & 0<\xi<\pi ,\\
0 & \xi =0,\pi \,.
\end{array} \right.
\end{equation}
More generally, the correct form of $\pl_\xi \hat{A}$ for any $\hat{A}$ can
be obtained from the form of $i[ \hat{P}_\s ,\hat{A} ]$ in the bulk
\begin{equation}
\pl_\xi \hat{A} = i[ \hat{P}_\s ,\hat{A} ] ,\quad 0< \xi <\pi  \label{Eq2.21}
\end{equation}
by smoothly extending this form to the boundary.

We conclude this section with a discussion of the {\it scalar twist-field
state} $|0\rangle_\s$
\begin{equation}
\label{Eq2.22}
\hj_{\nrm} (\mnrrs \geq 0) |0\rangle_\s = {}_\s \langle 0|\hj_\nrm (\mnrrs
\leq 0) =0 ,\quad \srange
\end{equation}
in sector $\s$ of $A_g^{open}(H)/H$. The conformal weight $\gscfwt$ of the
scalar twist-field
\begin{subequations}
\label{Eq2.23}
\begin{gather}
L_\s (m)\hc =L_\s (-m) \\
(L_\s (m\geq 0) -\de_{m,0} \gscfwt )|0\rangle_\s = {}_\s \langle 0|(L_\s
(m\leq 0) -\de_{m,0} \gscfwt )=0 \\
(\hat{H}_\s -\gscfwt )|0\rangle_\s = {}_\s \langle 0|(\hat{H}_\s -\gscfwt
)=0 \\
\gscfwt = \frac{x_\gfraks /2}{x_\gfraks +\tilde{h}_\gfraks} \sum_r
\srac{\bar{n}(r)}{2\r(\s)} (1-\srac{\bar{n}(r)}{\r(\s)})
   \text{ dim} [\bar{n}(r)]  \label{Eq 2.13d}
\end{gather}
\end{subequations}
follows from the defining relations \eqref{Eq2.22} and Eq.~\eqref{Eq A.3a}.
The general formula for $\gscfwt$ is given in Eq.~\eqref{Eq A.3e}, while
the simplified form given in Eq.~\eqref{Eq 2.13d} holds when the underlying
untwisted theory is permutation-invariant
\begin{gather}
g=\oplus_I \gfrak^I ,\quad \gfrak^I \simeq \text{simple }\gfrak ,\quad k_I
=k  \label{Eq2.24}
\end{gather}
which includes all the basic types of WZW orbifolds. In Eq.~\eqref{Eq2.23},
the quantities $\tilde{h}_\gfraks$ and $x_\gfraks$ are respectively the
dual Coxeter number of $\gfrak$ and the invariant level of affine $\gfrak$,
while $dim[\bar{n}(r)]$ is the degeneracy of the eigenvalue $E_{n(r)}(\s)$
in the $H$-eigenvalue problem \eqref{Eq 2.2a}. Further evaluation of
$\gscfwt$ is given for specific cases in Refs.~\cite{Big,Big',Perm,so2n}.

\section{Classical Description of Twisted Open WZW Strings}

We turn next to the classical theory of the open-string WZW orbifold
$A_g^{open}(H)/H$, beginning with the classical limit of the operator
relations above. Using intuition gained from the classical results of this
section, our discussion of the quantum theory will resume in Sec.~4.

\subsection{The Classical Strip Currents and Stress Tensors}

The classical (high-level) limit of the theory is obtained from the quantum
theory above by dropping all normal ordering, omitting the central terms
in Eq.~\eqref{Eq2.15} and carrying out the following replacements
\begin{subequations}
\label{Eq3.1}
\begin{gather}
[ \,,\,] \rightarrow \{ \,,\, \} \label{Eq 3.1a}  \\
{\cL}_{\sgb (\s)}^{\nrm;\nsn} (\s) \,\underset{\tyny{k \rightarrow
\infty}}{\longrightarrow} \,\frac{1}{2} \sG^{\nrm;\nsn} (\s)
\end{gather}
\end{subequations}
where $\{ \,,\,\}$ are (rescaled) Poisson brackets. Here
$\sG^{\bullet}(\s)$ is the inverse of the twisted tangent-space metric
defined in
Eq.~\eqref{Eq 2.2a}.

This gives for example the classical form of the open-string stress tensors
\begin{subequations}
\label{Eq3.2}
\begin{gather}
\hat{T}_\s^{(\pm)} (\xi,t) \equiv \frac{1}{4\pi} \sG^{\nrm;\mnrn} (\s)
\hj_\nrm^{(\pm)} (\xi,t,\s) \hj_\mnrn^{(\pm)} (\xi,t,\s) ,\quad 0\leq \xi
\leq \pi \\
\hat{T}_\s^{(+)} (\xi,t) = \hat{T}_\s^{(-)}(\xi,t) \,\,\,\text{at }
\xi=0,\pi ,\quad \pl_\mp \hat{T}_\s^{(\pm)} (\xi,t) =0
\end{gather}
\end{subequations}
where $\hj^{(\pm )}$ are now the classical strip currents. We list below
the remainder of those classical relations which do not follow directly
from the bracket substitution \eqref{Eq 3.1a} alone:
\begin{subequations}
\label{Eq3.3}
\begin{gather}
\{\hat{T}_\s^{(+)} (\xi,t) ,\hat{T}_\s^{(\pm)} (\eta,t)\}= i
(\hat{T}_\s^{(+)} (\xi,t) \!+\!\hat{T}_\s^{(\pm)} (\eta,t)) \pl_\xi \de
(\xi \mp \eta)  \\
\{\hat{T}_\s^{(-)} (\xi,t) ,\hat{T}_\s^{(\pm)} (\eta,t)\}= -i
(\hat{T}_\s^{(-)} (\xi,t) \!+\!\hat{T}_\s^{(\pm)} (\eta,t)) \pl_\xi \de
(\xi \pm \eta)
\end{gather}
\end{subequations}
\begin{subequations}
\label{Eq3.4}
\begin{align}
\hat{H}_\s &= \int_0^{\pi} \!\!d\xi (\hat{T}_\s^{(+)} (\xi,t)
+\hat{T}_\s^{(-)} (\xi,t) ) \label{Eq 3.4a} \\
&= \!\frac{1}{4\pi} \!\int_0^{\pi} \!\!d\xi \sG^{\nrm;\mnrn} (\s) \left(
\hj_\nrm^{(+)} (\xi,t) \hj_\mnrn^{(+)} (\xi,t) \!+\!\hj_\nrm^{(-)} (\xi,t)
   \hj_\mnrn^{(-)} (\xi,t) \right) \quad \nn \\
\hat{P}_\s &= \int_0^{\pi} \!\!d\xi (\hat{T}_\s^{(+)} (\xi,t)
-\hat{T}_\s^{(-)} (\xi,t) )  \label{Eq 3.4b} \\
&= \!\frac{1}{4\pi} \!\int_0^{\pi} \!\!d\xi \,\sG^{\nrm;\mnrn} (\s) \left(
\hj_\nrm^{(+)} (\xi,t) \hj_\mnrn^{(+)} (\xi,t) \!-\!\hj_\nrm^{(-)} (\xi,t)
   \hj_\mnrn^{(-)} (\xi,t) \right) \quad \nn
\end{align}
\end{subequations}
In particular, the equal-time current algebra \eqref{Eq2.7} and the
Hamiltonian equations of motion in Eqs.~\eqref{Eq2.17}, \eqref{Eq2.19} have
the
same form in the classical theory.

\subsection{Phase-Space Realization of the Strip Currents}

To go beyond a theory of currents, we now postulate the following {\it
phase-space realization} \cite{Giusto} of the twisted strip currents at time $t$:
\begin{subequations}
\label{Eq3.5}
\begin{align}
\!\!&\hj_\nrm^{(+)} (\xi)  \equiv \hj_\nrm^{(+)} (\xi,t,\s) \\
& \quad \equiv 2\pi \he^{-1} (\hx(\xi))_\nrm {}^{\!\ntd} \hp_\ntd (\hb
,\xi) + \frac{1}{2} \pl_\xi \hx_\s^\ntd (\xi)
  \he (\hx(\xi))_\ntd {}^{\!\nsn} \sG_{\nsn;\nrm} (\s) \quad \quad \nn \\
\!\!&\hj_\nrm^{(-)} (\xi) \equiv \hj_\nrm^{(-)} (\xi,t,\s) \\
& \quad \equiv 2\pi \heb^{-1} (\hx(\xi))_\nrm {}^{\!\ntd} \hp_\ntd (\hb
,\xi) - \frac{1}{2} \pl_\xi \hx_\s^\ntd (\xi)
  \heb (\hx(\xi))_\ntd {}^{\!\nsn} \sG_{\nsn;\nrm} (\s) \quad \quad \nn
\end{align}
\begin{gather}
\hp_\nrm (\hb ,\xi) \equiv \hp^\s_\nrm + \frac{1}{4\pi} \hb_{\nrm;\nsn}
(\hx(\xi)) \pl_\xi \hx_\s^\nsn (\xi) \\
 \srange \,.
\end{gather}
\end{subequations}
Here $\{\hx_\s \}$ is a set of Einstein coordinates and $\{\hp^\s \}$ is a
set of momenta which are expected to be canonically conjugate to $\hx_\s$
in
the bulk. The twisted vielbeins $\he ,\heb$ and twisted B field $\hB$ which
appear in \eqref{Eq3.5} are defined in terms of so-called {\it group
orbifold
elements} $\hg$:
\begin{subequations}
\label{Eq3.6}
\begin{gather}
\hg (\st,\xi,t,\s) = e^{i\hx_\s^\nrm (\xi,t) \st_\nrm (T,\s)} \label{Eq
3.6a} \\
\he_\nrm (\st) \equiv -i \hg^{-1} (\st) \hpl_\nrm \hg (\st) \equiv \he_\nrm
{}^\nsn \st_\nsn ,\quad \hpl_\nrm (\xi) \equiv
   \srac{\pl}{\pl \hx^\nrm (\xi)} \label{Eq 3.6b} \\
\heb_\nrm (\st) \equiv -i\hg (\st) \hpl_\nrm \hg^{-1} (\st) \equiv
\heb_\nrm {}^\nsn \st_\nsn \label{Eq 3.6c} \\
\he_\nrm{}^\ntd \he^{-1}_\ntd {}^\nsn = \heb_\nrm{}^\ntd \heb^{-1}_\ntd
{}^\nsn = \de_\nrm{}^\nsn = \de_\m^\n \de_{n(r)-n(s),0\, \text{mod }
\r(\s)}
\end{gather}
\begin{gather}
\hat{H}_{\nrm;\nsn;\ntd} (\hx) \equiv \hpl_\nrm \hb_{\nsn;\ntd} (\hx) \!+\!
\hpl_\nsn \hb_{\ntd;\nrm} (\hx) \!+\! \hpl_\ntd \hb_{\nrm;\nsn} (\hx) \quad
\nn \\
= -i \widehat{Tr} \left( \sm (\st,\s) \he_\nrm (\st,\hx) [\he_\nsn
(\st,\hx) ,\he_\ntd (\st,\hx)] \right) \\
\quad \quad \quad= \he (\hx)_\nrm{}^{n(r')\m'} \he(\hx)_\nsn {}^{n(s')\n'}
\he(\hx)_\ntd {}^{n(t')\de'} \scf_{n(r')\m';n(s')\n' ;n(t')\de'} (\s) \,.
\label{Eq 3.6f}
\end{gather}
\end{subequations}
As in closed-string orbifold theory, the group orbifold element $\hg
(\st,\xi,t,\s)$ is locally a group element, where the group is generated by
the
orbifold Lie algebra \eqref{Eq 2.2f} of twisted representation matrices
$\st$. Moreover, we have chosen $\he (0) \!=\!\thickone ,\, \heb(0)
\!=\!-\thickone$.
The totally antisymmetric twisted structure constants $\scf_{\bullet} (\s)$
in \eqref{Eq 3.6f} are constructed by lowering the last index of $\scf(\s)$
using the twisted metric $\sG_{\bullet} (\s)$.

Here we have generalized the strategy of Ref.~\cite{Giusto} by modelling the
realizations in Eq.~\eqref{Eq3.5} after the known phase-space realizations
of $\hj$ and $\hjb$ in closed-string WZW orbifold theory \cite{Geom}, where the
same geometric quantities and relations \eqref{Eq3.6} also appear. However,
as we will see below, the time dependence of these quantities in the
open-string orbifolds is not the same as that in the closed-string
orbifolds.

The phase-space form of the Hamiltonian
\begin{subequations}
\label{Eq3.7}
\begin{gather}
\hat{H}_\s = \int_0^\pi \!\!d\xi \hat{\sh}_\s (\hx (\xi) ,\hp (\xi)) ,\quad
\srange \\
\hat{\sh}_\s (\hx(\xi),\hp(\xi)) \!=\! \frac{1}{4\pi} \sG^{\nrm;\nsn} (\s)
\left( \hj_\nrm^{(+)} (\xi,t) \hj_\nsn^{(+)} (\xi,t) \!+\!
   \hj_\nrm^{(-)} (\xi,t) \hj_\nsn^{(-)} (\xi,t) \right) \nn \\
\quad = 2\pi \hG^{\nrm;\nsn} (\hx) \hp_\nrm (\hb) \hp_\nsn (\hb) \!+\!
\frac{1}{8\pi} \pl_\xi \hx_\s^\nrm \pl_\xi \hx_\s^\nsn \hG_{\nrm;\nsn}
(\hx) \label{Eq 3.7b}
\end{gather}
\begin{gather}
\hG_{\nrm;\nsn} (\hx) \equiv \he_\nrm{}^{\!\ntd} \he_\nsn{}^{\!\nue}
\sG_{\ntd;\nue} (\s) = \heb_\nrm{}^{\!\ntd} \heb_\nsn{}^{\!\nue}
\sG_{\ntd;\nue} (\s) \label{Eq 3.7c} \\
\hG^{\nrm;\nsn} (\hx) \equiv \sG^{\ntd;\nue} (\s) \hei_\ntd {}^{\!\nrm}
\hei_\nue {}^{\!\nsn} \\
\hG_{\nrm;\ntd} (\hx) \hG^{\ntd;\nsn} (\hx) = \de_\nrm {}^\nsn \label{Eq 3.7e}
\end{gather}
\end{subequations}
follows from Eq.~\eqref{Eq 3.4a} and the phase-space realization. In
Eq.~\eqref{Eq3.7}, the quantities $\hG_{\bullet}$ and $\hG^{\bullet}$ are
the twisted Einstein metric and its inverse respectively. More general
phase-space realizations of the currents, with the same Hamiltonian (i.~e.
T-dual formulations), will be discussed in a future paper.

As another application of the phase-space realization \eqref{Eq3.5}, we
obtain the following boundary conditions in terms of the phase-space
variables
\begin{subequations}
\label{Eq3.8}
\begin{gather}
4\pi (\heb^{-1}_\nrm{}^\nsn - \he^{-1}_\nrm {}^\nsn ) \hp_\nsn (\hb) =
\bigspc \bigspc \nn \\
\bigspc =\pl_\xi \hx_\s^{\nsn} (\heb_\nsn {}^\ntd + \he_\nsn {}^\ntd
)\sG_{\ntd;\nrm} (\s) \text{  at } \xi=0 \\
4\pi ( e^{-\tp\nrrs} \heb^{-1}_\nrm {}^\nsn - \he^{-1}_\nrm {}^\nsn
)\hp_\nsn (\hb) = \bigspc \bigspc \nn \\
\bigspc =\pl_\xi \hx_\s^\nsn (e^{-\tp\nrrs} \heb_\nsn{}^\ntd +
\he_\nsn{}^\ntd )\sG_{\ntd;\nrm} (\s) \text{  at } \xi=\pi.
\end{gather}
\end{subequations}
from the current boundary conditions in Eq.~\eqref{Eq2.6}.

\subsection{The Brackets of $\hj^{(\pm)}$ with $\hx$}

In the method of Ref.~\cite{Giusto}, all phase-space brackets are obtained by
solving partial differential equations derived from the current algebra and
the postulated phase-space realization of the currents.

Following this method, we begin our analysis of the open-string phase-space
brackets by writing down the so-called inverse relations
\begin{subequations}
\label{Eq3.9}
\begin{gather}
\pl_\xi \hx_\s^\nrm (\xi) = \hj_\nsn^{(+)} (\xi) \sG^{\nsn;\ntd} (\s)
\he^{-1} (\xi)_\ntd{}^\nrm \bigspc \bigspc \nn \\
\bigspc -\hj_\nsn^{(-)} (\xi) \sG^{\nsn;\ntd} (\s) \heb^{-1}
(\xi)_\ntd{}^\nrm ,\quad 0\leq \xi \leq \pi \label{Eq 3.9a} \\
\hp_\nrm (\hb,\xi) = \frac{1}{4\pi} \left( \he(\xi)_\nrm{}^\nsn
\hj_\nsn^{(+)} (\xi) + \heb(\xi)_\nrm {}^\nsn \hj_\nsn^{(-)} (\xi) \right)
\label{Eq 3.9b}
\end{gather}
\end{subequations}
which follow directly from the phase-space realization in
Eq.~\eqref{Eq3.5}. We also give the spatial derivative of $\hg$
\begin{subequations}
\label{Eq3.10}
\begin{gather}
\pl_\xi \hg(\st,\xi) = \pl_\xi \hx_\s^\nrm \hpl_\nrm \hg(\st,\xi) = i
\left( \hg(\st,\xi) \hj^{(+)}(\st,\xi) +\hj^{(-)}(\st,\xi) \hg(\st,\xi)
\right) \label{Eq 3.10a} \\
\hj^{(\pm)} (\st,\xi) \equiv \hj_\nrm^{(\pm)} (\xi) \sG^{\nrm;\nsn} (\s)
\st_\nsn \label{Eq 3.10b} \\
   \hj_\nrm^{(\pm)} (\xi) = \widehat{Tr} \left( \sm(\st,\s) \hj^{(\pm)}
(\st,\xi) \st_\nrm \right)
\end{gather}
\end{subequations}
which follows via the chain rule from Eqs.~\eqref{Eq3.9} and \eqref{Eq3.6}.

In analogy with Eq.~$(4.3)$ of Ref.~\cite{Giusto}, we may now derive
differential equations for the $\{ \hj ,\hx \}$ bracket
\begin{subequations}
\label{Eq3.11}
\begin{align}
&\pl_\eta \{ \hj_\nrm^{(+)} (\xi,t,\s), \hx_\s^\nsn (\eta,t) \} = \{
\hj_\nrm^{(+)} (\xi,t,\s) ,\pl_\eta \hx_\s^\nsn (\eta,t) \} \bigspc \nn \\
&\quad= \{ \hj^{(+)} (\xi) ,\hj^{(+)} (\eta) \sG^{\bullet} (\s) \he^{-1}
(\eta) - \hj^{(-)}(\eta) \sG^{\bullet} (\s) \heb^{-1} (\eta) \} \\
&\quad =\smal{ \!\tp \left( \scf_{\nrm;\ntd}{}^{\!\!\nue} (\s)
\hj_\nue^{(+)} (\eta) \!+\! \sG_{\nrm;\ntd}(\s) \pl_\xi \right) \times }
\nn \\
& \bigspc \bigspc \bigspc \smal{ \times \de_\nrrsf (\xi \!-\!\eta)
\sG^{\ntd;n(v)\kappa}(\s) \he^{-1} (\eta)_{n(v)\kappa}{}^{\!\!\nsn} } \nn \\
&\quad \quad \smal{ -\tp \left( \scf_{\nrm;\ntd}{}^{\!\!\nue}(\s)
\hj_\nue^{(-)} (\eta) \!+\! \sG_{\nrm;\ntd}(\s) \pl_\xi \right) \times }
\nn \\
& \bigspc \bigspc \bigspc \smal{ \times \de_\nrrsf (\xi \!+\!\eta)
\sG^{\ntd;n(v)\kappa}(\s) \heb^{-1} (\eta)_{n(v)\kappa}{}^{\!\!\nsn} } \nn
\\
&\quad \quad \smal{ +\{ \hj_\nrm^{(+)} (\xi) ,\hx^\ntd_\s (\eta) \} } (
\smal{\hpl_\ntd \he^{-1} (\eta)_\nue{}^\nsn \sG^{\nue;n(v) \kappa}(\s)
\hj_{n(v)\kappa}^{(+)} (\eta) } \nn \\
&\bigspc \bigspc \bigspc  \smal{ -\hpl_\ntd \heb^{-1} (\eta)_\nue{}^\nsn
\sG^{\nue;n(v)\kappa}(\s) \hj_{n(v)\kappa}^{(-)} (\eta) } )  \label{Eq
3.11b}
\end{align}
\end{subequations}
where we have used the twisted current algebra \eqref{Eq2.7}, Eq.~\eqref{Eq
3.9a} and the chain rule. The corresponding differential equation for the
bracket $\{ \hj^{(-)}(\xi) ,\hx(\eta) \}$ can be obtained by replacing $\xi
\rightarrow -\xi$ everywhere in Eq.~\eqref{Eq3.11}.

We wish to find a solution of these differential equations which is both
locally twisted WZW and consistent with the strip boundary conditions
for $\hj^{(\pm)} (\xi)$ in Eq.~\eqref{Eq2.6}. A natural candidate for the
solution is:
\begin{subequations}
\label{Eq3.12}
\begin{gather}
\{ \hj_\nrm^{(+)} (\xi,t,\s) ,\hx_\s^\nsn (\eta,t) \} = \bigspc \bigspc
\bigspc \bigspc \quad \quad \nn \\
  \bigspc -\tp \Big{(} \he^{-1} (\eta)_\nrm{}^{\!\nsn} \de_\nrrsf(\xi
\!-\!\eta) \!+\! \heb^{-1} (\eta)_\nrm{}^{\!\nsn} \de_\nrrsf (\xi
\!+\!\eta) \Big{)}  \\
\{ \hj_\nrm^{(-)} (\xi,t,\s) ,\hx_\s^\nsn (\eta,t) \} = \bigspc \bigspc
\bigspc \bigspc \quad \quad \nn \\
  \bigspc -\tp \Big{(} \heb^{-1}(\eta)_\nrm{}^{\!\nsn} \de_{-\nrrsf}(\xi
\!-\!\eta) \!+\!\he^{-1} (\eta)_\nrm{}^{\!\nsn} \de_{-\nrrsf}(\xi
\!+\!\eta) \Big{)}
\end{gather}
\begin{gather}
\s=0,\ldots ,N_c -1 \,.
\end{gather}
\end{subequations}
According to Eqs.~(B$.13$d,e) and the concluding paragraph of App.~B, this
candidate solution is locally twisted WZW. Moreover this candidate is
consistent with the strip boundary conditions \eqref{Eq2.6} because the
relations
\begin{equation}
\de_\nrrsf (-\eta) = \de_{-\nrrsf} (\eta) ,\quad \de_\nrrsf (\pi \mp \eta)
= e^{-\tp \nrrs} \de_{-\nrrsf} (\pi \pm \eta)  \label{Eq3.13}
\end{equation}
follow from the definition of the phase-modified delta functions.

Finally, one finds after considerable algebra that the candidate in
Eq.~\eqref{Eq3.12} indeed solves the differential equations. To verify
this, one inserts
both Eq.~\eqref{Eq3.12} and the phase-space form \eqref{Eq3.5} of
$\hj^{(\pm )}$ into Eq.~\eqref{Eq3.11}, following a line similar to the
corresponding untwisted computation in Ref.~\cite{Giusto}. In particular, the
reader will find the following identities useful:
\begin{subequations}
\label{Eq3.14}
\begin{gather}
\heb_\nrm{}^\nsn = -\he_\nrm{}^\ntd \ho_\ntd{}^\nsn \\
\hei_\nrm{}^\nue \hpl_\nue \ho_\nsn{}^\ntd = -\scf_{\nrm;\nsn}{}^\nue (\s)
\ho_\nue{}^\ntd \\
\hebi_\nrm{}^\nue \hpl_\nue \ho_\nsn{}^\ntd = -\ho_\nsn{}^\nue
\scf_{\nue;\nrm}{}^\ntd (\s)
\end{gather}
\begin{gather}
\ho_\nrm{}^\ntd \ho_\nsn{}^\nue \sG_{\ntd;\nue} (\s) = \sG_{\nrm;\nsn} (\s) \\
\ho_\nrm{}^\nue \ho_\nsn{}^\nvk \scf_{\nue;\nvk}{}^\ntd (\s) =
\scf_{\nrm;\nsn}{}^\nue (\s) \ho_\nue{}^\ntd \\
\heb_\nrm{}^\nue \heb_\nsn{}^\nvk \scf_{\nue;\nvk}{}^\nwl (\s)
\hebi_\nwl{}^\ntd \bigspc \bigspc \nn \\
  \bigspc =-\he_\nrm{}^\nue \he_\nsn{}^\nvk \scf_{\nue;\nvk}{}^\nwl (\s)
\hei_\nwl{}^\ntd
\end{gather}
\begin{gather}
\hpl_\nrm \he_\nsn{}^\ntd - \hpl_\nsn \he_\nrm{}^\ntd = \he_\nrm{}^\nue
\he_\nsn{}^\nvk \scf_{\nue;\nvk}{}^\ntd (\s) \\
\hei_\nrm{}^\ntd \hpl_\ntd \hei_\nsn{}^\nue - \hei_\nsn{}^\ntd \hpl_\ntd
\hei_\nrm{}^\nue = \scf_{\nsn;\nrm}{}^\ntd (\s)
   \hei_\ntd{}^\nue \,.
\end{gather}
\end{subequations}
In this list, the quantity $\ho$ is the twisted adjoint action
\begin{gather}
\hg (\st,\xi,t,\s) \st_\nrm \hg^{-1} (\st,\xi,t,\s) \equiv \ho
(\hx(\xi))_\nrm{}^\nsn \st_\nsn \label{Eq3.15}
\end{gather}
and the relations ($3.14$g,h) are respectively the twisted Cartan-Maurer
and inverse twisted Cartan-Maurer identities (the CM identities
also hold for $\he \rightarrow \heb$). These identities and others of this
type also appear in the geometry \cite{Geom} of closed-string WZW orbifolds.

The general solution of the inhomogeneous differential equations contains
an additional term which satisfies the corresponding homogeneous equations,
but (as in Ref.~\cite{Giusto}) one finds that such a term would violate the
requirement that the strip algebra is locally twisted WZW.

In the solution \eqref{Eq3.12}, the extra boundary terms proportional to
$\de_{\pm \nrrsf} (\xi+\eta)$ can be interpreted with Ref.~\cite{Giusto} as the
interaction of a non-abelian charge at $\xi$ with a non-abelian {\it image
charge} at $-\eta$.

As an application of the $\{ \hj ,\hx \}$ brackets, we may now use the
chain rule and Eqs.~($3.6$b,c) to obtain the brackets of the twisted strip
currents
with the group orbifold elements
\begin{subequations}
\label{Eq3.16}
\begin{align}
\{ \hj_\nrm^{(+)} (\xi,t,\s) ,\hg(\st,\eta,t,\s) \} =& 2\pi \Big{(}(
\hg(\st,\eta) \st_\nrm \de_\nrrsf (\xi \!-\!\eta)  \nn \\
   &\quad \quad  - \st_\nrm \hg(\st,\eta) \de_\nrrsf (\xi \!+\!\eta) \Big{)} \\
\{ \hj_\nrm^{(-)} (\xi,t,\s) ,\hg(\st,\eta,t,\s) \} =& 2\pi \Big{(}
-\st_\nrm \hg(\st,\eta) \de_{-\nrrsf} (\xi \!-\!\eta)  \nn \\
   &\quad \quad +\hg(\st,\eta) \st_\nrm \de_{-\nrrsf} (\xi \!+\!\eta) \Big{)}
\end{align}
\end{subequations}
and we have checked with Eqs.~\eqref{Eq2.7} and \eqref{Eq 2.2f} that all
$\hj,\hj,\hg$ bracket Jacobi identities for
this algebra are satisfied. We note in App.~B that the result
\eqref{Eq3.16} is also locally twisted WZW, that is, isomorphic in the bulk
to the
corresponding rectified WZW orbifold result \eqref{EqB.14}.

The result \eqref{Eq3.16} is easily quantized because it is linear in
$\hg$, and the operator form of this result will play a central role in our
discussion of the quantum theory of open-string WZW orbifolds (see Sec.~$4$).

Finally, using the integral identities \eqref{EqC.3} and the $\{\hj ,\hg
\}$ brackets in Eq.~\eqref{Eq3.16}, we may also compute the
action of the bulk momentum \eqref{Eq 3.4b} on the group orbifold elements
\begin{align}
&i \{ \hat{P}_\s (t) ,\hg(\st,\xi,t) \} = \nn \\
& \bigspc =\left\{ \begin{array}{ll} \!i \Big{(} \hg(\st,\xi,t) \hj^{(+)}
(\st,\xi,t) +\hj^{(-)} (\st,\xi,t) \hg(\st,\xi,t) \Big{)}
   & {\rm for}\,\; 0 <\xi <\pi \\ 0 & {\rm for}\,\; \xi=0,\pi \ \end{array}
\right.  \label{Eq3.17}
\end{align}
and we note that this result agrees in the bulk with the form of $\pl_\xi
\hg$ given in Eq.~\eqref{Eq 3.10a}.

\subsection{Coordinate Space}

We now move to coordinate space, beginning with the computation of $\pl_t
\hx_\s$:
\begin{subequations}
\label{Eq3.18}
\begin{align}
\bkspc \pl_t \hx_\s^\nrm (\xi,t) &= i\{ \hat{H}_\s (t),\hx_\s^\nrm (\xi,t)
\} \nn \\
&=\! \sG^{\nsn;\ntd} (\s) \left( \hei (\xi)_\nsn{}^{\!\!\nrm}
\hj_\ntd^{(+)} (\xi) + \!\heb^{-1}(\xi)_\nsn{}^{\!\!\nrm} \hj_\ntd^{(-)}
(\xi) \right) \label{Eq 3.18a} \\
&= 4\pi \hG^{\nrm;\nsn} (\hx(\xi,t)) \hp_\nsn (\hB ,\xi,t)  \label{Eq 3.18b} \\
\bkspc \hp_\nrm (\hB,\xi,t) &= \frac{1}{4\pi} \hG_{\nrm;\nsn} (\hx(\xi,t))
\pl_t \hx_\s^\nsn (\xi,t) \,. \label{Eq 3.18c}
\end{align}
\end{subequations}
Here we have used the form of the Hamiltonian in Eq.~\eqref{Eq3.7} and the
$\{ \hj ,\hx \}$ brackets \eqref{Eq3.12}, as well as Eqs.~\eqref{EqC.3},
\eqref{Eq 3.7c} and \eqref{Eq 3.9b}. These relations between $\pl_t \hx_\s$
and $\hp (\hB)$ are locally twisted WZW, i.~e.~they are the same as
those found in closed-string WZW orbifold theory \cite{Geom}. This is not
surprising because the locally twisted WZW requirement also played a role
in
determining the form \eqref{Eq3.12} of the $\{ \hj ,\hx \}$
brackets\footnote{In fact, the locally twisted WZW property of the results
\eqref{Eq3.18}
tells us that our solution \eqref{Eq3.12} for $\{ \hj ,\hx \}$ is {\it a
posteriori} correct -- even without consideration of the rectification
problem (see App.~B).}, which were used in this computation.

As an application of the result \eqref{Eq 3.18a}, we use the chain rule and
Eq.~\eqref{Eq3.6} to derive an equation for the time derivative of the
group orbifold element
\begin{subequations}
\label{Eq3.19}
\begin{gather}
\pl_t \hg(\st,\xi,t,\s) = i\left( \hg(\st,\xi,t,\s) \hj^{(+)}
(\st,\xi,t,\s) - \hj^{(-)} (\st,\xi,t,\s) \hg(\st,\xi,t,\s) \right)
   \label{Eq 3.19a} \\
\pl_\xi \hg(\st,\xi,t,\s) = i\left( \hg(\st,\xi,t,\s) \hj^{(+)}
(\st,\xi,t,\s) + \hj^{(-)}(\st,\xi,t,\s) \hg(\st,\xi,t,\s) \right) \\
\pl_+ \hg(\st,\xi,t,\s) = 2i\hg(\st,\xi,t.\s) \hj^{(+)} (\st,\xi,t,\s) \\
\pl_- \hg(\st,\xi,t,\s) = -2i\hj^{(-)} (\st,\xi,t,\s) \hg(\st,\xi,t,\s)
\end{gather}
\end{subequations}
where the matrix currents $\hj^{(\pm)} (\st)$ are defined in Eq.~\eqref{Eq
3.10b}. For symmetry, we have also included the corresponding spatial
derivative \eqref{Eq 3.10a} of $\hg$. Note that these equations of motion
are the same as those of closed-string WZW orbifold theory \cite{Big,Geom}
and hence are locally twisted WZW.

Similarly, we may substitute Eq.~\eqref{Eq 3.18c} into the phase-space
realization \eqref{Eq3.5} to obtain the coordinate-space form of the
twisted
strip currents:
\begin{subequations}
\label{Eq3.20}
\begin{gather}
\hj_\nrm^{(+)} (\xi,t,\s) = \frac{1}{2} \pl_+ \hx_\s^\nsn (\xi,t)
\he(\xi,t)_\nsn{}^\ntd \sG_{\ntd;\nrm} (\s) \\
\hj_\nrm^{(-)} (\xi,t,\s) = \frac{1}{2} \pl_- \hx_\s^\nsn (\xi,t) \heb(\xi,t)_\nsn{}^\ntd \sG_{\ntd;\nrm} (\s) \\ \hj^{(+)} (\st,\xi,t,\s) 
= -\frac{i}{2} \hg^{-1} (\st,\xi,t,\s) \pl_+ \hg(\st,\xi,t,\s) \\
\hj^{(-)} (\st,\xi,t,\s) = -\frac{i}{2} \hg(\st,\xi,t,\s) \pl_- \hg^{-1}
(\st,\xi,t,\s) \\
\pl_- \left( \hg^{-1} (\st) \pl_+ \hg(\st) \right) = \pl_+ \left( \hg(\st)
\pl_- \hg^{-1} (\st) \right) =0 \,. \label{Eq 3.20e}
\end{gather}
\end{subequations}
The relations in Eqs.~($3.20$c,d) may also be obtained by inverting
Eqs.~($3.19$c,d). The results in Eq.~\eqref{Eq 3.20e} follow from
Eqs.~($3.20$c,d) and the conservation \eqref{Eq 2.5b} of the currents.

We may also define a {\it bulk Lagrange density} for each twisted open WZW
string by the usual Legendre transformation
\begin{subequations}
\label{Eq3.21}
\begin{align}
& \hat{{\cL}}_\s \equiv \pl_t \hx_\s^\nrm \hp_\nrm - \hat{\sh}_\s \nn \\
& \bigspc \quad \quad = \frac{1}{8\pi} (\hG_{\nrm;\nsn} \!+\!
\hB_{\nrm;\nsn}) \pl_+ \hx^\nrm_\s \pl_- \hx^\nsn_\s ,\quad \quad 0< \xi
<\pi \label{Eq 3.21a} \\
& \frac{1}{8\pi} \hG_{\nrm;\nsn} \pl_+ \hx^\nrm_\s \pl_- \hx^\nsn_\s =
-\frac{1}{8\pi} \widehat{Tr}
  \left( \sm (\st,\s) \hg^{-1} (\st) \pl_+ \hg(\st) \hg^{-1} (\st) \pl_-
\hg(\st) \right) \label{Eq 3.21b}
\end{align}
\end{subequations}
where $\sh_\s$ is the Hamiltonian density in Eq.~\eqref{Eq 3.7b}, and $\sm
(\st,\s)$ in Eq.~\eqref{Eq 3.21b} is the twisted data matrix encountered
in Eq.~\eqref{Eq 2.2g}. As expected, this Lagrange density is locally
twisted WZW, i.~e.~isomorphic in the bulk to the twisted sigma model
form of the closed-string WZW orbifold Lagrange density \cite{Geom}.

A complete action formulation of {\it untwisted} open WZW strings (in
terms of group elements and extra boundary variables) was given in Ref.~\cite{GT-TNB},
but the complete action formulation of the basic class of twisted open WZW
 strings is an open problem. We note in passing, however, that the complete
action formulation is in fact known for the twisted open-string sectors
of the WZW orientation orbifolds \cite{Orient2}. This action is expressed in
terms of group orbifold elements alone on the solid half cylinder, but we
remind the reader that the open-string orientation orbifold sectors are not
 included in the basic class of twisted open strings.

\subsection{The Generalized WZW Branes of $A_g^{open}(H)/H$}

Using Eq.~\eqref{Eq 3.18c}, we find that the phase-space form \eqref{Eq3.8}
of the boundary conditions can be written in the following two
coordinate-space forms
\begin{gather}
\hj_\nrm^{(+)} (0,t,\s) = \hj_\nrm^{(-)} (0,t,\s) ,\quad \hj_\nrm^{(+)}
(\pi,t,\s) = e^{-\tp \nrrs} \hj_\nrm^{(-)} (\pi,t,\s) \label{Eq3.22}
\end{gather}
\begin{subequations}
\label{Eq3.23}
\begin{align}
&\pl_t \hx_\s^{\nrm}(\xi,t) \Big{(} \heb (\xi,t)_\nrm{}^{\!\nsn} -
\he(\xi,t)_\nrm{}^{\!\nsn} \Big{)} \nn \\
&   \bigspc  =\pl_\xi \hx^\nrm (\xi,t) \Big{(} \heb (\xi,t)_\nrm{}^{\!\nsn}
+\he (\xi,t)_\nrm{}^{\!\nsn} \Big{)} \,\,\, \text{at } \xi=0\\
&\pl_t \hx_\s^{\nrm}(\xi,t) \Big{(} e^{\tp \srac{n(s)}{\r(\s)}} \heb
(\xi,t)_\nrm{}^{\!\nsn} - \he(\xi,t)_\nrm{}^{\!\nsn} \Big{)} \nn \\
&   \bigspc =\pl_\xi \hx^\nrm (\xi,t) \Big{(} e^{\tp \srac{n(s)}{\r(\s)}}
\heb (\xi,t)_\nrm{}^{\!\nsn} +\he (\xi,t)_\nrm{}^{\!\nsn} \Big{)}
\,\,\,\text{at } \xi=\pi
\end{align}
\end{subequations}
where the form in Eq.~\eqref{Eq3.22} appeared earlier in Eq.~\eqref{Eq2.6}.
The equivalence of Eqs.~\eqref{Eq3.22} and \eqref{Eq3.23} also follows
directly from Eq.~\eqref{Eq3.20}.

These coordinate-space boundary conditions can be reexpressed in terms of
the matrix currents and/or the group orbifold elements:
\begin{subequations}
\label{Eq3.24}
\begin{gather}
\hj^{(+)} (\st,0,t,\s) \!=\! \hj^{(-)} (\st,0,t,\s) ,\,\,\,\,
\hj^{(+)}(\st,\pi,t,\s) \!=\! E(T,\s) \hj^{(-)}(\st ,\pi,t,\s) E(T,\s)^\ast
\label{Eq 3.24a}
\end{gather} \vspace{-0.1in}
\begin{align}
\label{Eq 3.24b}
\!\!\!&\!\!\!\hg^{-1} (\st,\xi,t,\s) \pl_+ \hg(\st,\xi,t,\s)  \nn \\
& \quad \quad =\left \{ \begin{array}{ll} \hg(\st,\xi,t,\s) \pl_- \hg^{-1}
(\st,\xi,t,\s) & \text{at } \xi=0 \\
 E(T,\s) \hg(\st,\xi,t,\s) \pl_- \hg^{-1} (\st,\xi,t,\s) E(T,\s)^\ast &
\text{at } \xi=\pi \end{array} \right. \\
&\bigspc \bigspc \quad \quad \srange \,.
\end{align}
\end{subequations}
These results follow from Eqs.~\eqref{Eq 3.10b}, ($3.20$c,d),
\eqref{Eq3.22} and the selection rule \eqref{Eq 2.2e} for the twisted
representation
matrices. We remind that $E(T,\s)$ is the eigenvalue matrix of the extended
$H$-eigenvalue problem in Eq.~\eqref{Eq2.3}.

The first part of Eq.~\eqref{Eq 3.24b} tells us that each twisted open WZW
string ends at $\xi=0$ on an ordinary WZW brane \cite{AS,Giusto}. The second
part of Eq.~\eqref{Eq 3.24b} tells us however that the open string ends at
$\xi =\pi$ on what may be called a {\it generalized WZW brane} -- different
from the ordinary WZW brane at $\xi=0$. Indeed, the twisting of our open
WZW strings can be understood as a consequence of having different branes
at each end of the string\footnote{The prototype of this situation -- a
twisted open string with Neumann boundary conditions at one end and
Dirichlet
at the other -- was first studied by Siegel in Ref.~\cite{Siegel}.}.

We also give the form of the generalized WZW branes in terms of matrix elements
\begin{align}
\!\!\!&\!\!\!\left( \hg^{-1} (\st,\xi,t,\s) \pl_+ \hg(\st,\xi,t,\s)
\right)_\Nrm {}^{\Nsn} \nn \\
& \bigspc =\left\{ \begin{array}{ll} \left( \hg(\st,\xi,t,\s) \pl_-
\hg^{-1}(\st,\xi,t,\s) \right)_\Nrm{}^{\Nsn} & \text{at } \xi=0 \\
     e^{-\tp \frac{N(r)-N(s)}{R(\s)}} \left( \hg(\st,\xi,t,\s) \pl_-
\hg^{-1} (\st,\xi,t,\s) \right)_\Nrm{}^{\Nsn} &
     \text{at } \xi=\pi  \end{array} \right. \label{Eq3.25}
\end{align}
where we have used the form of $E(T,\s)$ in Eq.~\eqref{Eq2.3} and the fact
that the index structure of $\hg(\st)$ is the same as that of $\st$.

The description of the generalized WZW branes in Eqs.~\eqref{Eq3.24},
\eqref{Eq3.25} is a central result of this paper.

As an explicit example, we give the matrix element form of the branes in
sector $\s$ of the general {\it open-string WZW permutation orbifold}:
\begin{subequations}
\label{Eq3.26}
\begin{gather}
\nrm \rightarrow \hat{j}aj ,\quad \Nrm \rightarrow \hat{j}\a j ,\quad \nrrs
= \srac{N(r)}{R(\s)} =\srac{\hat{j}}{f_j(\s)} \\
\st_{\hat{j}aj}(T,\s) = T_a t_{\hat{j}j}(\s) ,\quad t_{\hat{j}j}
(\s)_{\hat{l}l}{}^{\hat{m}m} =\de_{jl} \de_l{}^m \de_{\hat{j}+\hat{l}
-\hat{m},0\,
   \text{mod }f_j(\s)} \label{Eq 3.26b} \\
[T_a ,T_b] =if_{ab}{}^c T_c ,\quad t_{\hat{j}j} (\s) t_{\hat{l}l}(\s) =
\de_{jl} t_{\hat{j} +\hat{l},j} (\s) \\
\hg(\st,\xi,t,\s)_{\hat{j}\a j}{}^{\hat{l}\be l} =\de_j{}^l \hg_j
(\st,\xi,t,\s)_{\hat{j}\a}{}^{\hat{l}\be} \label{Eq 3.26d} \\
\left( \hg^{-1} (\st,\xi,t,\s) \pl_+ \hg(\st,\xi,t,\s) \right)_{\hat{j}\a
j}{}^{\hat{l}\be l} =\de_j{}^l \left( \hg_j^{-1} (\st,\xi,t,\s) \pl_+
   \hg_j(\st,\xi,t,\s) \right)_{\hat{j}\a}{}^{\hat{l}\be} \bigspc \nn \\
\bigspc = \de_j{}^l \left\{ \begin{array}{ll} \left( \hg_j(\st,\xi,t,\s)
\pl_- \hg_j^{-1}(\st,\xi,t,\s) \right)_{\hat{j}\a}{}^{\hat{l}\be} &
\text{at } \xi=0 \\
  e^{-\tp \frac{\hat{j}-\hat{l}}{f_j (\s)}}  \left( \hg_j(\st,\xi,t,\s)
\pl_- \hg_j^{-1}(\st,\xi,t,\s) \right)_{\hat{j}\a}{}^{\hat{l}\be} &
\text{at } \xi=\pi
   \end{array} \right. \\
g =\oplus_I \gfrak^I ,\quad \gfrak^I \simeq \gfrak ,\quad T^I \simeq T \\
a= 1,\ldots ,\text{dim }\gfrak ,\,\,\, \a =1,\ldots ,\text{dim }T ,\,\,\,
\bar{\hat{j}} =0 ,\ldots ,f_j(\s) \!-\!1 ,\,\,\, \srange \,.
\end{gather}
\end{subequations}
Here $T$ is any irrep of Lie $\gfrak$ which labels an integrable irrep at
level $k$ of affine $\gfrak$. This twisted open string is obtained when we
appropriate our initial data from sector $\s$ of the general closed-string
WZW permutation orbifold \cite{Big,Big',Perm} on semisimple $g$. In this case,
the twisted current algebra of sector $\s$
\begin{gather}
[\hj_{\hat{j}aj} (m\!+\!\srac{\hat{j}}{f_j(\s)}) , \hj_{\hat{l}bl}
(n\!+\!\srac{\hat{l}}{f_l(\s)})] =\de_{jl} \Big{(} if_{ab}{}^c
   \hj_{\hat{j} +\hat{l},cj} (m\!+\!n \!+\!\srac{\hat{j}+\hat{l}}{f_j(\s)})
+ \bigspc \nn \\
\bigspc \bigspc \bigspc +\eta_{ab} kf_j(\s) (m\!+\!\srac{\hat{j}}{f_j(\s)})
\de_{m+n+\frac{\hat{j}+\hat{l}}{f_j(\s)} ,0} \Big{)} \label{Eq3.27}
\end{gather}
is known as a general orbifold affine algebra [1,3,5-7,9]. The results \eqref{Eq3.26}, \eqref{Eq3.27} 
are given in the cycle notation
for permutation orbifolds \cite{Big',Perm}, where $f_j(\s)$ is the length of
cycle $j$ and $\hat{j}$ labels the position in each cycle, e.~g.~:
\begin{subequations}
\begin{align}
& \Zint_\l :   f_j(\s)\! = \!\rho(\s), \,\, \bar{\hat{j}} = 0,\ldots,
\rho(\s)\! -\!1, \,\, j = 0,\ldots,
   \srac{\lambda}{\rho(\s)}\! -\!1, \,\,\s = 0,\ldots, \rho(\s)\!-\!1
\label{Eq 3.28a} \\
& \Zint_\l, \,\,\l = \text{prime}: \quad \r(\s) = \l , \quad
\bar{\hat{j}}\! =\! 0,\dots, \l \!-\!1,
\quad  j\!=\!0, \quad \s =1, \ldots, \l -1 \, \\
& S_N :\,\, f_j(\s) = \s_j, \quad \s_{j+1} \leq \s_j, \quad j = 0, \dots,
n(\vec{\s})-1,
\quad\sum_{j=0}^{n(\vec{\s})-1} \s_j = N \,. \label{Eq 3.28c}
\end{align}
\end{subequations}
The cycle-diagonal form of the group orbifold element $\hg$ in
Eq.~\eqref{Eq 3.26d} follows as in Ref.~\cite{Perm} from the form of $\st$ in
Eq.~\eqref{Eq 3.26b} and the exponential form \eqref{Eq 3.6a}.

Similarly, the explicit data \cite{Big,Big',Perm,so2n} for the various
closed-string orbifolds on simple $g$ can be substituted in Eq.~\eqref{Eq
3.24b} to obtain the
branes of the corresponding open-string WZW orbifolds.

\subsection{The Non-Commutative Geometry of $A_g^{open}(H)/H$}

In this subsection, we continue our phase-space construction to determine
the equal-time coordinate brackets
\begin{equation}
\hD_\s^{\nrm;\nsn} (\xi,\eta) \equiv \hD^{\nrm;\nsn} (\xi,\eta,t,\s) = \{
\hx_\s^\nrm (\xi,t) ,\hx_\s^\nsn (\eta,t) \}   \label{Eq3.29}
\end{equation}
which will generalize the open-string non-commutative geometry of
Ref.~\cite{Giusto}.

To begin, we use the inverse relations \eqref{Eq3.9} and the $\{ \hj, \hx
\}$ brackets \eqref{Eq3.12} to find the following partial differential
equation for $\hD_\s$:
\begin{subequations}
\label{Eq3.30}
\begin{align}
&\pl_\eta \hD_\s^{\nrm;\nsn} (\xi,\eta) \nn \\
& \,\,\, =\!\tp \sG^{\ntd;\nue}(\s) \!\Big{(} \hebi(\xi)_\ntd{}^{\!\!\nrm}
\hei(\eta)_\nue{}^{\!\!\nsn}  \nn \\
& \quad \quad -\hei(\xi)_\nue{}^{\!\!\nrm} \hebi(\eta)_\ntd{}^{\!\!\nsn}
\Big{)} \de_{\frac{n(t)}{\r(\s)}} (\xi \!+\!\eta)
   +\!\hD_\s^{\nrm;\ntd}(\xi,\eta) \hL_\s (\eta)_\ntd{}^\nsn \\
& \,\,\, =\tp \hPs_\s^{\nrm;\nsn} (\xi,\eta) \de (\xi+\eta) +
\hD_\s^{\nrm;\ntd} (\xi,\eta) \hL_\s (\eta)_\ntd{}^\nsn \label{Eq 3.30b}
\end{align}
\begin{align}
&\hPs_\s^{\nrm;\nsn}(\xi,\eta) \equiv \hPs^{\nrm;\nsn} (\xi,\eta,t,\s) \\
   &\,\,\, =\sG^{\ntd;\nue}(\s) e^{-i \frac{n(t)}{\r(\s)}(\xi+\eta)} \times
\nn \\
   &\bigspc \quad \quad \quad \times \left( \hebi(\xi)_\ntd{}^{\!\!\!\nrm}
\hei(\eta)_\nue{}^{\!\!\!\nsn}
   \!-\hei(\xi)_\nue{}^{\!\!\!\nrm} \hebi(\eta)_\ntd{}^{\!\!\!\nsn} \right)
\nn \\
&\hL_\s (\eta)_\ntd{}^\nsn \equiv \hL (\eta,t,\s)_\ntd{}^\nsn \nn \\
   &\,\,\, =\sG^{\nue;\nvk}(\s) \left( \hj_\nue^{(+)} (\eta) \hpl_\ntd
\hei(\eta)_\nvk{}^{\!\!\!\nsn} \!-\! \hj_\nue^{(-)} (\eta) \hpl_\ntd
     \hebi(\eta)_\nvk{}^{\!\!\!\nsn} \right) \,.
\end{align}
\end{subequations}
The corresponding $\pl_\eta \hD_\s$ equation is obtained from this result
by noting that the bracket $\hD_\s$ is totally anti-symmetric under the
exchange
$\xi \leftrightarrow \eta ,\, \nrm \leftrightarrow \nsn$.

In a compact matrix notation \cite{Giusto}, these PDEs read
\begin{subequations}
\label{Eq3.31}
\begin{gather}
\pl_\eta \hD_\s (\xi,\eta) = \tp \hPs_\s (\xi,\eta) \de(\xi+\eta) + \hD_\s
(\xi,\eta) \hL_\s (\eta) \\
\pl_\xi \hD_\s (\xi,\eta) = \tp \hPs_\s (\xi,\eta) \de(\xi+\eta) + \hL_\s^t
(\xi) \hD_\s (\xi,\eta) \\
\hPs_\s^t (\eta,\xi) = - \hPs_\s (\xi,\eta)
\end{gather}
\end{subequations}
where $t$ is matrix transpose. The integrability condition for this system is:
\begin{subequations}
\label{Eq3.32}
\begin{gather}
\pl_\xi \pl_\eta \hD_\s (\xi,\eta) = \pl_\eta \pl_\xi \hD_\s (\xi,\eta)
\,\,\,\text{ iff } \\
\left( \pl_\eta \hPs_\s(\xi,\eta) -\hPs_\s (\xi,\eta) \hL_\s(\eta) \right)
\de(\xi+\eta) =\left( \pl_\xi \hPs_\s(\xi,\eta) -\hL_\s^t (\xi)
   \hPs_\s (\xi,\eta) \right) \de(\xi+\eta) \,.
\end{gather}
\end{subequations}
Using the definitions of
$\hPs$ and $\hL$ in \eqref{Eq3.30}, the phase-space realization
\eqref{Eq3.5} of the strip currents and the identities in \eqref{Eq3.14},
we have been
able to confirm (after considerable algebra) that the integrability
condition is satisfied. The inhomogeneous terms in Eq.~\eqref{Eq3.31} are
boundary terms associated to the interaction between a non-abelian charge
at $\xi$ and a non-abelian {\it image charge} at $-\eta$. Note that,
because of
these terms, the closed-string WZW orbifold bracket $\hD_{WZW}
(\xi,\eta,t,\s) =0$ is not a solution of the $\{ \hx ,\hx \}$ PDEs.

We may now follow Ref.~\cite{Giusto} to find the following integral
representation for the solution of Eq.~\eqref{Eq3.31}:
\begin{subequations}
\label{Eq3.33}
\begin{align}
\hD_\s (\xi,\eta) =& \hU_\s^t (\xi) \hD_\s (0,0) \hU_\s(\eta)  \nn \\
&+ \!\pi i \! \int_0^\eta \!d\eta' \left( \hPs_\s (\xi,\eta')
\de(\xi+\eta')+ \hU_\s^t (\xi) \hPs_\s (0,\eta') \de (\eta') \right)
   \hU_\s^{-1} (\eta') \hU_\s(\eta) \nn \\
&+ \! \pi i\hU_\s^t (\xi) \!\int_0^\xi \!d\xi' \, \hU_\s^{-1 t} (\xi')
\left( \hPs_\s (\xi',\eta) \de(\xi'+\eta) + \hPs_\s (\xi',0)
    \de(\xi') \hU_\s (\eta) \right)
\end{align}
\begin{gather}
\pl_\xi \hU_\s (\xi) = \hU_\s (\xi) \hL_\s(\xi) ,\quad \pl_\eta \hU_\s^t
(\eta) = \hL_\s^t (\eta) \hU_\s^t (\eta) \\
\hU_\s(0) =\hU_\s^t (0) =\one ,\quad \hU_\s(\xi) \equiv \hU(\xi,t,\s)
,\quad \hU_\s^t (\eta) \equiv \hU^t (\eta,t,\s) \,.
\end{gather}
\end{subequations}
In this solution, we have introduced the ordered product $\hU_\s$ in each
sector $\s$, and only the quantity $\hD_\s(0,0) = \hD(0,0,t,\s)$ is
undetermined. With the integrability condition \eqref{Eq3.32}, it is
straightforward to check directly that Eq.~\eqref{Eq3.33} solves
Eq.~\eqref{Eq3.31}.
Moreover, the required antisymmetry of the coordinate bracket
\begin{equation}
\hD_\s^t (\eta,\xi) = -\hD_\s (\xi,\eta) \label{Eq3.34}
\end{equation}
is guaranteed by the corresponding antisymmetric choice $\hD_\s^t (0,0)
=-\hD_\s (0,0)$.

We may also evaluate the solution \eqref{Eq3.33} explicitly, using the
following integral identity
\begin{equation}
\int_0^\eta d\eta' f(\xi,\eta') \de(\xi+\eta') = \left\{ \begin{array}{ll}
\srac{1}{2} f(0,0) & \text{if } \xi=0 ,\, \eta \neq 0 \\
\srac{1}{2} f(\pi,\pi) & \text{if } \xi=\eta=\pi \\
0 & \text{otherwise } \end{array} \right. \label{Eq3.35}
\end{equation}
to obtain the result:
\begin{equation}
\hD_\s (\xi,\eta) = \left\{ \begin{array}{ll} \hD_\s(0,0) & \text{if }
\xi=\eta=0, \\
\hU_\s^t (\pi) ( \hD_\s(0,0)+i\pi \hPs_\s (0,0) )\hU_\s(\pi) + i\pi
\hPs_\s(\pi,\pi) & \text{if } \xi=\eta=\pi, \\
\hU_\s^t (\xi) (\hD_\s(0,0) +i\pi \hPs_\s(0,0)) \hU_\s(\eta) &
\text{otherwise.} \end{array} \right. \label{Eq3.36}
\end{equation}
As in Ref.~\cite{Giusto}, this expression for the coordinate bracket is suitable
for the computation of $\xi$-derivatives in the bulk, but one must
return to the form given in Eq.~\eqref{Eq3.33} in order to compute
$\xi$-derivatives at the boundary.

The preferred choice for $\hD_\s(0,0)$ is therefore
\begin{equation}
\hD_\s(0,0) = -i\pi \hPs_\s(0,0)  \label{Eq3.37}
\end{equation}
because in this case the twisted equal-time coordinate brackets {\it vanish
in the bulk}, and are therefore locally twisted WZW:
\begin{subequations}
\label{Eq3.38}
\begin{align}
& \{ \hx_\s^\nrm (\xi) ,\hx_\s^\nsn(\eta) \} = \hD_\s^{\nrm;\nsn}(\xi,\eta)
=\left\{ \begin{array}{ll} -i\pi\hPs_\s^{\nrm;\nsn} (0,0)
   &\text{if } \xi=\eta=0, \\
   i\pi \hPs_\s^{\nrm;\nsn} (\pi,\pi) & \text{if } \xi=\eta=\pi, \\ 0 &
\text{otherwise} \end{array} \right.
\end{align}
\begin{align}
\!\!\!&\hPs_\s^{\nrm;\nsn}(\xi,\xi) = -\hPs_\s^{\nsn;\nrm} (\xi,\xi) \\
&\, =\!\sG^{\ntd;\nue}(\s) e^{-2i\frac{n(t)}{\r(\s)}\xi} \!\left(
\hebi(\xi)_\ntd{}^{\!\!\!\nrm} \hei(\xi)_\nue{}^{\!\!\!\nsn}
   \!-\! \hei(\xi)_\nue{}^{\!\!\!\nrm} \hebi(\xi)_\ntd{}^{\!\!\!\nsn}
\right) \quad \quad \quad \nn \\
&\,\, =\!\hei (\xi)_\nue{}^{\!\!\!\nrm} \sG^{\nue;\ntd} (\s) \times \nn \\
& \bigspc \times \Big{(} e^{-2i \frac{n(t)}{\r(\s)}\xi}
\ho^{-1}(\xi)_\ntd{}^{\!\!\!\nvk}
   \!-\!\ho (\xi)_\ntd{}^{\!\!\!\nvk} e^{2i\frac{n(v)}{\r(\s)}\xi} \Big{)}
\hei(\xi)_\nvk{}^{\!\!\!\nsn}  \\
& \bigspc \bigspc \bigspc \bigspc \srange \,.
\end{align}
\end{subequations}
These twisted coordinate brackets, which describe the new {\it twisted
non-commutative geometry} of the open-string WZW orbifold
$A_g^{open}(H)/H$, are
a central result of this paper.

In the untwisted sector ($\s=0$) of $A^{open}_g (H)/H$, the twisted
coordinate brackets reduce to the following
\begin{subequations}
\label{Eq3.39}
\begin{gather}
\sG (\s=0) = G ,\quad \he (\xi,\s=0) = e(\xi) ,\quad \hPs_0 (\xi,\eta)
=\Psi (\xi,\eta) ,\quad \hD_0 (\xi,\eta) = \Delta (\xi,\eta) \\
\{ x^i (\xi) ,x^j (\eta) \} = \Delta^{ij} (\xi,\eta) = \left\{
\begin{array}{ll} -i\pi \Psi^{ij} (0,0) & \text{if } \xi=\eta=0, \\
   i\pi \Psi^{ij} (\pi,\pi) & \text{if } \xi=\eta=\pi, \\ 0 &
\text{otherwise,} \end{array} \right. \\
\Psi^{ij} (\xi,\xi) = \bar{e} (\xi)_a{}^i G^{ab} e(\xi)_b{}^j -
e(\xi)_a{}^i G^{ab} \bar{e} (\xi)_b{}^j \nn \\
\quad \quad \quad \quad = e(\xi)_a{}^i G^{ac} (\Omega(\xi)^{-1}
-\Omega(\xi) )_c{}^b e(\xi)_b{}^j
\end{gather}
\end{subequations}
which is recognized as the untwisted non-commutative geometry of
Ref.~\cite{Giusto}, now for the special case of the untwisted open WZW string
$A_g^{open} (H)$ with a symmetry $H$.

Using Eqs.~\eqref{Eq3.6}, \eqref{Eq3.38} and the chain rule, we have also
computed the equal-time brackets of the group orbifold elements among
themselves
\begin{subequations}
\label{Eq3.40}
\begin{align}
&\{ \hg (\st,\xi)_\Nrm{}^{\!\!\Nsn} ,\hg (\st' ,\eta)_\Nrpm{}^{\!\!\Nspn}
\} = \bigspc \bigspc \\
&\quad \left\{ \begin{array}{ll} \smal{ \!i\pi (\hg(\st,0) \st_\ntd
)_\Nrm{}^{\!\!\Nsn} (\hg(\st',0) \st^\prime_\nue )_\Nrpm{}^{\!\!\Nspn}
  \times }& \\
  \quad \quad \smal{\sG^{\ntd;\nvk}(\s)
(\ho^{-1}(0)-\ho(0))_\nvk{}^{\!\!\nue} } &\smal{ \text{if }\xi=\eta=0 ,} \\
\smal{ \!-i\pi (\hg(\st,\pi) \st_\ntd )_\Nrm{}^{\!\!\Nsn} (\hg(\st',\pi)
\st^\prime_\nue )_\Nrpm{}^{\!\!\Nspn} \times } & \\
   \quad \quad \smal{ \sG^{\ntd;\nvk}(\s) (e^{-\tp \frac{n(v)}{\r(\s)}}
\ho^{-1}(\pi)_\nvk{}^{\!\!\nue} -\ho(\pi)_\nvk{}^{\!\!\nue}
   e^{\tp \frac{n(u)}{\r(\s)}} ) } & \smal{\text{if } \xi=\eta=\pi ,} \\
\smal{\!0} & \smal{\text{otherwise}} \end{array} \right. \nn
\end{align}
\begin{gather}
\hg(\st,\xi) \equiv \hg(\st,\xi,t,\s)
\end{gather}
\end{subequations}
where $\ho$ is the adjoint action defined in Eq.~\eqref{Eq3.14}.

All other phase-space brackets
\begin{gather}
\{ \hx^\nrm_\s (\xi,t) ,\hp_\nsn (\hb,\eta,t) \} ,\quad \{ \hx^\nrm_\s
(\xi,t) ,\hp^\s_\nsn (\eta,t) \} ,\quad \{ \hj_\nrm^{(\pm)} (\xi,t),
   \hp_\nsn (\hb,\eta,t) \}, \nn \\
\{ \hp_\nrm (\hb,\xi,t) ,\hp_\nsn (\hb,\eta,t) \} ,\quad \{ \hp^\s_\nrm
(\xi,t) ,\hp^\s_\nsn (\eta,t) \} \label{Eq3.41}
\end{gather}
can now be straightforwardly computed (in the order listed) from the
inverse relations \eqref{Eq3.9} and the known brackets
\begin{equation}
\label{Eq3.42}
\{ \hj_\nrm^{(\pm)} (\xi) ,\hj_\nsn^{(\pm)}(\eta) \} ,\quad
\{\hj_\nrm^{(\pm)} (\xi) ,\hx^\nsn_\s (\eta) \} ,\quad \{\hx^\nrm_\s
   (\xi) ,\hx^\nsn_\s (\eta) \}
\end{equation}
without solving any additional differential equations. The corresponding
results are given for the untwisted case in Ref.~\cite{Giusto} (our
sector $\s=0$), but owing to the length of these expressions, we will not
reproduce the twisted analogues here.

Some explicit examples of the new twisted non-commutative geometry are
given in App.~D, and the following subsection works out the relatively
simple quasi-canonical algebra of $\hx_\s ,\hp^\s$ for a large class of
open-string free-bosonic orbifolds.

\subsection{Example: Twisted Free-Bosonic Open Strings}

Applications of the orbifold program to closed-string free-bosonic
orbifolds (including their twisted vertex operators and classical
formulations)
can be found in Refs.~\cite{Big',Perm, Geom}. In the following paragraph, we
briefly review the classical development of closed-string free-bosonic
orbifolds.

One begins with the general untwisted left-mover free-boson sector on {\it
abelian} $g$
\begin{subequations}
\label{Eq3.43}
\begin{gather}
f_{ab}{}^c =0 ,\quad e_i{}^a = - \bar{e}_i{}^a = \de_i{}^a ,\quad B_{ij}
=H_{ijk} =0  \label{Eq 3.43a} \\
\{ J_a(m) ,J_b (n)\} =mG_{ab} \de_{m+n,0} ,\quad [T_a ,T_b] =0
\end{gather}
\end{subequations}
which may be considered as a formal abelian limit\footnote{The automorphism
group of the abelian current algebra is of course larger than the
automorphism group of the non-abelian starting point; for example, $\ws
=-\thickone$ appears in the abelian case.} of the
untwisted WZW model $A_g(H)$. Under a general automorphism $\ws \!\in \!H
\!\subset \!Aut(g)$, one finds the twisted current algebra of sector $\s$
of a closed-string free-bosonic orbifold \cite{Big',Geom}. This gives us in
particular the following left-mover input data for sector $\s$ of the
corresponding open-string free-bosonic orbifold
\begin{subequations}
\label{Eq3.44}
\begin{gather}
\scf_{\nrm;\nsn}{}^\ntd (\s) =0 ,\quad [\st_\nrm ,\st_\nsn]=0 \\
\{ \hj_\nrm (m\!+\!\nrrs ),\hj_\nsn (n\!+\!\nsrs )\} =(m\!+\!\nrrs )
\de_{m+n +\frac{n(r)+n(s)}{\r(\s)},0} \sG_{\nrm;\mnrn}(\s) \label{Eq 3.44b}
\\
\{ \hj_\nrm (m\!+\!\nrrs ) ,\hg (\st,\xi,t) \} =\hg (\st,\xi,t) \st_\nrm
e^{i(m+\nrrsf)(t+\xi)} \bigspc \nn \\
  \bigspc \bigspc \bigspc \quad -\st_\nrm \hg (\st,\xi,t) e^{i(m+\nrrsf
)(t-\xi)} \\
\he_\nrm{}^\nsn = -\heb_\nrm{}^\nsn = \de_\m^\n \de_{n(r)-n(s) ,0\,
\text{mod } \r(\s)}  \\
\hG_{\nrm;\nsn} = \sG_{\nrm;\nsn} (\s) \label{Eq 3.44e} \\
\hat{B}_{\nrm;\nsn} =\hat{H}_{\nrm;\nsn;\ntd} =0 ,\quad \s=0,\ldots ,N_c -1
\label{Eq 3.44f}
\end{gather}
\end{subequations}
which also may be considered as a formal abelian limit of the left-mover
WZW orbifold data. In these results, Eq.~\eqref{Eq 3.44e} records that, for
free bosons,
the twisted Einstein metric $\hG$ is equal \cite{Geom} to the twisted tangent
space metric $\sG$ (see Eqs.~\eqref{Eq 2.2a} and \eqref{Eq 3.7c}). As is
clear from
Eqs.~\eqref{Eq 3.43a} and \eqref{Eq 3.44f}, we are not treating the most
general twisted free-bosonic open string in the basic class, which can also
involve $\hB
\neq 0$. Similarly, we will not discuss any particular target-space
compactification of the twisted open-string coordinates $\hx_\s$ below.

Further classical description of these twisted free-bosonic open strings is
obtained by substituting the data \eqref{Eq3.44} into the more general WZW
results
above. For example, the phase-space realization \eqref{Eq3.5} of the
twisted open string reduces to the following:
\begin{subequations}
\label{Eq3.45}
\begin{gather}
\hj_\nrm^{(+)} (\xi,t) =2\pi \hat{p}_\nrm^\s (\xi,t) +\srac{1}{2} \pl_\xi
\hx^\nsn_\s (\xi,t) \sG_{\nsn ;\nrm}(\s) \\
\hj_\nrm^{(-)} (\xi,t) =-2\pi \hat{p}_\nrm^\s (\xi,t) +\srac{1}{2} \pl_\xi
\hx^\nsn_\s (\xi,t) \sG_{\nsn ;\nrm}(\s) \\
\hj_\nrm^{(\pm )} (\xi,t) =\sum_{m\in \Zint} \hj_\nrm (m\!+\!\nrrs ) e^{-i
(m+\nrrsf )(t \pm \xi)}  \,. \label{Eq 3.45c}
\end{gather}
\end{subequations}
As another example, the boundary conditions in Eq.~\eqref{Eq3.23} now take
the simple form
\begin{subequations}
\label{Eq3.46}
\begin{gather}
\pl_t \hx_\s^\nrm (\xi,t) = \left\{ \begin{array}{ll} 0 & \text{at } \xi=0 \\
   i\tan (\srac{ \pi n(r)}{\r(\s)} ) \pl_\xi \hx_\s^\nrm (\xi,t) & \text{at
} \xi=\pi  \end{array} \right. \\
\{ \bar{n}(r) \} \subset \{ 0,\ldots ,\r(\s)-1 \} ,\quad \s =0,\ldots ,N_c -1
\end{gather}
\end{subequations}
which describe the {\it branes} at each end of the twisted free-bosonic
open string. Each sector $\s$ of the open-string free-bosonic orbifold
comes equipped
(via the $H$-eigenvalue problem \eqref{Eq 2.4a}) with a specific {\it set}
$\{ n(r) \}$ of spectral indices and hence a corresponding specific set of
boundary conditions \eqref{Eq3.46} on the coordinates $\hx_\s^\nrm$.

In the untwisted sector ($\s \!=\!\bar{n}(r) \!=\!0$) we have $\nrm
\rightarrow i$ and these abelian results reduce to Dirichlet-Dirichlet
(D-D) strings
(with $\pl_t x^i =0$ at both ends), as discussed in Ref.~\cite{Giusto}. More
generally, all the twisted coordinates are Dirichlet at $\xi=0$ in every
sector $\s$.
One also finds that the coordinates $\hx_\s^{0\m} (\xi,t)$ always have D-D
boundary conditions, and furthermore coordinates $\hx_\s^\nrm (\xi,t)$ with
$\srac{\bar{n}(r)}{\r(\s)} =\srac{1}{2}$ have Dirichlet-Neumann (D-N)
boundary conditions (with $\pl_\xi \hx_\s =0$ at $\xi=\pi$). As an
illustration,
the simple $\Zint_2$ permutation symmetry $x^1 \leftrightarrow x^2$ with
$G_{ab} =\de_{ab}$ leads to the abelian orbifold affine algebra
\begin{subequations}
\label{Eq3.47}
\begin{gather}
\{ \hj_{n(r)} (m\!+\!\srac{n(r)}{2}) , \hj_{n(s)} (n\!+\!\srac{n(s)}{2}) \}
= (m\!+\!\srac{n(r)}{2}) 2\de_{m+n+\frac{n(r)+n(s)}{2} ,0} \\
\r(\s) =2 ,\quad \bar{n}(r) ,\bar{n}(s) \in \{ 0,1 \}
\end{gather}
\end{subequations}
and the corresponding open-string coordinates $\hx^{n(r)}$ of this twisted
sector have boundary conditions D-D for $\bar{n}(r) =0$ and D-N for
$\bar{n}(r) =1$. However, in more general examples (with other values of
$\sbnrrs$), the sets of boundary conditions in each sector are larger and
have more
complicated (mixed) behavior at $\xi =\pi$, as seen in
Eq.~\eqref{Eq3.46}. Mixed boundary conditions with $\hB =0$ have also
been observed in the open-string sectors of free-bosonic orientation
orbifolds \cite{Orient2}.

In the remainder of this subsection, we will obtain the full
quasi-canonical algebra of these twisted free-bosonic open strings,
including in particular the
non-commutative geometry of the $\{ \hx ,\hx \}$ brackets.

The non-commutative geometry of sector $\s$ of the open-string free-bosonic
orbifold follows immediately in the abelian limit of the
$\{ \hx ,\hx \}$ brackets \eqref{Eq3.38}
\begin{subequations}
\label{Eq3.48}
\begin{gather}
\hL_\s (\xi,t) =0 ,\quad \hU_\s (\xi,t) =1 \\
\hPs^{\nrm;\nsn} (\xi,\xi,t,\s) = 2i \sG^{\nrm;\nsn} (\s) \sin (2\nrrs \xi)
,\quad \s=0 ,\ldots ,N_c -1
\end{gather}
\begin{align}
&\{ \hx_\s^\nrm (\xi,t) ,\hx_\s^\nsn (\eta,t) \} = \nn \\
& \bigspc -2\pi \de_{n(r)+n(s) ,0\, \text{mod }\r(\s)} \sG^{\nrm;\mnrn} (\s)
   \left\{ \begin{array}{ll} 0 & \text{if } \xi=\eta=0, \\ \sin (\srac{2\pi
n(r)}{\r(\s)} ) & \text{if } \xi=\eta=\pi, \\
   0 & \text{otherwise} \end{array} \right.  \label{Eq 3.48c}
\end{align}
\end{subequations}
where $\sG^{\bullet} (\s)$ is the inverse of the twisted tangent space
metric in Eq.~\eqref{Eq2.2}. We remark in particular that this geometry
becomes commutative when $\sbnrrs =0$ or $\srac{1}{2}$ (i.~e.~D-D or D-N
coordinates), which includes the untwisted sector $\s=0$ (see
Ref.~\cite{Giusto}).

To find the other brackets among $\hx_\s$ and $\hp^\s$, we begin with the
equations of motion
\begin{subequations}
\label{Eq3.49}
\begin{gather}
\pl_t \hx_\s^\nrm (\xi,t) =\sG^{\nrm ;\nsn}(\s) \Big{(} \hj_\nsn^{(+)}
(\xi,t) -\hj_\nsn^{(-)} (\xi,t) \Big{)} \\
\pl_\xi \hx_\s^\nrm (\xi,t) =\sG^{\nrm ;\nsn}(\s) \Big{(} \hj_\nsn^{(+)}
(\xi,t) +\hj_\nsn^{(-)} (\xi,t) \Big{)}
\end{gather}
\end{subequations}
which are the abelian limits of Eqs.~\eqref{Eq 3.9a} and \eqref{Eq 3.18a}.
Note that the boundary conditions in Eq.~\eqref{Eq3.46} also
follow with the mode expansions \eqref{Eq 3.45c} from these equations of
motion.

The solution of the equations of motion for the twisted coordinates is:
\begin{subequations}
\label{Eq3.50}
\begin{gather}
\hx^{0\m}_\s (\xi,t) = \hat{q}^{0\m} +2 \sG^{0\m ;0\n}(\s) \Big{(}
\hj_{0\n}(0) \xi +\sum_{m\neq 0} \frac{\hj_{0\n}(m)}{m} e^{-imt} \sin (m\xi
) \Big{)} \\
\bar{n}(r) \!\neq \!0\!: \,\,\hx^\nrm_\s (\xi,t) =\hat{q}^\nrm \!+ \bigspc
\bigspc \bigspc \bigspc \bigspc \quad \quad \quad \nn \\
  \quad \quad \bigspc +2\sG^{\nrm;\nsn}(\s) \sum_{m\in \Zint}
\frac{\hj_\nsn (m\!+\!\frac{n(s)}{\r(\s)})} {m+\frac{n(s)}{\r(\s)}}
  e^{-i(m+\frac{n(s)}{\r(\s)})t} \sin (m\!+ \!\nsrs ) \xi \,.
\end{gather}
\end{subequations}
At this point, it is helpful to note the following algebraic requirement
\begin{gather}
\{ \hj_\nrm (m\!+\! \nrrs) ,\hx^\nsn (\xi,t) \}= 2\de_\nrm{}^\nsn
e^{i(m+\nrrsf )t} \sin (m\!+\!\nrrs )\xi \label{Eq3.51}
\end{gather}
which follows from Eq.~\eqref{Eq 3.45c} and the abelian limit of
Eq.~\eqref{Eq3.12}. We find that the mode expansions in Eq.~\eqref{Eq3.50}
are consistent
with Eq.~\eqref{Eq3.51} only when we choose:
\begin{gather}
\{\hj_\nrm (m\!+\!\nrrs ), \hat{q}^\nsn \}=0 \,.  \label{Eq3.52}
\end{gather}
Jacobi identities then tell us that the bracket of any two $\hat{q}$'s is
at most a constant, which we will tentatively assume to be zero
\begin{gather}
\{\hat{q}^\nrm ,\hat{q}^\nsn \}=0  \label{Eq3.53}
\end{gather}
so that the $\hat{q}$'s are simply c-numbers.

We will also need the free-bosonic momenta
\begin{gather}
\hp_\nrm^\s (\xi,t) =\srac{1}{4\pi} \Big{(} \hj_\nrm^{(+)} (\xi,t)
-\hj_\nrm^{(-)} (\xi,t) \Big{)} \label{Eq3.54}
\end{gather}
which are obtained in the abelian limit of Eq.~\eqref{Eq 3.9b}. With these
momenta, the mode expansions \eqref{Eq 3.45c}, \eqref{Eq3.50} and the
current
algebra \eqref{Eq 3.44b}, we may now compute the full {\it quasi-canonical
algebra} of $\hx_\s ,\,\hp^\s$
\begin{subequations}
\label{Eq3.55}
\begin{gather}
\{ \hp_\nrm^\s (\xi,t) ,\hp_\nsn^\s (\eta,t) \} =0 \\
\{ \hx_\s^\nrm (\xi,t) ,\hp_\nsn^\s (\eta,t) \} = i\de_\nsn{}^\nrm \Big{(}
\de (\xi-\eta) - \cos (\nsrs (\xi+\eta)) \de
   (\xi+\eta) \Big{)}
\end{gather}
\begin{align}
&\{ \hx_\s^\nrm (\xi,t) ,\hx_\s^\nsn (\eta,t) \} = \nn \\
& \bigspc -2\pi \de_{n(r)+n(s) ,0\, \text{mod }\r(\s)} \sG^{\nrm;\mnrn} (\s)
   \left\{ \begin{array}{ll} 0 & \text{if } \xi=\eta=0, \\ \sin (\srac{2\pi
n(r)}{\r(\s)} ) & \text{if } \xi=\eta=\pi, \\
   0 & \text{otherwise} \end{array} \right.  \label{Eq 3.55c} \\
&\bigspc \bigspc \bigspc \quad \s=0,\ldots ,N_c -1
\end{align}
\end{subequations}
in each sector of each open-string free-bosonic orbifold. To obtain these
results, we found that the following identities cite{Orient1} were useful:
\begin{subequations}
\label{Eq3.56}
\begin{gather}
\sin (\nrrs (\xi -\eta)) \de(\xi -\eta) =0 ,\quad 0\leq \xi,\eta \leq \pi \\
\cos (\nrrs (\xi -\eta)) \de(\xi -\eta) =\de(\xi-\eta) ,\quad 0\leq
\xi,\eta \leq \pi \\
\sin (\nrrs (\xi +\eta)) \de(\xi +\eta) =0 \,\text{ except at } \xi=\eta=\pi
\end{gather}
\begin{align}
&\sum_{\begin{array}{c} \smal{m\in \Zint } \\ \smal{m\!+\! \nrrsf \neq 0}
\end{array}} \bkspc \!\!\! \frac{1}{m\!+\!\nrrsf} \sin ((m\!+\!\nrrs )\xi )
   \,\sin ((m\!+\!\nrrs )\eta )= \bigspc \quad \quad \nn \\
&\bigspc  = \pi \int_{\xi-\eta}^{\xi +\eta} \!d\eta ' \sin (\nrrs \eta ')
\de (\eta ') = \left\{ \begin{array}{cc} 0&\text{if } \xi =\eta =0 \\
   \srac{\pi}{2} \sin (\frac{2\pi n(r)}{\r(\s)} ) & \text{if } \xi=\eta=
\pi \\ 0 & \text{otherwise } \,. \end{array} \right. \label{Eq 3.56d}
\end{align}
\end{subequations}
Following standard procedure in distribution theory, the sum in
Eq.~\eqref{Eq 3.56d} was defined by temporarily inserting a smearing
function
$\exp (-\ep m^2) ,\, \ep \rightarrow 0^+$.

We call the algebras in Eq.~\eqref{Eq3.55} quasi-canonical because they are
canonical in the bulk, and we also note that the result for the
$\{\hx ,\hx \}$ bracket in Eq.~\eqref{Eq 3.55c} is in agreement with
Eq.~\eqref{Eq 3.48c}. Allowing constant non-zero $\{ \hat{q} ,\hat{q} \}$
brackets
gives rise to overall additive constants in the $\{\hx ,\hx\}$ brackets, so
the assumption in Eq.~\eqref{Eq3.53} is equivalent to our earlier
requirement that $\{\hx ,\hx\}$ vanishes in the bulk.

\section{Conformal Field Theory of Twisted Open WZW Strings}

In this section, we further develop (see Sec.~2) the {\it quantum} theory
of our basic class of twisted open WZW strings. This discussion culminates
in
the derivation of the twisted open-string KZ equations in Subsec.~$4.6$.

\subsection{The Quantum Hamiltonian and the Twisted Affine Primary Fields}

Combining Eqs.~\eqref{Eq A.3a} and \eqref{Eq2.16}, we obtain the quantum
Hamiltonian of sector $\s$ of $A_g^{open}(H)/H$:
\begin{subequations}
\label{Eq4.1}
\begin{align}
\bkspc &\bkspc \hat{H}_\s \!=\!L_\s(0) = \nn \\
&\,\, =\!\srac{1}{2\pi} \!\int_0^{\pi} \!\!\!d\xi
{\cL}_{\sgb(\s)}^{\nrm;\mnrn}(\s) \Big{(} \!:\!\hj_\nrm^{(+)}(\xi,t)
\hj_\mnrn^{(+)}
  (\xi,t) \!+\!\hj_\nrm^{(-)}(\xi,t) \hj_\mnrn^{(-)}(\xi,t)\!: \!\Big{)} \nn \\
&\,\, = {\cL}_{\sgb(\s)}^{\nrm;\mnrn}(\s) \Big{(} \sum_{p\in \Zint}
:\!\hj_\nrm (p\!+\!\nrrs) \hj_{\mnrn} (-p\!-\!\nrrs) \!:_M  \nn \\
&\bigspc \bigspc \bigspc -i\srac{\bar{n}(r)}{\r(\s)}
\scf_{\nrm;\mnrn}^{0,\de}(\s) \hj_{0,\de} (0) \Big{)} +\gscfwt \label{Eq
4.1a}
\end{align}
\begin{align}
\bkspc&\bkspc:\!\hj_\nrm (m\!+\!\nrrs) \hj_\nsn (n\!+\!\nsrs)\! :_M
\,\equiv \theta (m\!+\!\nrrs \geq 0) \hj_\nsn (n\!+\!\nsrs) \hj_\nrm (m\!+
  \!\nrrs) \nn \\
& \bigspc \bigspc + \theta (m\!+\!\nrrs <0) \hj_\nrm (m\!+\!\nrrs) \hj_\nsn
(n\!+\!\nsrs)
\end{align}
\begin{align}
\bkspc \!\! [\hj_\nrm(m\!+\!\srac{n(r)}{\r(\s)}),
\hj_\nsn(n\!+\!\srac{n(s)}{\r(\s)}) ] \!&=\!
i\scf_{\nrm;\nsn}{}^{\!\!\!\!\!n(r)+
n(s),\d}(\s) \hj_{n(r)+n(s),\d} (m\!+\!n\!+\!\srac{n(r)+n(s)}{\r(\s)}) \nn\\
 &+ (m\!+\!\srac{n(r)}{\r(\s)})
\;\d_{m+n+\frac{n(r)+n(s)}{\r(\s)},\,0}\;\sG_{\nrm;-\nrn}(\s)
\end{align}
\begin{gather}
\pl_t \hat{A}(\xi,t) = i[\hat{H}_\s ,\hat{A}(\xi,t)] \,. \label{Eq 4.1d}
\end{gather}
\end{subequations}
Here we have also collected Eqs.~\eqref{Eq A.3c}, \eqref{EqA.1} and
\eqref{Eq 2.17a} for reference. Explicit formulae for the conformal weight
$\gscfwt$ of the scalar twist-field state are given in Eqs.~\eqref{Eq A.3e}
and \eqref{Eq 2.13d}.

Using the classical theory and in particular the $\{ \hj ,\hg \}$ brackets
in Eq.~\eqref{Eq3.16} as a guide, we may augment the quantum theory
with the following
equal-time commutators
\begin{subequations}
\label{Eq4.2}
\begin{align}
[\hj_\nrm^{(+)} (\xi,t) ,\hg(\st,\eta,t)] =& 2\pi \Big{(} \hg(\st,\eta,t)
\st_\nrm \de_\nrrsf (\xi \!-\!\eta)  \nn \\
& \bigspc  -\st_\nrm \hg(\st,\eta,t) \de_\nrrsf (\xi \!+\!\eta) \Big{)}
\label{Eq 4.2a} \\
[\hj_\nrm^{(-)} (\xi,t) ,\hg(\st,\eta,t)] =&2\pi \Big{(} -\st_\nrm
\hg(\st,\eta,t) \de_{-\nrrsf}(\xi \!-\!\eta) \nn \\
& \bigspc +\hg(\st,\eta,t) \st_\nrm \de_{-\nrrsf} (\xi \!+\!\eta) \Big{)}
\label{Eq 4.2b}
\end{align}
\end{subequations}
where this $\hg (\st)$ is the open-string {\it twisted affine primary
field} in twisted rep $\st$. The (classical) group orbifold element $\hg
(\st)$
of Sec.~3 is the high-level limit of this operator field. As noted above
for the corresponding brackets in Eq.~\eqref{Eq3.16}, the commutators
\eqref{Eq4.2} satisfy the $\hj,\hj,\hg$ operator Jacobi identities.

Moreover, using the mode expansions \eqref{Eq2.5} of the twisted strip
currents, we find that the mode commutator
\begin{align}
[\hj_\nrm (m\!+\!\nrrs) ,\hg(\st,\xi,t)] =&\hg(\st,\xi,t) \st_\nrm
e^{i(m+\nrrs)(t+\xi)} \nn \\
&\quad -\!\st_\nrm \hg(\st,\xi,t) e^{i(m+\nrrs)(t-\xi)} \label{Eq4.3}
\end{align}
is equivalent to both Eqs.~\eqref{Eq 4.2a} and \eqref{Eq 4.2b}.

\subsection{Time Dependence of the Twisted Affine Primary Fields}

Using Eqs.~\eqref{Eq4.1} and \eqref{Eq4.2}, we find after some algebra the
time evolution of the open-string twisted affine primary fields
\begin{subequations}
\label{Eq4.4}
\begin{align}
\!&\!\pl_t \hg(\st,\xi,t) = i[ \hat{H}_\s ,\hg(\st,\xi,t)] \\
& \, =\!2i{\cL}_{\sgb(\s)}^{\nrm;\mnrn}(\s) \Big{(}
\!:\!\hj_\nrm^{(+)}(\xi,t) \hg(\st,\xi,t) \st_\mnrn \!-\!\st_\nrm
\hj_\mnrn^{(-)}
   (\xi,t) \hg(\st,\xi,t) \!:_M \!\!\Big{)} \quad \quad \nn \\
& \quad \quad +i{\cL}_{\sgb(\s)}^{0,\m;0,\n}(\s) \left( \hg(\st,\xi,t)
\st_{0,\m} \st_{0,\n} +\st_{0,\m} \st_{0,\n} \hg(\st,\xi,t)
   -2\st_{0,\m} \hg(\st,\xi,t) \st_{0,\n} \right) \nn \\
& \quad \quad +{\cL}_{\sgb(\s)}^{\nrm;\mnrn}(\s) \srac{\bar{n}(r)}{\r(\s)}
\scf_{\nrm;\mnrn}^{0,\de}(\s) \left( \hg(\st,\xi,t)
   \st_{0,\de} -\st_{0,\de} \hg(\st,\xi,t) \right) \label{Eq 4.4b}
\end{align}
\end{subequations}
which provides our first example of a {\it twisted vertex operator
equation}. The mode normal ordering $:\cdot :_M$ in Eq.~\eqref{Eq4.4}
is defined as follows
\begin{subequations}
\label{Eq4.5}
\begin{gather}
:\! \hj_\nrm^{(\pm)} (\xi,t) \hg(\st,\xi,t) \!:_M \equiv \hj_\nrm^{(\pm)-}
(\xi,t) \hg(\st,\xi,t) +\hg(\st,\xi,t) \hj_\nrm^{(\pm)+}(\xi,t) \\
\hj_\nrm^{(\pm)+} (\xi,t) \equiv \sum_{m \geq 0} \hj_{\bar{n}(r)\m} (m\!+\!
\srac{\bar{n}(r)}{\r(\s)}) e^{-i(m+ \frac{\bar{n}(r)}{\r(\s)})
   (t\pm \xi)} \\
\hj_\nrm^{(\pm)-} (\xi,t) \equiv \sum_{m \leq -1} \hj_{\bar{n}(r)\m}
(m\!+\! \srac{\bar{n}(r)}{\r(\s)}) e^{-i(m+ \frac{\bar{n}(r)}{\r(\s)})
   (t\pm \xi)} \\
\hj_\nrm^{(\pm)+} (\xi,t) +\hj_\nrm^{(\pm)-} (\xi,t) =\hj_\nrm^{(\pm)} (\xi,t)
\end{gather}
\end{subequations}
where $\hj^{(\pm)\pm}$ are called the partial currents. Since ${\cL}
\rightarrow \sG^{\bullet}/2$ and $\hj =\Ord (k^{\frac{1}{2}})$ in the
high-level limit, the classical equation of motion in Eq.~\eqref{Eq 3.19a}
is indeed the classical limit of the twisted vertex operator
equation \eqref{Eq4.4}.

An alternate form of the open-string twisted vertex operator equation
\eqref{Eq 4.4b} is the following
\begin{subequations}
\label{Eq4.6}
\begin{align}
\!&\!\pl_t \hg(\st,\xi,t) =\nn \\
& \quad \!2i{\cL}_{\sgb(\s)}^{\nrm;\mnrn}(\s) \left(
:\!\hj_\nrm^{(+)}(\xi,t) \hg(\st,\xi,t) \st_\mnrn
   \!-\!\st_\nrm \hj_\mnrn^{(-)}(\xi,t) \hg(\st,\xi,t) \!:_M \right) \nn \\
& \quad \,\,\, +i [\Dg (\st) ,\hg(\st,\xi,t) ]_+
-2i{\cL}_{\sgb(\s)}^{0,\m;0,\n} (\s) \st_{0,\m} \hg(\st,\xi,t) \st_{0,\n}
\nn \\
& \quad \,\,\, -2i{\cL}_{\sgb(\s)}^{\nrm;\mnrn} (\s) \left(
\srac{\bar{n}(r)}{\r(\s)} \hg(\st,\xi,t) \st_\nrm \st_\mnrn \!+\!
   \srac{\overline{-n(r)}}{\r(\s)} \st_\nrm \st_\mnrn \hg(\st,\xi,t) \right) \\
& \bigspc \Dg (\st) \!\equiv {\cL}_{\sgb(\s)}^{\nrm;\mnrn}(\s) \st_\nrm
\st_\mnrn \!=U(T,\s) D_g (T) U\hc (T,\s) \label{Eq 4.6b} \\
& \bigspc \bigspc \bigspc \quad \quad D_g (T) =L_g^{ab} T_a T_b
\end{align}
\end{subequations}
where $\overline{-n(r)}$, which is the pullback of $-n(r)$, is formally
defined in Eq.~\eqref{Eq A.3f}. The quantity $\Dg (\st)$ is called the
twisted
conformal weight matrix \cite{Big} and $D_g(T)$ is the untwisted conformal
weight matrix of rep $T$ of $g$. To obtain this result, we used the
identities:
\begin{subequations}
\label{Eq4.7}
\begin{align}
& 2{\cL}_{\sgb(\s)}^{\nrm;\mnrn} \sbnrrs \st_\nrm \st_\mnrn = \nn \\
& \quad \quad ={\cL}_{\sgb(\s)}^{\nrm;\mnrn} (\s) \left( \st_\nrm \st_\mnrn
(1-\de_{\bar{n}(r),0}) +i\sbnrrs \scf_{\nrm;\mnrn}{}^{0,\de}(\s)
   \st_{0,\de} \right) \label{Eq 4.7a} \\
& 2{\cL}_{\sgb(\s)}^{\nrm;\mnrn} \sbnrrs \st_\mnrn \st_\nrm = \nn \\
& \quad \quad ={\cL}_{\sgb(\s)}^{\nrm;\mnrn} (\s) \left( \st_\nrm \st_\mnrn
(1-\de_{\bar{n}(r),0}) -i\sbnrrs \scf_{\nrm;\mnrn}{}^{0,\de}(\s)
   \st_{0,\de} \right) \,. \label{Eq 4.7b}
\end{align}
\end{subequations}
The identity \eqref{Eq 4.7a} was given in Ref.~\cite{Big}, and Eq.~\eqref{Eq
4.7b} follows easily from Eq.~\eqref{Eq 4.7a} and the
orbifold Lie algebra \eqref{Eq 2.2f}.

We consider next the correlators $\hat{A}_\s$ of the open-string twisted
affine primary fields in the scalar twist-field state (see
Eq.~\eqref{Eq2.22}).
With Eq.~\eqref{Eq4.3}, we immediately obtain the global Ward identities
for these correlators
\begin{subequations}
\label{Eq4.8}
\begin{gather}
\hat{A}_\s (\st,\xi,t) \equiv \langle \hg(\st^{(1)},\xi_1,t_1) \ldots
\hg(\st^{(n)},\xi_n,t_n) \rangle_\s \bigspc \nn \\
\bigspc \equiv {}_\s \langle 0| \hg(\st^{(1)},\xi_1,t_1) \ldots
\hg(\st^{(n)},\xi_n,t_n) |0\rangle_\s \\
\langle [\hj_{0,\m}(0) ,\hg(\st^{(1)},\xi_1,t_1) \ldots
\hg(\st^{(n)},\xi_n,t_n)] \rangle_\s =0 \bigspc \\
\bigspc \Rightarrow \hat{A}_\s(\st,\xi,t) \sum_{i=1}^n \st_{0,\m}^{(i)} -
\sum_{i=1}^n \st_{0,\m}^{(i)} \hat{A}_\s (\st,\xi,t) =0
\end{gather}
\end{subequations}
which generalize the untwisted open-string Ward identities of Ref.~\cite{Giusto}.

Towards obtaining twisted open-string KZ equations for these correlators,
we give the following properties of the scalar twist-field state
\begin{gather}
\hj_\nrm^{(\pm)+} (\xi,t) |0\rangle_\s = {}_\s \langle 0| \hj_\nrm^{(\pm)-}
(\xi,t) =0 \label{Eq4.9}
\end{gather}
which follow immediately from Eq.~\eqref{Eq2.22} and the definition of the
partial currents in Eqs.~($4.5$b,c). With these properties and the twisted
vertex operator equation \eqref{Eq4.4}, we may obtain the time derivatives
of the open-string correlators $\hat{A}_\s$:
\begin{subequations}
\label{Eq4.10}
\begin{align}
&\pl_{t_i} \hat{A}_\s = 2i{\cL}_{\sgb(\s)}^{\nrm;\mnrn}(\s) \times \bigspc
\bigspc \bigspc \quad \quad \nn \\
& \quad \quad  \times \sum_{j \neq i} \left\{ \frac{e^{i\bnrrs (\phi_j
-\phi_i)}}{1- e^{i(\phi_j -\phi_i)}} \hat{A}_\s \st_{\nrm}^{(j)}
\st_\mnrn^{(i)}
   - \frac{e^{i\bnrrs (\bar{\phi}_j -\phi_i)}}{1- e^{i(\bar{\phi}_j
-\phi_i)}} \st_\nrm^{(j)} \hat{A}_\s \st_\mnrn^{(i)} \right. \nn \\
&\left. \bigspc -\frac{e^{i\bnrrs (\phi_j -\bar{\phi}_i)}}{1- e^{i(\phi_j
-\bar{\phi}_i)}} \st_\mnrn^{(i)} \hat{A}_\s \st_\nrm^{(j)}
   +\frac{e^{i\bnrrs (\bar{\phi}_j -\bar{\phi}_i)}}{1- e^{i(\bar{\phi}_j
-\bar{\phi}_i)}} \st_\mnrn^{(i)} \st_{\nrm}^{(j)} \hat{A}_\s \right\} \nn \\
&  \quad +i [\Dg (\st^{(i)}) ,\hat{A}_\s ]_+
-2i{\cL}_{\sgb(\s)}^{0,\m;0,\n}(\s) \st_{0,\m}^{(i)} \hat{A}_\s
\st_{0,\n}^{(i)} \nn \\
&  \quad -2i{\cL}_{\sgb(\s)}^{\nrm;\mnrn}(\s) \left( \sbnrrs \hat{A}_\s
\st_\nrm^{(i)} \st_\mnrn^{(i)} +\srac{\overline{-n(r)}}{\r(\s)}
\st_\nrm^{(i)}
   \st_\mnrn^{(i)} \hat{A}_\s \right)
\end{align}
\begin{gather}
\phi_i \equiv t_i +\xi_i ,\quad \bar{\phi}_i \equiv t_i -\xi_i ,\quad i,j
=1,\ldots ,n \,.
\end{gather}
\end{subequations}
Tensor products are assumed in the result \eqref{Eq4.10}, for example
\begin{subequations}
\label{Eq4.11}
\begin{gather}
(\hat{A}_\s \st^{(i)} \st^{(j)} )_{N(r_i)\m_i ,N(r_j)\m_j}^{\quad \quad
N(s_i)\n_i ,N(s_j)\n_j} \equiv (\hat{A}_\s \st^{(i)} \otimes
   \st^{(j)} )_{N(r_i)\m_i ,N(r_j)\m_j}^{\quad \quad N(s_i)\n_i
,N(s_j)\n_j} \quad \nn \\
\quad = (\hat{A}_\s)_{N(r_i)\m_i ,N(r_j)\m_j}^{\quad \quad N(t_i)\de_i
,N(t_j)\de_j} (\st^{(i)} )_{N(t_i)\de_i}{}^{\bkspc N(s_i)\n_i}
    (\st^{(j)} )_{N(t_j)\de_j}{}^{\bkspc N(s_j)\n_j} ,\quad i\neq j \\
[ \st_\nrm^{(i)} ,\st_\nsn^{(j)}] = i\de^{ij} \scf_{\nrm;\nsn}{}^{\ntd}
(\s) \st_\ntd^{(i)}
\end{gather}
\end{subequations}
although $\st^{(i)} \st^{(i)}$ is standard matrix multiplication.

In the derivation of Eq.~\eqref{Eq4.10}, we also needed the commutators:
\begin{subequations}
\label{Eq4.12}
\begin{align}
[\hj_\nrm^{(+)\pm} (\xi_i,t_i) ,\hg(\st^{(j)},\xi_j,t_j)] =& \pm
\frac{e^{i\bnrrs (\phi_j -\phi_i)}}
   {1- e^{i(\phi_j -\phi_i)}} \hg(\st^{(j)} ,\xi_j,t_j) \st_\nrm^{(j)}
\bigspc \nn \\
&\bigspc \mp \frac{e^{i\bnrrs (\bar{\phi}_j -\phi_i)}}{1- e^{i(\bar{\phi}_j
-\phi_i)}} \st_\nrm^{(j)} \hg(\st^{(j)} ,\xi_j,t_j)  \\
[\hj_\nrm^{(-)\pm} (\xi_i,t_i) ,\hg(\st^{(j)},\xi_j,t_j)] =&\pm
\frac{e^{i\bnrrs (\phi_j -\bar{\phi}_i)}}
    {1- e^{i(\phi_j -\bar{\phi}_i)}} \hg(\st^{(j)},\xi_j,t_j)
\st_\nrm^{(j)}  \bigspc \nn \\
& \bigspc \mp \frac{e^{i\bnrrs (\bar{\phi}_j -\bar{\phi}_i)}}{1-
e^{i(\bar{\phi}_j -\bar{\phi}_i)}} \st_\nrm^{(j)} \hg(\st^{(j)} ,\xi_j,t_j)
\,.
\end{align}
\end{subequations}
The infinite sums evaluated here are conditionally convergent, for example
\begin{gather}
[\hj_\nrm^{(+)+} (\xi_i,t_i) ,\hg(\st^{(j)},\xi_j ,t_j)] =e^{i\bnrrs
(\phi_j -\phi_i)} \hg (\st^{(j)} ,\xi_j ,t_j ) \st_\nrm^{(j)} \sum_{m\geq
0}
   e^{im(\phi_j -\phi_i)} \bigspc \nn \\
\bigspc \bigspc \quad \quad -e^{i\bnrrs (\bar{\phi}_j -\phi_i)}
\st_\nrm^{(j)} \hg (\st^{(j)} ,\xi_j ,t_j ) \sum_{m\geq 0}
e^{im(\bar{\phi}_j -\phi_i)}
\label{Eq4.13}
\end{gather}
and we used the untwisted open-string prescription of Ref.~\cite{Giusto} to
evaluate these sums.

The global Ward identities in Eq.~\eqref{Eq4.8} and the differential
equations in Eq.~\eqref{Eq4.10} are the first two components of the desired
open-string KZ system for $A_g^{open}(H)/H$, a complete form of which can
be found in Subsec.~$4.6$.

\subsection{Constituent Twisted Affine Primary Fields}

In this subsection, we introduce the {\it constituent} twisted affine
primary fields which provide an alternate derivation of the twisted vertex
operator
equation \eqref{Eq 4.4b} for $\pl_t \hg$, and will also be helpful in
determining the corresponding twisted vertex operator equation for $\pl_\xi
\hg$.

Following Refs.~\cite{Witten,H+O,Giusto,Big,Perm}, we define the constituent
twisted affine primary fields $\hg_\pm (\st)$ by {\it factorization} of
the twisted affine primary fields:
\begin{subequations}
\label{Eq4.14}
\begin{align}
\hg(\st,\xi,t) &\equiv \hg_- (\st,\xi,t) \hg_+ (\st,\xi,t) \\
[\hj_\nrm (\mnrrs) ,\hg_+ (\st,\xi,t)] &=\hg_+ (\st,\xi,t) \st_\nrm
e^{i(\mnrrs )(t+\xi)} \\
\quad \quad [\hj_\nrm (\mnrrs) ,\hg_- (\st,\xi,t)] &=-\st_\nrm \hg_-
(\st,\xi,t) e^{i(\mnrrs )(t-\xi)} \,.
\end{align}
\end{subequations}
The simpler commutators in Eq.~($4.14$b,c), reminiscent of chiral primary
fields, reproduce the algebra \eqref{Eq4.3} of the current modes with the
full
twisted affine primary fields $\hg (\st)$.

By direct computation with Eq.~\eqref{Eq4.14} and the mode form \eqref{Eq
4.1a} of the Hamiltonian, we find after some algebra the twisted vertex
operator equations of the constituents
\begin{subequations}
\label{Eq4.15}
\begin{gather}
\pl_t \hg_{\pm} (\st,\xi,t) = i[\hat{H}_\s ,\hg_{\pm} (\st,\xi,t)]
\end{gather}
\begin{align}
&\pl_t \hg_+ (\st,\xi,t)= 2i{\cL}_{\sgb(\s)}^{\nrm;\mnrn} (\s)
:\!\hj_\nrm^{(+)} (\xi,t) \hg_+ (\st,\xi,t) \st_\mnrn \!:_M \nn \\
&\quad +i\hg_+ (\st,\xi,t) \Dg (\st) -2i{\cL}_{\sgb(\s)}^{\nrm;\mnrn}(\s)
\sbnrrs \hg_+ (\st,\xi,t) \st_\nrm \st_\mnrn \\
&\pl_t \hg_- (\st,\xi,t)= -2i{\cL}_{\sgb(\s)}^{\nrm;\mnrn} (\s) :\!\st_\nrm
\hj_\mnrn^{(-)} (\xi,t) \hg_- (\st,\xi,t)\!:_M \nn \\
&\quad +i\Dg (\st) \hg_-(\st,\xi,t) -2i{\cL}_{\sgb(\s)}^{\nrm;\mnrn}(\s)
\srac{\overline{-n(r)}}{\r(\s)} \st_\nrm \st_\mnrn \hg_- (\st,\xi,t)
\end{align}
\end{subequations}
where we have again used the identities in Eq.~\eqref{Eq4.7} to simplify
these results. The mode normal ordering here is the same as that defined in
Eq.~\eqref{Eq4.5} with $\hg \rightarrow \hg_{\pm}$.

As a check on the consistency of the factorization \eqref{Eq4.14}, we may
use Eqs.~($4.15$b,c) to successfully reconstruct the vertex operator
equation
\eqref{Eq4.6} of the {\it full} twisted affine primary field:
\begin{subequations}
\label{Eq4.16}
\begin{align}
&\pl_t \hg(\st,\xi,t) = \pl_t \hg_- (\st,\xi,t) \hg_+ (\st,\xi,t) + \hg_-
(\st,\xi,t) \pl_t \hg_+ (\st,\xi,t) \\
&\,\, =2i{\cL}_{\sgb(\s)}^{\nrm;\mnrn}(\s) \left( \hg_-(\st,\xi,t) :\!
\hj_\nrm^{(+)}(\xi,t) \hg_+(\st,\xi,t) \st_\mnrn \!:_M \right. \nn \\
& \quad \,\, \left. - :\!\st_\nrm \hj_\mnrn^{(-)} (\xi,t) \hg_-(\st,\xi,t)
\!:_M \hg_+(\st,\xi,t) \right) +i[\Dg (\st),\hg(\st,\xi,t) ]_+ \nn \\
& \quad \,\, -2i{\cL}_{\sgb(\s)}^{\nrm;\mnrn} (\s) \left(
\srac{\bar{n}(r)}{\r(\s)} \hg(\st,\xi,t) \st_\nrm \st_\mnrn \!+\!
   \srac{\overline{-n(r)}}{\r(\s)} \st_\nrm \st_\mnrn \hg(\st,\xi,t)
\right) \label{Eq 4.16b} \\
& \,\, =2i{\cL}_{\sgb(\s)}^{\nrm;\mnrn}(\s) \left( :\!\hj_\nrm^{(+)}(\xi,t)
\hg(\st,\xi,t) \st_\mnrn
   \!-\!\st_\nrm \hj_\mnrn^{(-)}(\xi,t) \hg(\st,\xi,t) \!:_M \right) \nn \\
& \quad \,\, +i [\Dg (\st) ,\hg(\st,\xi,t) ]_+
-2i{\cL}_{\sgb(\s)}^{0,\m;0,\n} (\s) \st_{0,\m} \hg(\st,\xi,t) \st_{0,\n}
\nn \\
& \quad \,\, -\!2i{\cL}_{\sgb(\s)}^{\nrm;\mnrn} (\s) \left(
\srac{\bar{n}(r)}{\r(\s)} \hg(\st,\xi,t) \st_\nrm \st_\mnrn \!+\!
   \srac{\overline{-n(r)}}{\r(\s)} \st_\nrm \st_\mnrn \hg(\st,\xi,t)
\right) \,. \label{Eq 4.16c}
\end{align}
\end{subequations}
In this computation, the expression in Eq.~\eqref{Eq 4.16b} was not
completely normal-ordered, so we used the following identities
\begin{subequations}
\label{Eq4.17}
\begin{gather}
[\hj_\nrm^{(+)-} (\xi,t) ,\hg_-(\st,\xi,t)] = \frac{e^{-2i\bnrrs
\xi}}{1-e^{-2i\xi}} \st_\nrm \hg_-(\st,\xi,t) 
\end{gather}
\begin{gather}
[\hj_\nrm^{(-)+} (\xi,t) ,\hg_+(\st,\xi,t)] = \frac{e^{2i\bnrrs
\xi}}{1-e^{2i\xi}} \hg_+(\st,\xi,t) \st_\nrm \\
\frac{e^{2i\frac{\overline{-n(r)}}{\r(\s)}\xi}}{1-e^{2i\xi}}
+\frac{e^{-2i\bnrrs \xi}}{1-e^{-2i\xi}} =\de_{\bar{n}(r),0} \label{Eq 4.17c}
\end{gather}
\end{subequations}
to obtain the completely normal-ordered result in Eq.~\eqref{Eq 4.16c}.
With Ref.~\cite{Giusto}, we note the singularities at $\xi =0$ and $\pi$ of the
commutators in Eq.~\eqref{Eq4.17}. Although (as seen in Eq.~\eqref{Eq
4.17c}) these singularities cancel in the computation of $\pl_t \hg$, we
shall see
in Subsec.~$4.5$ that the singularities persist in $\pl_\xi \hg$.

Using the mode-ordered form \eqref{Eq A.3a}, the algebra \eqref{Eq4.14} and
the twisted vertex operator equations, we may also give the action of
the Virasoro generators $L_\s (m)$ on the constituent twisted affine
primary fields:
\begin{subequations}
\label{Eq4.18}
\begin{gather}
[L_\s (m),\hg_+ (\st,\xi,t)] =\hg_+ (\st,\xi,t) (-i \lpl_{\!\!t} +m\Dg
(\st)) e^{im(t+\xi)} \\
[L_\s (m),\hg_- (\st,\xi,t)] =e^{im(t-\xi)} (-i\pl_t +m\Dg (\st)) \hg_-
(\st,\xi,t) \,.
\end{gather}
\end{subequations}
We have checked that these commutators satisfy the $L,L,\hg_{\pm}$ Jacobi
identities.

\subsection{The Constituents are Chiral}

In the untwisted results of Ref.~\cite{Giusto}, the constituent affine primary
fields $g_+(T),g_-(T)$ of $A_g^{open}$ are in fact chiral and anti-chiral
respectively. As we will see below, the same properties hold for the
twisted constituent fields
 $\hg_+(\st) ,\hg_-(\st)$ of the open-string orbifold $A_g^{open}(H)/H$.

Following Ref.~\cite{Giusto}, we begin this discussion by returning to the bulk
momentum operator \eqref{Eq2.18}
\begin{gather}
\hat{P}_\s (t) = -\frac{2i}{\pi} \sum_{m \in \Zint}
\frac{e^{-i(2m+1)t}}{2m+1} L_\s (2m+1)  \label{Eq4.19}
\end{gather}
which we have now expressed in terms of the open-string Virasoro modes. By
direct computation with the algebra \eqref{Eq4.18}, we then find that
\begin{subequations}
\label{Eq4.20}
\begin{align}
i[\hat{P}_\s (t) ,\hg_+ (\st,\xi,t)] =& 4 \left( \int_0^{\xi} \!\!d\eta
\,e^{i\eta} \de(2\eta) \right) \pl_t \hg_+ (\st,\xi,t) \nn \\
&+ 4e^{i\xi} \de (2\xi) \hg_+(\st,\xi,t) \Dg (\st) \label{Eq 4.20a} 
\end{align}
\begin{align}
i[\hat{P}_\s (t) ,\hg_- (\st,\xi,t)] =& -4\left( \int_0^{\xi} \!\!d\eta
\,e^{-i\eta} \de(2\eta) \right) \pl_t \hg_- (\st,\xi,t) \nn \\
&+ 4e^{-i\xi} \de (2\xi) \Dg (\st) \hg_-(\st,\xi,t) \label{Eq 4.20b} \\
4\int_0^{\xi} \!\!d\eta \,e^{i\eta} \de(2\eta) =& 4\int_0^{\xi} \!\!d\eta
\,e^{-i\eta} \de(2\eta) = \left\{ \begin{array}{ll} 1& \text{if }
   0<\xi<\pi , \\ 0 & \text{if } \xi=0,\pi  \end{array} \right.
\end{align}
\end{subequations}
where the twisted conformal weight matrix $\Dg (\st)$ is defined in
Eq.~\eqref{Eq 4.6b}. The summation identities \cite{Giusto}
\begin{gather}
\sum_{m \in \Zint} \frac{e^{\pm i(2m+1)\xi}}{2m +1} =\pm \tp \!\int_0^{\xi}
\!\! d\eta \,e^{\pm i\eta} \de(2\eta)  \label{Eq4.21}
\end{gather}
were used to obtain these results.

The last terms in Eqs.~\eqref{Eq 4.20a} and \eqref{Eq 4.20b} are quantum
effects which contribute only at the boundary, so that the result
\begin{gather}
\pl_\xi \hg_\pm (\st,\xi,t) =i[ \hat{P}_\s (t),\hg_\pm (\st,\xi,t)] =\pm
\pl_t \hg(\st,\xi,t) \quad 0< \xi <\pi  \label{Eq4.22}
\end{gather}
is obtained in the bulk. Following the classical intuition developed in
Subsec.~$3.2$, we may smoothly extend this result to include the boundary
\begin{gather}
\pl_- \hg_+(\st,\xi,t) =\pl_+ \hg_-(\st,\xi,t) =0 ,\quad \pl_\pm =\pl_t \pm
\pl_\xi ,\quad 0\leq \xi \leq \pi \label{Eq4.23}
\end{gather}
which tells us that, as in Ref.~\cite{Giusto}, the operators $\hg_+(\st)$ and
$\hg_-(\st)$ are respectively chiral and anti-chiral.

In fact, we may use the twisted vertex operator equations \eqref{Eq4.15}
and the chiralities \eqref{Eq4.23} in the form $\srac{1}{2} \pl_\pm \hg_\pm
=\pl_t \hg_\pm$ to obtain the following set of light-cone twisted vertex
operator equations:
\begin{subequations}
\label{Eq4.24}
\begin{align}
&\srac{1}{2} \pl_+ \hg_+ (\st,\xi,t)= 2i{\cL}_{\sgb(\s)}^{\nrm;\mnrn} (\s)
:\!\hj_\nrm^{(+)} (\xi,t) \hg_+ (\st,\xi,t) \st_\mnrn \!:_M \nn \\
&\quad +i\hg_+ (\st,\xi,t) \Dg (\st) -2i{\cL}_{\sgb(\s)}^{\nrm;\mnrn}(\s)
\sbnrrs \hg_+ (\st,\xi,t) \st_\nrm \st_\mnrn \\
&\srac{1}{2} \pl_- \hg_- (\st,\xi,t)= -2i{\cL}_{\sgb(\s)}^{\nrm;\mnrn} (\s)
:\!\st_\nrm \hj_\mnrn^{(-)} (\xi,t) \hg_- (\st,\xi,t)\!:_M \nn \\
&\quad +i\Dg (\st) \hg_-(\st,\xi,t) -2i{\cL}_{\sgb(\s)}^{\nrm;\mnrn}(\s)
\srac{\overline{-n(r)}}{\r(\s)} \st_\nrm \st_\mnrn \hg_- (\st,\xi,t) \,.
\end{align}
\end{subequations}
It is easily checked that these equations are consistent
\begin{gather}
\pl_+ \pl_- \hg_\pm (\st,\xi,t) = \pl_- \pl_+ \hg_\pm (\st,\xi,t) =0
\label{Eq4.25}
\end{gather}
because $\hg_+(\st) ,\hj^{(+)} (\xi)$ are chiral and $\hg_-(\st) ,\hj^{(-)}
(\xi)$ are anti-chiral.

Together, the chirality conditions \eqref{Eq4.23} and the light-cone vertex
operator equations \eqref{Eq4.24} will determine the dynamics of the full
open-string primary fields $\hg =\hg_- \hg_+$. We defer this analysis to
the following subsection however, focussing here on further properties of
the constituent fields.

As an example, we may use the differential equations \eqref{Eq4.24} and the
commutator identities
\begin{subequations}
\label{Eq4.26}
\begin{gather}
[\hj_\nrm^{(+)\pm}(\xi_i,t_i) ,\hg_+ (\st^{(j)},\xi_j ,t_j)] =\pm
\frac{e^{i\bnrrs (\phi_j -\phi_i)}}{1-e^{i(\phi_j-\phi_i)}} \hg_+
(\st^{(j)},\xi_j,t_j) \st_\nrm^{(j)} \\
[\hj_\nrm^{(-)\pm}(\xi_i,t_i) ,\hg_- (\st^{(j)},\xi_j ,t_j)] =\mp
\frac{e^{i\bnrrs (\bar{\phi}_j
-\bar{\phi}_i)}}{1-e^{i(\bar{\phi}_j-\bar{\phi}_i)}} \st_\nrm^{(j)} \hg_-
(\st^{(j)},\xi_j ,t_j)
\end{gather}
\end{subequations}
to find the twisted KZ equations for the chiral and anti-chiral correlators
$\hat{A}_\s^{\pm}$:
\begin{gather}
\hg_\pm (i) \equiv \hg_\pm (\st^{(i)} ,\xi_i ,t_i) ,\quad \hat{A}^\pm_\s
\equiv \langle \hg_\pm (1) \cdots \hg_\pm (n) \rangle_\s ,\quad
  \pl_{i-} \hat{A}_\s^+ = \pl_{i+} \hat{A}_\s^- =0   \label{Eq4.27}
\end{gather}
\begin{subequations}
\label{Eq4.28}
\begin{align}
\srac{1}{2} \pl_{i+} \hat{A}_\s^+ =& 2i{\cL}_{\sgb(\s)}^{\nrm;\mnrn}(\s)
\sum_{j\neq i}
   \frac{e^{i\bnrrs (\phi_j -\phi_i)}}{1-e^{i(\phi_j-\phi_i)}} \hat{A}_\s^+
\st_\nrm^{(j)} \st_\mnrn^{(i)} \nn \\
&+i\hat{A}_\s^+ \D_{\sgb(\s)}(\st^{(i)})
-2i{\cL}_{\sgb(\s)}^{\nrm;\mnrn}(\s) \sbnrrs \hat{A}_\s^+ \st_\nrm^{(i)}
\st_\mnrn^{(i)} 
\end{align}
\begin{align}
\srac{1}{2} \pl_{i-} \hat{A}_\s^- =& 2i{\cL}_{\sgb(\s)}^{\nrm;\mnrn}(\s)
\sum_{j\neq i}
   \frac{e^{i\bnrrs (\bar{\phi}_j
-\bar{\phi}_i)}}{1-e^{i(\bar{\phi}_j-\bar{\phi}_i)}} \st_\mnrn^{(i)}
\st_\nrm^{(j)} \hat{A}_\s^-  \nn \\
&+i\D_{\sgb(\s)}(\st^{(i)}) \hat{A}_\s^-
-2i{\cL}_{\sgb(\s)}^{\nrm;\mnrn}(\s) \srac{\overline{-n(r)}}{\r(\s)}
\st_\nrm^{(i)}\st_\mnrn^{(i)} \hat{A}_\s^-
\end{align}
\begin{gather}
\pl_{i\pm} \equiv \pl_{t_i} \pm \pl_{\xi_i} ,\quad \st^{(i)} =\st(T^{(i)})
,\quad i=1\ldots n \,.
\end{gather}
\end{subequations}
These twisted KZ equations are similar in form to the chiral and
anti-chiral twisted KZ equations of closed-string orbifold theory
\cite{Big,Big',Perm,so2n}.
In contrast to closed-string orbifold theory, however, the open-string
correlators $\hat{A}_\s$ of the full twisted affine primary fields
{\it cannot} be factorized into the open-string chiral and anti-chiral
correlators $\hat{A}_\s^\pm$:
\begin{subequations}
\label{Eq4.29}
\begin{gather}
\hg(i) \equiv \hg(\st^{(i)},\xi_i,t_i) = \hg_-(i) \hg_+(i) \\
\hat{A}_\s =\langle \hg(1) \cdots \hg(n) \rangle_\s \neq \langle \hg_-(1)
\cdots \hg_-(n) \rangle_\s \langle \hg_+(1) \cdots \hg_+(n) \rangle_\s =
  \hat{A}_\s^- \hat{A}_\s^+ \,.
\end{gather}
\end{subequations}
This follows because in open WZW theory \cite{Giusto}, twisted or untwisted, the
{\it single set} of current modes $\hj$ (or $J$) acts on {\it both}
chiral constituents of the primary fields, so that the chiral constituents
are not independent.

Finally, we may use the chirality \eqref{Eq4.23} to recast the commutators
\eqref{Eq4.18} in the form
\begin{subequations}
\label{Eq4.30}
\begin{gather}
[L_\s (m),\hg_+ (\st,\xi,t)] =\hg_+ (\st,\xi,t) (-\frac{i}{2} \lpl_{\!\!+}
+m\Dg (\st)) e^{im(t+\xi)} \\
[L_\s (m),\hg_- (\st,\xi,t)] =e^{im(t-\xi)} (-\frac{i}{2}\pl_- +m\Dg (\st))
\hg_- (\st,\xi,t)
\end{gather}
\end{subequations}
and these commutators also satisfy the $L,L ,\hg_\pm$ Jacobi identities, as
expected.

\subsection{The Full Twisted Vertex Operator Equations}

In this subsection, we assemble the chiral and anti-chiral constituent
information above to obtain the complete set of vertex operator equations
$\pl_t \hg,
\,\pl_\xi \hg$ for the full twisted affine primary fields $\hg$ of the
open-string orbifold.

We begin by writing down the following set of light-cone twisted vertex
operator equations for $\hg$:
\begin{subequations}
\label{Eq4.31}
\begin{gather}
\srac{1}{2} \pl_+ \hg(\st,\xi,t) =2i{\cL}_{\sgb(\s)}^{\nrm ;\mnrn}(\s)
:\!\hj_\nrm^{(+)} (\xi,t) \hg(\st,\xi,t) \st_\mnrn \!:_M \bigspc \nn \\
  \quad \quad +i\hg(\st,\xi,t) \D_{\sgb(\s)} (\st)
-2i{\cL}_{\sgb(\s)}^{\nrm ;\mnrn}(\s) \sbnrrs \hg(\st,\xi,t) \st_\nrm
\st_\mnrn \nn \\
  \quad \quad -2i{\cL}_{\sgb(\s)}^{\nrm ;\mnrn}(\s) \frac{e^{-2i\bnrrs
\xi}}{1-e^{-2i\xi}} \st_\nrm \hg(\st,\xi,t) \st_\mnrn \bigspc 
\end{gather}
\begin{gather}
\srac{1}{2} \pl_- \hg(\st,\xi,t) =-2i{\cL}_{\sgb(\s)}^{\nrm ;\mnrn}(\s)
:\!\st_\nrm \hj_\mnrn^{(-)} (\xi,t) \hg(\st,\xi,t)\!:_M \bigspc \nn \\
  \quad \quad +i\D_{\sgb(\s)}(\st) \hg(\st,\xi,t) -2i{\cL}_{\sgb(\s)}^{\nrm
;\mnrn}(\s) \srac{\overline{-n(r)}}{\r(\s)} \st_\nrm \st_\mnrn
\hg(\st,\xi,t) \nn \\
  \quad \quad -2i{\cL}_{\sgb(\s)}^{\nrm ;\mnrn}(\s)
\frac{e^{2i\frac{\overline{-n(r)}}{\r(\s)} \xi}}{1-e^{2i\xi}} \st_\nrm
\hg(\st,\xi,t) \st_\mnrn \,.\bigspc
\end{gather}
\end{subequations}
To obtain these equations, we used the chiralities \eqref{Eq4.23} and the
constituent vertex operator equations \eqref{Eq4.24} -- and moreover, as in
Subsec.~$4.3$, the commutators \eqref{Eq4.17} were used to obtain the fully
normal-ordered form given here. The consistency of this system
\begin{gather}
[ \pl_+ ,\pl_- ] \hg(\st,\xi,t) =0 ,\quad \pl_\pm =\pl_t \pm \pl_\xi
\label{Eq4.32}
\end{gather}
follows by construction from $\hg =\hg_- \hg_+$.

Using the light-cone vertex operator equations \eqref{Eq4.31}, we then
immediately obtain the vertex operator equations for the $t$- and
$\xi$-derivatives separately:
\begin{subequations}
\label{Eq4.33}
\begin{align}
\!&\!\pl_t \hg(\st,\xi,t) \!=\nn \\
& \,\,\,\, =\!2i{\cL}_{\sgb(\s)}^{\nrm;\mnrn}(\s) \left(
:\!\hj_\nrm^{(+)}(\xi,t) \hg(\st,\xi,t) \st_\mnrn
   \!-\!\st_\nrm \hj_\mnrn^{(-)}(\xi,t) \hg(\st,\xi,t) \!:_M \right) \,\,\,
\nn \\
& \quad \, +i [\Dg (\st) ,\hg(\st,\xi,t) ]_+
-2i{\cL}_{\sgb(\s)}^{0,\m;0,\n} (\s) \st_{0,\m} \hg(\st,\xi,t) \st_{0,\n}
\nn \\
& \quad \, -2i{\cL}_{\sgb(\s)}^{\nrm;\mnrn} (\s) \left(
\srac{\bar{n}(r)}{\r(\s)} \hg(\st,\xi,t) \st_\nrm \st_\mnrn \!+\!
   \srac{\overline{-n(r)}}{\r(\s)} \st_\nrm \st_\mnrn \hg(\st,\xi,t) \right) \\
\!&\!\pl_\xi \hg(\st,\xi,t) \!=\nn \\
& \,\,\,\, =\! 2i{\cL}_{\sgb(\s)}^{\nrm;\mnrn}(\s) \left(
:\!\hj_\nrm^{(+)}(\xi,t) \hg(\st,\xi,t) \st_\mnrn \!+\!
   \st_\nrm \hj_\mnrn^{(-)}(\xi,t) \hg(\st,\xi,t) \!:_M \right) \,\,\, \nn \\
& \quad \, +\!i[ \Dg (\st) ,\hg(\st,\xi,t)] \nn \\
& \quad \, +\! 2i{\cL}_{\sgb(\s)}^{\nrm;\mnrn} (\s) \Big{(} i\cot \xi
\de_{\bar{n}(r),0}-\! \frac{2e^{-2i\bnrrs \xi}}{1-e^{-2i\xi}}
   (1\!-\!\de_{\bar{n}(r),0}) \Big{)} \st_\nrm \hg(\st,\xi,t) \st_\mnrn \nn \\
&\quad \, -2i{\cL}_{\sgb(\s)}^{\nrm;\mnrn} (\s) \Big{(}
\srac{\bar{n}(r)}{\r(\s)} \hg(\st,\xi,t) \st_\nrm
   \st_\mnrn \!-\! \srac{\overline{-n(r)}}{\r(\s)} \st_\nrm \st_\mnrn
\hg(\st,\xi,t) \Big{)} \,. \label{Eq 4.33b}
\end{align}
\end{subequations}
We remind that the $\pl_t$ equation appeared earlier in Eq.~\eqref{Eq 4.16c}.

As anticipated in Subsec.~$4.3$, the singularities in Eq.~($4.17$a,b) at
$\xi=0$ and $\pi$ persist in the $\pl_\xi \hg$ equation \eqref{Eq 4.33b} --
and hence
in the twisted affine primary fields themselves. Following Ref.~\cite{Giusto},
we emphasize that these singularities are closely related to the
non-factorization phenomenon in
Eq.~\eqref{Eq4.29}: Both phenomena arise because the chiral constituents
$\hg_+$ and $\hg_-$ of open WZW strings do not live in independent
subspaces. On
the other hand, the affine primary fields of {\it closed} strings
(including ordinary orbifolds) have independent chiral constituents
$\hg_\pm$ -- and therefore have no such singularities.

We finally note that the chiral commutators in Eq.~\eqref{Eq4.30} and
the factorization \eqref{Eq4.14} give the following commutator for the
full twisted affine primary fields
\begin{subequations}
\label{Eq4.34}
\begin{align}
&[L_\s (m),\hg(\st,\xi,t)] =\hg (\st,\xi,t) (-\frac{i}{2} \lpl_{\!\!+}
+m\Dg (\st)) e^{im(t+\xi)} \bigspc \nn \\
& \bigspc \bigspc \bigspc +e^{im(t-\xi)} (-\frac{i}{2}\pl_- +m\Dg (\st))
\hg (\st,\xi,t) \\
&\quad \quad = e^{imt} \left( -i\cos (m\xi) \pl_t +\sin(m\xi) \pl_\xi
\right) \hg(\st,\xi,t) \nn \\
& \bigspc \bigspc +me^{imt} \left( e^{im\xi} \hg(\st,\xi,t) \Dg (\st)
+e^{-im\xi} \Dg (\st) \hg(\st,\xi,t) \right)
\end{align}
\end{subequations}
where the twisted conformal weight matrix $\Dg (\st)$ was given in
Eq.~\eqref{Eq 4.6b}.

\subsection{The Twisted Open-String KZ Systems of $A_g^{open}(H)/H$}

Using the light-cone twisted vertex operator equations \eqref{Eq4.31} and
the commutators \eqref{Eq4.12}, we may now obtain the full
twisted open-string KZ system
\begin{subequations}
\label{Eq4.35}
\begin{gather}
\hat{A}_\s = {}_\s \langle 0| \hg (\st^{(1)} ,\xi_1 ,t_1) \ldots \hg
(\st^{(n)},\xi_n ,t_n) |0 \rangle_\s ,\quad \srange \\
\srac{1}{2} \pl_{+i} \hat{A}_\s =2i{\cL}_{\sgb(\s)}^{\nrm ;\mnrn}(\s)
\BIG{(} \sum_{j\neq i} \frac{e^{i\bnrrs (\phi_j -\phi_i)}}{1-e^{i(\phi_j
-\phi_i)}}
   \hat{A}_\s \st_\nrm^{(j)} \st_\mnrn^{(i)} \bigspc \bigspc \bigspc \nn \\
\quad -\!\sum_j \frac{e^{i\bnrrs (\bar{\phi}_j
-\phi_i)}}{1-e^{i(\bar{\phi}_j -\phi_i)}} \st_\nrm^{(j)} \hat{A}_\s
\st_\mnrn^{(i)}
  \!-\!\sbnrrs \hat{A}_\s \st_\nrm^{(i)} \st_\mnrn^{(i)} \BIG{)}
+i\hat{A}_\s \Dg (\st^{(i)}) \\
\srac{1}{2} \pl_{-i} \hat{A}_\s =2i{\cL}_{\sgb(\s)}^{\nrm ;\mnrn}(\s)
\BIG{(} \sum_{j\neq i} \frac{e^{i\bnrrs (\bar{\phi}_j -\bar{\phi}_i)}}
   {1- e^{i(\bar{\phi}_j -\bar{\phi}_i)}} \st_\mnrn^{(i)} \st_\nrm^{(j)}
\hat{A}_\s \bigspc \bigspc \bigspc \nn \\
\quad -\!\sum_j \frac{e^{i\bnrrs (\phi_j -\bar{\phi}_i)}}{1- e^{i(\phi_j
-\bar{\phi}_i)}} \st_\mnrn^{(i)} \hat{A}_\s \st_\nrm^{(j)}
  \!-\!\srac{\overline{-n(r)}}{\r(\s)} \st_\nrm^{(i)} \st_\mnrn^{(i)}
\hat{A}_\s \BIG{)} +i\Dg (\st^{(i)}) \hat{A}_\s 
\end{gather}
\begin{gather}
\hat{A}_\s \sum_{i=1}^n \st_{0,\m}^{(i)} - \sum_{i=1}^n \st_{0,\m}^{(i)}
\hat{A}_\s =0 \\
\pl_{\pm i} = \pl_{t_i} \pm \pl_{\xi_i} ,\quad \phi_i =t_i +\xi_i ,\quad
\bar{\phi}_i =t_i -\xi_i ,\quad \st^{(i)} =\st(T^{(i)},\s) ,\quad i=1\ldots
n
\end{gather}
\end{subequations}
in sector $\s$ of each open-string WZW orbifold. The explicit formulae for
the inverse inertia tensor ${\cL}$ and the twisted representation matrices
$\st$ are given in Eqs.~($2.2$c,d), while the integers $\bar{n}(r),\r(\s)$
are defined in the $H$-eigenvalue problem \eqref{Eq 2.4a} of the underlying
untwisted theory.

The twisted open-string KZ system in Eq.~\eqref{Eq4.35} is another central
result of this paper. For brevity we will not give the $\pl_{t_i},
\pl_{\xi_i}$ form of this system, which is easily obtained by linear
combination.

As an explicit example, we give the information needed to realize the
twisted KZ systems of the open-string WZW permutation orbifolds:
\begin{subequations}
\label{Eq4.36}
\begin{gather}
\nrm \rightarrow \hat{j} aj ,\quad \sbnrrs = \srac{\bar{\hat{j}}}{f_j(\s)}
,\quad g =\oplus_I \gfrak^I ,\quad \gfrak^I \simeq \text{ simple } \gfrak \\
{\cL}_{\sgb(\s)}^{\nrm;\nsn}(\s) \rightarrow {\cL}_{\sgb(\s)}^{\hat{j}aj
;\hat{l}bl}(\s) = \de^{jl} \frac{1}{f_j(\s)} \frac{\eta^{ab}}{2k
+Q_{\gfraks}} \\
\st_\nrm (T,\s) \rightarrow \st_{\hat{j}aj} (T,\s) = T_a t_{\hat{j}j}(\s) \\
\bar{\hat{j}} = 0,\ldots ,f_j(\s) -1 ,\quad a =1,\ldots \text{dim } \gfrak \,.
\end{gather}
\end{subequations}
This class of open strings arises when we appropriate our initial data
from the sectors of any closed-string WZW permutation orbifold
\cite{Big,Big',Perm} on semisimple $g$. The cycle
notation in Eq.~\eqref{Eq4.36} and the branes of these permutation-twisted
open WZW strings were discussed in Subsec.~$3.5$.

Similarly, the explicit data \cite{Big,Big',Perm,so2n} for the various
closed-string orbifolds on simple $g$ can be substituted into
Eq.~\eqref{Eq4.35} to obtain
the twisted KZ systems of the corresponding open-string WZW orbifolds.

\subsection{The One-Sided Form of the Twisted KZ Systems}

In the {\it two-sided} notation above, the twisted representation matrices
$\st$ act on both sides of the open-string primary fields $\hg$ and the
open-string correlators $\hat{A}_\s$. The open-string dynamics is however
more transparent when expressed in an equivalent {\it one-sided} notation
\cite{Giusto,Orient1}:
\begin{subequations}
\label{Eq4.37}
\begin{gather}
\tilde{\hg}(\st,\bz,z,\s)^{\Nsn ;\Nrm} \equiv
\hg(\st,\bz,z,\s)_{\Nrm}{}^{\Nsn} \label{Eq 4.37a} \\
(B \hg(\st,\bz,z,\s) C )_{\Nrm}{}^{\Nsn} = B_{\Nrm}{}^{\Ntd}
\hg_{\Ntd}{}^{N(u)\ep} C_{N(u)\ep}{}^{\Nsn} \bigspc \bigspc \nn \\
\bigspc =\!-\tilde{\hg}^{N(u)\ep;\Ntd} C_{N(u)\ep}{}^{\!\!\!\Nsn}
(\bar{B})_{\Ntd}{}^{\!\!\!\Nrm} \label{Eq 4.37b} \\
=\!-(\tilde{\hg}(\st,\bz,z,\s) C \otimes \bar{B} )^{\Nsn;\Nrm} \quad \,
\label{Eq 4.37c} \\
\bar{B}=-B^t ,\quad (B^t )_{\Nrm}{}^{\Nsn} \equiv B_{\Nsn}{}^{\Nrm} \,.
\end{gather}
\end{subequations}
Here $t$ is matrix transpose and $B^t$ is the image on the right of $B$ on
the left. In our application, we will then need the barred\footnote{The
last
forms of $\bar{\st}$ in \eqref{Eq 4.38a} use the eigenvector matrix
relation $U(\bar{T},\s) =U(T,\s)^\ast$ chosen for untwisted rep $\bar{T}$
in
Ref.~\cite{Orient1}.} matrices $\bar{\st}$
\begin{subequations}
\label{Eq4.38}
\begin{gather}
\bar{\st}_\nrm (T,\s) \equiv -\st_\nrm (T,\s)^t \bigspc \bigspc \bigspc
\bigspc \nn \\
\bigspc \bigspc \,\,\, =\schisig_\nrm U(\s)_\nrm{}^a U(\bar{T},\s) \bT_a
U\hc (\bar{T},\s) =\st_\nrm (\bar{T},\s) \label{Eq 4.38a} \\
[\bar{\st}_\nrm ,\bar{\st}_\nrm ] =i\scf_{\nrm;\nsn}{}^{n(r)+n(s),\de}(\s)
\bar{\st}_{n(r)+n(s),\de}
\end{gather}
\end{subequations}
which are the image of $\st$ and moreover satisfy the same orbifold Lie
algebra \eqref{Eq 2.2f}. We remind the reader that, as shown in
Eqs.~\eqref{Eq 4.37b},
\eqref{Eq 4.37c}, $\bar{\st}$ always acts on the right indices of
$\tilde{\hg}$, while $\st$ acts on the left indices. This $C\!\otimes
\!\bar{B}$
bookkeeping should be born in mind even though we sometimes neglect the
ordering in the tensor product
\begin{gather}
\bar{\st} \!\otimes \!\st  \simeq \st \!\otimes \!\bar{\st}  \label{Eq4.39}
\end{gather}
for notational convenience.

It is straightforward to reexpress all the results above in the one-sided
notation, but we limit ourselves here to the one-sided form of the
twisted open-string KZ system
\begin{subequations}
\label{Eq4.40}
\begin{gather}
\tilde{\hat{A}}_\s (\st,\xi,t) \equiv {}_\s \langle 0| \tilde{\hg}
(\st^{(1)},\xi_1 ,t_1) \ldots \tilde{\hg} (\st^{(n)} ,\xi_n ,t_n)
|0\rangle_\s \\
\srac{1}{2} \pl_{+i} \tilde{\hat{A}}_\s (\st,\xi,t) = \tilde{\hat{A}}_\s
(\st,\xi,t) \hat{W}_i (\xi,t,\s) ,\quad \srac{1}{2} \pl_{-i}
\tilde{\hat{A}}_\s
  (\st,\xi,t) = \tilde{\hat{A}}_\s (\st,\xi,t) \hat{\bar{W}}_i (\xi,t,\s) \\
\hat{W}_i (\xi,t,\s) = 2i{\cL}_{\sgb(\s)}^{\nrm;\mnrn}(\s) \Big{(}
\sum_{j\neq i} \frac{e^{i\bnrrs (\phi_j -\phi_i)}}{1-e^{i(\phi_j -\phi_i)}}
   \st_\nrm^{(j)} + \sum_j \frac{e^{i\bnrrs (\bar{\phi}_j -\phi_i)}}{1-
e^{i(\bar{\phi}_j -\phi_i)}} \bar{\st}_\nrm^{(j)} \bigspc \nn \\
  \bigspc \bigspc -\sbnrrs \st_\nrm^{(i)} \Big{)} \st_\mnrn^{(i)} +i \Dg
(\st^{(i)}) 
\end{gather}
\begin{gather}
\hat{\bar{W}}_{\!i} (\xi,t,\s) =2i{\cL}_{\sgb(\s)}^{\nrm;\mnrn}(\s) \Big{(}
\sum_{j\neq i} \frac{e^{i\bnrrs (\bar{\phi}_j -\bar{\phi}_i)}}
  {1-e^{i(\bar{\phi}_j -\bar{\phi}_i)}} \bar{\st}_\nrm^{(j)} + \sum_j
\frac{e^{i\bnrrs (\phi_j -\bar{\phi}_i)}}{1-e^{i(\phi_j -\bar{\phi}_i)}}
  \st_\nrm^{(j)} \bigspc \nn \\
  \bigspc \bigspc -\sbnrrs \bar{\st}_\nrm^{(i)} \Big{)}
\bar{\st}_\mnrn^{(i)} +i \Dg (\bar{\st}^{(i)}) \\
\tilde{\hat{A}}_\s (\st,\xi,t) \sum_{i=1}^n (\st_{0,\m}^{(i)} \otimes \one
+ \one \otimes \bar{\st}_{0,\m}^{(i)} )=0 \\
\Dg (\bar{\st}) = {\cL}_{\sgb(\s)}^{\nrm;\mnrn} \bar{\st}_\nrm
\bar{\st}_\mnrn =\Dg (\st)^t =\Dg(\st)
  \label{Eq 4.40f}
\end{gather}
\end{subequations}
which holds in sector $\s$ of $A_g^{open}(H)/H$. Here $\hat{W},
\hat{\bar{W}}$ are the twisted KZ connections of the system. In the
$\sbnrrs \bar{\st}
\bar{\st}$ term of $\hat{\bar{W}}\!,$ we have used an $\nrm \leftrightarrow
\mnrn$ exchange to order the twisted representation matrices as shown.
We remind that, in spite of the numerical equality \eqref{Eq 4.40f}, the
twisted conformal weight matrix $\Dg (\bar{\st}^{(i)})$ operates on the
right indices of $\tilde{\hg}(\st^{(i)})$.

A more familiar form of the twisted open-string KZ system is obtained in
terms of the reduced correlators $\tilde{\hat{F}}_\s$
\begin{subequations}
\label{Eq4.41}
\begin{gather}
\tilde{\hat{F}}_\s (\st,\bz,z) \equiv \prod_{i=1}^n \tilde{\hat{A}}_\s
(\st,\xi,t) z_i^{-\Dg (\st^{(i)})} \otimes \bz_i^{-\Dg (\bar{\st}^{(i)})} \\
\pl_i \tilde{\hat{F}}_\s (\st,\bz,z) = \tilde{\hat{F}}_\s (\st,\bz,z)
\hat{W}_i (\bz,z,\s) ,\quad \bpl_i \tilde{\hat{F}}_\s (\st,\bz,z) =
\tilde{\hat{F}}_\s (\st,\bz,z)
   \hat{\bar{W}}_i (\bz,z,\s) \\
\hat{W}_i (\bz,z,\s) =2{\cL}_{\sgb(\s)}^{\nrm;\mnrn}(\s) \Big{(}
\sum_{j\neq i} \left( \frac{z_j}{z_i} \right)^\bnrrs \frac{\st_\nrm^{(j)}
\otimes
  \st_\mnrn^{(i)}}{z_i -z_j} \bigspc \bigspc \nn \\
  \bigspc +\sum_j \left( \frac{\bz_j}{z_i} \right)^\bnrrs
\frac{\bar{\st}_\nrm^{(j)} \otimes \st_\mnrn^{(i)}}{z_i -\bz_j} -\sbnrrs
\frac{1}{z_i}
  \st_\nrm^{(i)} \st_\mnrn^{(i)} \Big{)} 
\end{gather}
\begin{gather}
\hat{\bar{W}}_{\!i} (\bz,z,\s) =2{\cL}_{\sgb(\s)}^{\nrm;\mnrn}(\s) \Big{(}
\sum_{j\neq i} \left( \frac{\bz_j}{\bz_i} \right)^\bnrrs
\frac{\bar{\st}_\nrm^{(j)}
  \otimes \bar{\st}_\mnrn^{(i)}}{\bz_i -\bz_j} \bigspc \bigspc \nn \\
  \bigspc +\sum_j \left( \frac{z_j}{\bz_i} \right)^\bnrrs
\frac{\st_\nrm^{(j)} \otimes \bar{\st}_\mnrn^{(i)}}{\bz_i -z_j} -\sbnrrs
\frac{1}{\bz_i}
  \bar{\st}_\nrm^{(i)} \bar{\st}_\mnrn^{(i)} \Big{)} \\
\tilde{\hat{F}}_\s (\st,\bz,z) \sum_{i=1}^n (\st_{0,\m}^{(i)} \otimes \one
+ \one \otimes \bar{\st}_{0,\m}^{(i)} )=0 \\
z_i \equiv e^{i\phi_i} ,\quad \bz_i \equiv e^{i\bar{\phi}_i} ,\quad \pl_i
\equiv \srac{\pl}{\pl z_i} =-\srac{i}{2z_i} \pl_{+i} ,\quad \bpl_i
   \equiv \srac{\pl}{\pl \bz_i} =-\srac{i}{2\bz_i} \pl_{-i} 
\end{gather}
\begin{gather}
\left( \frac{z_j}{z_i} \right)^\bnrrs \equiv e^{i\bnrrs (\phi_j -\phi_i)}
,\quad \left( \frac{\bz_j}{z_i} \right)^\bnrrs \equiv e^{i\bnrrs
(\bar{\phi}_j -\phi_i)} \\
\left( \frac{z_j}{\bz_i} \right)^\bnrrs \equiv e^{i\bnrrs (\phi_j
-\bar{\phi}_i)} ,\quad \left( \frac{\bz_j}{\bz_i} \right)^\bnrrs \equiv
   e^{i\bnrrs (\bar{\phi}_j -\bar{\phi}_i)} \label{Eq 4.41h} \\
\s =0,\ldots ,N_c -1
\end{gather}
\end{subequations}
where the variables $\bz,z$ are the Minkowski-space analogues of the usual
Euclidean variables on the sphere. An analytic continuation of this system
to Euclidean space would presumably maintain the phase conventions in
Eqs.~($4.41$g,h).

As emphasized for the untwisted case in Ref.~\cite{Giusto} and expected
generally in open strings, the system \eqref{Eq4.41} shows interactions
among
charges $\st^{(i)}$ at $z_i$ and image charges $\bar{\st}^{(i)}$ at $\bz_i$.

In fact, the twisted open-string KZ system \eqref{Eq4.41} can be
understood as a ``doubled" but {\it ordinary} chiral orbifold KZ system
\cite{Big,Big',Perm,so2n}
on $2n$ variables:
\begin{subequations}
\label{Eq4.42}
\begin{gather}
\hat{F}_\s (\st ,\{ z\}) \equiv \tilde{\hat{F}}_\s (\st,\bz,z) ,\quad
\pl_\kappa \hat{F}_\s (\st,\{z\}) =\hat{F}_\s (\st,\{z\}) \hat{W}_\kappa
(\st,\{z\},\s)
\end{gather}
\begin{align}
\hat{W}_\kappa (\st,\{z\},\s) &= 2{\cL}_{\sgb(\s)}^{\nrm;\mnrn}(\s) \BIG{[}
\!\sum_{\r \neq \kappa} \left( \frac{z_\r}{z_\kappa}
  \right)^{\!\nrrsf} \!\!\frac{1}{z_{\kappa \r}} \st_\nrm^{(\r)} \otimes
\st_{\mnrn}^{(\kappa)} \bigspc \nn \\
& \bigspc \bigspc \bigspc \quad \quad -\frac{1}{z_\kappa} \nrrs
\st_\nrm^{(\kappa)} \st_{\mnrn}^{(\kappa)} \BIG{]}
\end{align}\vspace{-0.25in}
\begin{gather}
\hat{F}_\s (\st,\{z\}) \sum_{\kappa =1}^{2n} \st_{0,\m}^{(\kappa )} =0
,\quad \forall \m \\
\pl_\kappa \equiv \frac{\pl}{\pl z_\kappa} ,\quad z_{\kappa \r} \equiv
z_\kappa -z_\r ,\quad z_\kappa \equiv \left\{ \begin{array}{ll} z_\kappa ,
   &\kappa =1\ldots n, \\ \bz_{\kappa -n} ,&\kappa =n+1 \ldots 2n
\end{array} \right. 
\end{gather}
\begin{gather}
\st_\nrm^{(\kappa)} \!\equiv \!\left\{ \begin{array}{ll} \st_\nrm
(T^{(\kappa)}\!,\s) &\kappa =1\ldots n \\
   \st_\nrm (\bT^{(\kappa -n)}\!,\s) & \kappa =n+1\ldots 2n \,. \end{array}
\right.
\end{gather}
\end{subequations}
It follows that the $n$-point correlators of our twisted open WZW strings
have the same general structure as the $2n$-point correlators of
closed-string orbifold theory. This structure for the correlators of
open-string CFTs \cite{Cardy1,Cardy2} was emphasized in the untwisted open WZW theory of
Ref.~\cite{Giusto} and in the open-string sectors of the WZW orientation
orbifolds \cite{Orient1}.

\section{Discussion}

In this paper, we have constructed what we call the {\it basic class} of
twisted open WZW strings. This class consists of all the sectors
$\s =0\ldots N_c -1$ of the general open-string orbifold $A_g^{open}(H)/H$,
which is an orbifold of the Giusto-Halpern open string $A_g^{open}(H)$
by a symmetry $H \subset Aut(g)$.

Our construction generalizes the untwisted construction of Ref.~\cite{Giusto}, and
 all of our results reduce in the untwisted sector $\s=0$ to the
 corresponding results of that reference, including e.~g.~the untwisted
open-string KZ equations
\begin{subequations}
\label{Eq5.1}
\begin{gather}
\tilde{F} (\bz,z) \equiv \prod_{i=1}^n \tilde{A}(T,\xi,t) z_i^{-\Delta
(T^{(i)})} \otimes \bz_i^{-\Delta (T^{(i)})} \\
\pl_i \tilde{F}(\bz,z) =\tilde{F}(\bz,z) W_i (\bz,z) ,\quad \bpl_i
\tilde{F}(\bz,z) =\tilde{F}(\bz,z) \bar{W}_i (\bz,z)
\end{gather}
\begin{gather}
W_i (\bz,z) =2L_g^{ab} \left( \sum_{j\neq i} \frac{T_a^{(j)} \otimes
T_b^{(i)}}{z_i -z_j} +\sum_j \frac{\bT_a^{(j)} \otimes T_b^{(i)}}{z_i
-\bz_j} \right) \\
\bar{W}_i (\bz,z) =2L_g^{ab} \left( \sum_{j\neq i} \frac{\bT_a^{(j)}
\otimes \bT_b^{(i)}}{\bz_i -\bz_j} +\sum_j \frac{T_a^{(j)} \otimes
\bT_b^{(i)}}
   {\bz_i -z_j} \right) \\
\tilde{F}(\bz,z) \sum_{i=1}^n (T_a^{(i)} \otimes \one + \one \otimes
\bT_a^{(i)}) =0 ,\quad \forall a=1\ldots \text{dim }g  \\
[T_a ,T_b] =if_{ab}{}^c T_c ,\quad [\bT_a ,\bT_b] =if_{ab}{}^c \bT_c
\end{gather}
\end{subequations}
where T is any irrep of g and $\bT =-T^t$. We remind the reader however
that these correlators
must be $H$-symmetrized because they now reside in sector $\s =0$ of an
(open-string)
orbifold.

As noted in the introduction, there is another known class of twisted open
WZW strings, namely the open-string sectors of the WZW orientation
orbifolds \cite{Orient1,Orient2}, and it is natural
to ask whether these twisted open strings are included among the
constructions in our basic class. Certainly one does not expect the basic
class to contain
{\it all} the open-string orientation-orbifold sectors, because the sectors
of each $A_g^{open}(H)/H$ are labelled by a complete set of conjugacy
classes
of $H$, whereas this is not always true for the open-string sectors of an
orientation orbifold. In fact, we can be certain that the basic class
contains
{\it no} open-string orientation-orbifold sectors for the following
technical reason: The twisted representation matrices associated to the
left- and
right-movers of $A_g^{open}(H)/H$ are $\st$ and $\bar{\st} =-\st^t$
respectively, while in the orientation orbifolds the left- and right-mover
twisted
representation matrices are {\it not} the bar of each other.

The construction of this paper should therefore be generalized in order to
include at least the open-string orientation-orbifold sectors,
 and presumably many other twisted open WZW strings. The general twisted
boundary state equation
 of  App.~A suggests the appropriate generalization:  The idea is to
introduce into the present construction an extra automorphism
$\tilde{\omega}(n(r),\s)$ of the twisted current algebra -- which is
expected to generate
new twisted open WZW strings T-dual to our basic class at $\tilde{\omega}
=1$.  It is of course well-known even at the untwisted level that N-N and D-D
strings are T-dual to each other, and are related by just such an
automorphism. Moreover, it is not difficult to check that the left- and
right-mover twisted representation matrices of the open-string WZW
orientation-orbifold sectors are indeed related by an automorphism of the
twisted current algebra. With these clues, we will return to discuss the
open-string
picture of the general twisted open WZW string in a future paper.

\vspace{-.02in}
\bigskip

\noindent
{\bf Acknowledgements}

For helpful discussions, we thank O.~Ganor, J.~Gomis, S.~Giusto,
P.~Ho\v{r}ava, N.~Ishibashi and C.~Schweigert.

This work was supported in part by the Director, Office of Energy Research,
Office of High Energy and Nuclear Physics, Division of High Energy Physics
of the U.S. Department of Energy under Contract DE-AC03-76SF00098 and in
part by the National Science Foundation under grant PHY00-98840.

\appendix

\section{General Twisted Boundary States}

The development of the text followed the standard {\it open-string
picture} \cite{Giusto} of open strings, in which the strings are described by a
{\it single}
set of current modes and a corresponding single Virasoro algebra. On the
other hand, in the standard {\it closed-string picture} of open strings
(see e.~g.
Refs.~\cite{Ishi,FS}), one describes the open string by a boundary state
equation in the space of left- and right-mover current modes.

In this appendix, we will discuss and extend what is known about the
boundary states in the closed-string picture of {\it twisted} open WZW
strings. In this
picture, we begin with the twisted left- and right-mover current algebras
\cite{Big} in sector $\s$ of any WZW orbifold $A_g(H)/H$
\begin{subequations}
\label{EqA.1}
\begin{align}
&\![\hj_\nrm(m\!+\!\srac{n(r)}{\r(\s)}),
\hj_\nsn(n\!+\!\srac{n(s)}{\r(\s)}) ] \!= \nn \\
&\bigspc \bigspc i\scf_{\nrm;\nsn}{}^{\!\!\!\!\!\!n(r)+n(s),\d}(\s)
\hj_{n(r)+n(s),\d}(m\!+\!n\!+\!\srac{n(r)+n(s)}{\r(\s)}) \nn \\
&\bigspc \bigspc \quad \quad + (m\!+\!\srac{n(r)}{\r(\s)})
\de_{m+n+\frac{n(r)+n(s)}{\r(\s)},\,0} \sG_{\nrm;-\nrn}(\s) \label{Eq A.1a}
\\
&\![\hjb_\nrm(m\!+\!\srac{n(r)}{\r(\s)}),
\hjb_\nsn(n\!+\!\srac{n(s)}{\r(\s)}) ] \!= \nn \\
&\bigspc \bigspc i\scf_{\nrm;\nsn}{}^{\!\!\!\!\!\!n(r)+n(s),\d}(\s)
\hjb_{n(r)+n(s),\d}(m\!+\!n\!+\!\srac{n(r)+n(s)}{\r(\s)}) \nn \\
& \bigspc \bigspc \quad \quad -(m\!+\!\srac{n(r)}{\r(\s)})
\de_{m+n+\frac{n(r)+n(s)}{\r(\s)},\,0} \sG_{\nrm;-\nrn}(\s) \label{Eq A.1b}
\end{align}
\begin{equation}
[\hj_\nrm(m\!+\!\srac{n(r)}{\r(\s)}), \hjb_\nsn(n\!+\!\srac{n(s)}{\r(\s)})]
= 0, \quad \srange
\end{equation}
\begin{gather}
\hj_{n(r) \pm \r(\s),\m} (m +\srac{n(r) \pm \r(\s)}{\r(\s)}) =\hj_\nrm (m
\pm 1+\nrrs) \\
\hjb_{n(r) \pm \r(\s),\m} (m +\srac{n(r) \pm \r(\s)}{\r(\s)}) =\hjb_\nrm
(m\pm 1+\nrrs)
\end{gather}
\end{subequations}
where $N_c$ is the number of conjugacy classes of $H\!\subset \!Aut(g)$.
The left-mover modes $\hj$ are the same as those of the text, and $\sG(\s)$
and
$\scf(\s)$ are the twisted metric and twisted structure constants of the
sector.

Because of the {\it sign reversal} of the central term\footnote{This
surprising feature follows directly from the WZW orbifold action, and has
also been explicitly checked at the level of characters \cite{Big}.} in the
right-mover current algebra \eqref{Eq A.1b}, Ref.~\cite{Big} was able to give
the following
consistent equation for the twisted open-string boundary states of twisted
sector $\s$
\begin{subequations}
\label{EqA.2}
\begin{gather}
\left( \hj_\nrm (\mnrrs) + \hjb_\nrm (\mnrrs) \right) |B \rangle_\s =0
,\quad \srange \label{Eq A.2a}\\
\left[ \Big{(} \hj +\hjb \Big{)}_\nrm (\mnrrs) , \Big{(} \hj + \hjb
\Big{)}_\nsn (\nnsrs) \right] = \bigspc \bigspc \nn \\
\bigspc = i \scf_{\nrm ;\nsn}{}^{n(r)+n(s) ,\de} (\s)  \Big{(}\hj +\hjb
\Big{)}_{n(r)+n(s),\de} (\mnnrnsrs)
\end{gather}
\end{subequations}
without any consideration of the {\it rectification problem} \cite{Big,Big',
Perm,so2n} reviewed in App.~B. We expect that these twisted boundary states
describe
the twisted open strings in the basic class of this paper, but we emphasize
that for $\s \neq 0$ no attempt has thus far been made to find explicit
solutions of this system.

As an extension, we consider next the corresponding action of the Virasoro
generators on this set of twisted boundary states. For this computation,
we will need the explicit forms \cite{Big} of the left- and right-mover Virasoro
generators in terms of twisted current modes:
\begin{subequations}
\label{EqA.3}
\begin{align}
L_\s (m) = & {\cL}_{\sgb (\s)}^{\nrm;\mnrn} (\s) \sum_{p \in \Zint}
:\!\hj_\nrm (p+\nrrs) \hj_\mnrn (m-p-\nrrs)\! : \nn \\
& ={\cL}_{\sgb (\s)}^{\nrm;\mnrn} (\s) \Big{(} \sum_{p \in \Zint}
:\!\hj_\nrm (p+\nrrs) \hj_\mnrn (m-p-\nrrs)\! :_M \nn \\
& - i\srac{\bar{n}(r)}{\r(\s)} \scf_{\nrm;\mnrn}{}^{0,\de} (\s) \hj_{0,\de}
(m) \Big{)} +\de_{m,0} \gscfwt \label{Eq A.3a}
\end{align}
\begin{align}
\bar{L}_\s (m) = & {\cL}_{\sgb (\s)}^{\nrm;\mnrn} (\s) \sum_{p\in \Zint}
:\!\hjb_\nrm (p+\nrrs) \hjb_\mnrn (-m-p-\nrrs)\! : \nn \\
& ={\cL}_{\sgb (\s)}^{\nrm;\mnrn} (\s) \Big{(} \sum_{p\in \Zint}
:\!\hjb_\nrm (p+\nrrs) \hjb_\mnrn (-m-p-\nrrs)\! :_{\bar{M}} \nn \\
& - i\srac{\overline{-n(r)}}{\r(\s)} \scf_{\nrm;\mnrn}{}^{0,\de} (\s)
\hjb_{0,\de} (-m) \Big{)} +\de_{m,0} \gscfwt
\end{align}
\begin{align}
&\!\!:\!\hj_\nrm (m\!+\!\nrrs) \hj_\nsn (n\!+\!\nsrs)\! :_M \equiv \!\theta
(m\!+\!\nrrs \!\geq 0) \hj_\nsn (n\!+\!\nsrs) \hj_\nrm (m\!+\!\nrrs) \quad
\quad \nn \\
& \bigspc \bigspc + \theta (m\!+\!\nrrs <0) \hj_\nrm (m\!+\!\nrrs) \hj_\nsn
(n\!+\!\nsrs) \label{Eq A.3c} \\
&\!\!:\!\hjb_\nrm (m\!+\!\nrrs) \hjb_\nsn (n\!+\!\nsrs) \!:_{\bar{M}}
\equiv \!\theta (m\!+\!\nrrs \!\leq 0) \hjb_\nsn (n\!+\!\nsrs) \hjb_\nrm
(m\!+\!\nrrs) \quad \quad \nn \\
& \bigspc \bigspc + \theta (m\!+\!\nrrs >0) \hjb_\nrm (m\!+\!\nrrs)
\hjb_\nsn (n\!+\!\nsrs)
\end{align}
\begin{gather}
\gscfwt \equiv \sum_{r,\mu,\nu} {\cL}_{\sgb (\s)}^{\nrm;\mnrn} (\s)
\srac{\bar{n}(r)}{2\r(\s)} (1-\srac{\bar{n}(r)}{\r(\s)})
  \sG_{\nrm;\mnrn} (\s) \label{Eq A.3e} \\
\srac{\overline{-n(r)}}{\r(\s)} = \left\{ \begin{array}{ll}
   1 -\srac{\bar{n}(r)}{\r(\s)} & {\rm for}\,\; \bar{n}(r) \neq 0 \\
   0 & {\rm for}\,\; \bar{n}(r)=0  \
   \end{array} \right. \label{Eq A.3f}
\end{gather}
\begin{gather}
[L_\s (m) ,L_\s (n)] =(m\!-\!n) L_\s (m\!+\!n) +\de_{m+n,0} \frac{c_g}{12}
m(m^2 -1) \\
[\bar{L}_\s (m) ,\bar{L}_\s (n)] =(m\!-\!n) \bar{L}_\s (m\!+\!n)
+\de_{m+n,0} \frac{c_g}{12} m(m^2 -1) \\
[L_\s (m) ,\bar{L}_\s (n)] =0 \,.
\end{gather}
\end{subequations}
Here $:\,\cdot \,:_M$ and $:\,\cdot \,:_{\bar{M}}$ are the standard mode
normal orderings of the orbifold program \cite{Dual,More,Big}, and $:\, \cdot
\,:$
is the mode form of operator-product normal ordering. The left-mover
Virasoro generators $L_\s (m)$ are the same as those of the text (see e.~g.
Eq.~\eqref{Eq 2.1a}), and $\gscfwt$ in Eq.~\eqref{Eq A.3e} is the conformal
weight of the scalar twist field.

We are interested in particular in the relation of $\bar{L}_\s
|B\rangle_\s$ to $L_\s |B\rangle_\s$, as implied by the boundary-state
equation \eqref{EqA.2}.
To study $\bar{L}_\s |B\rangle_\s$ we found the following identities useful
\begin{subequations}
\label{EqA.4}
\begin{align}
\bkspc&\!\! : \!\hjb_\nrm (\mnrrs) \hjb_\mnrn (n-\nrrs)
\!:_{\bar{M}}|B\rangle_\s = \label{Eq A.4a} \\
& \quad \Big{(} \!:\!\hj_\nrm (\mnrrs) \hj_\mnrn (n-\!\nrrs)\!:_M +i
\de_{m+\nrrsf ,0} \scf_{0,\mu;0,\nu}{}^{0,\de} (\s) \hj_{-n(r),\de}
  (n-\!\nrrs) \Big{)} |B\rangle_\s \nn
\end{align}
\begin{gather}
{\cL}_{\sgb (\s)}^{0,\mu;0,\nu} (\s) \scf_{0,\mu;0,\nu}{}^{0,\de} (\s)
={\cL}_{\sgb (\s)}^{\nrm;\mnrn} (\s) \scf_{\nrm;\mnrn}{}^{0,\de} (\s) =0
\label{Eq A.4b} \\
{\cL}_{\sgb(\s)}^{\nrm;\mnrn}(\s) \srac{\overline{-n(r)}}{\r(\s)}
\scf_{\nrm;\mnrn}{}^{0,\de}(\s) \hjb_{0,\de}(m) |B\rangle_\s \bigspc
\bigspc \bigspc \nn \\
  \bigspc ={\cL}_{\sgb(\s)}^{\nrm;\mnrn}(\s) \srac{\bar{n}(r)}{\r(\s)}
\scf_{\nrm;\mnrn}{}^{0,\de}(\s) \hj_{0,\de}(m) |B\rangle_\s  \label{Eq A.4c}
\end{gather}
\end{subequations}
where \eqref{Eq A.4a} follows from the boundary-state equation, the
mode-ordering definitions (A$.3$c,d) and the twisted current algebra. The
relation in
Eq.~\eqref{Eq A.4b} is obtained from the symmetries ($2.2$b,c) of ${\cL}$
and $\scf$, and Eq.~\eqref{Eq A.4c} then follows from Eq.~\eqref{Eq A.3f}
and the
boundary-state equation.

With the identities in Eq.~\eqref{EqA.4} and the Virasoro generators in
Eq.~\eqref{EqA.3}, we then verify the twisted open-string Virasoro
conditions on
the twisted boundary states:
\begin{subequations}
\label{EqA.5}
\begin{gather}
\left( L_\s (m) - \bar{L}_\s (-m) \right) |B\rangle_\s =0 ,\quad \srange
\label{Eq A.5a} \\
[ L_\s (m) -\bar{L}_\s (-m) ,L_\s (n) -\bar{L}_\s (-n)] = (m-n) \left( L_\s
(m+n) - \bar{L}_\s (-m-n) \right) \,.
\end{gather}
\end{subequations}
Such Virasoro conditions are familiar \cite{Ishi} in the closed string picture
of untwisted open WZW strings.

The boundary state equation in Eq.~\eqref{EqA.2} is in fact a special case
of the following {\it general twisted boundary-state equation}
\begin{subequations}
\label{EqA.6}
\begin{gather}
\left( \hj_\nrm (\mnrrs) + \pom (n(r),\s)_\m{}^\n \hjb_{n(r),\n} (\mnrrs)
\right) |B\rangle_\s =0 \label{Eq A.6a} \\
\s =0 ,\ldots ,N_c -1
\end{gather}
\end{subequations}
which is consistent when $\pom$ is any mode-number-preserving automorphism
of the right- or left-mover twisted mode algebra \eqref{EqA.1}:
\begin{subequations}
\label{EqA.7}
\begin{gather}
\pom (n(r),\s)_\m{}^\k \pom(n(s),\s)_\n{}^\l
\scf_{n(r),\k;n(s),\l}{}^{n(t),\ep} (\s) = \scf_{\nrm;\nsn}{}^\ntd (\s)
\pom(n(t),\s)_\de{}^\ep \\
\pom (n(r),\s)_\m{}^\k \pom(n(s),\s)_\n{}^\l
\sG_{n(r)\srange,\k;n(s),\l}(\s) = \sG_{\nrm;\nsn} (\s) \label{Eq A.7c} \\
\sG^{n(r),\k;n(s),\l}(\s) \pom (n(r),\s)_\k{}^\m \pom(n(s),\s)_\l{}^\n =
\sG^{\nrm;\nsn} (\s) \,. \label{Eq A.7d}
\end{gather}
\end{subequations}
We expect that such an $\pom$ modification can also be used to describe the
general twisted open WZW string in the
open-string picture of the text (see the discussion in Sec.~5).

Similarly, the Virasoro conditions
\begin{gather}
\left( L_\s (m) - \bar{L}_\s (-m) \right) |B\rangle_\s =0 ,\quad \srange \nn
\end{gather}
hold as well for the general boundary state in Eq.~\eqref{EqA.6}. To see
this, a helpful identity is
\begin{equation}
\label{EqA.8}
{\cL}_{\sgb(\s)}^{n(r),\k;n(s),\l} (\s) \pom (n(r),\s)_\k{}^\m
\pom(n(s),\s)_\l{}^\n = {\cL}_{\sgb(\s)}^{\nrm;\nsn} (\s)
\end{equation}
which follows from Eqs.~\eqref{Eq 2.2c} and \eqref{Eq A.7d} when the
underlying untwisted current algebra has the special form:
\begin{subequations}
\label{EqA.9}
\begin{gather}
g =\oplus_I \gfrak^I ,\quad \gfrak^I \simeq \text{ simple } g ,\quad k^I =k \\
\Rightarrow {\cL}_{\sgb (\s)}^{\nrm;\nsn} (\s)= \frac{k}{2k+Q}
\sG^{\nrm;\nsn} (\s) \,. \label{Eq A.9b}
\end{gather}
\end{subequations}
The proof for the general case follows because the general untwisted
current algebra can be considered as a direct sum of these special cases.

The general twisted boundary-state equation in Eq.~\eqref{EqA.6} reduces to
the standard\footnote{See for example Refs.~\cite{Ishi,FS}.} untwisted boundary
state
equation in sector $\s=0$
\begin{gather}
\Big{(} J_a (m) +\omega_a{}^b \bJ_b (-m) \Big{)} |B\rangle_0 =0 ,\quad
a,b=1\ldots \text{dim }g ,\quad \omega \in Aut (g)  \label{EqA.10}
\end{gather}
where $\hjb$ reduces \cite{Big} to $\bJ (-m)$ and $\pom \rightarrow \omega$ is
any mode-preserving automorphism of untwisted affine $g$.

\section{Rectification in the Closed-String WZW Orbifolds}

In the text, we noticed that the equal-time current algebra of the twisted
open WZW string is not isomorphic in the bulk to the twisted current algebra
of the corresponding closed-string orbifold -- but we claimed instead that
the open-string algebra was isomorphic to the {\it rectified} current
algebra
of the closed-string orbifold. For application in the text, we therefore
review and extend in this appendix the concept of {\it rectification}
\cite{Big,Big',Perm,so2n} in closed-string orbifold theory, beginning with the rectified
current algebra.

The left- and right-mover current algebras of ordinary closed-string
orbifolds $A_g(H)/H$ are given in Eq.~\eqref{EqA.1}, and we remind the
reader
that the right-mover commutators are {\it not} a copy of the left-mover
commutators due to the sign reversal of the central term \cite{Big}. In what
follows
we will assume that the right-mover algebra can however be {\it rectified}
into a copy of the left-mover algebra
\begin{subequations}
\label{EqB.1}
\begin{gather}
\hjbb_\nrm (\mnrrs) \equiv \theta (n(r),\s)_\m{}^\n \hjb_{-n(r),\n}
(-m-\nrrs) \label{Eq B.1a}
\end{gather}
\begin{align}
\![\hj_\nrm(m\!+\!\srac{n(r)}{\r(\s)}), \hj_\nsn(n\!+\!\srac{n(s)}{\r(\s)})
] \!&\!=\! i\scf_{\nrm;\nsn}{}^{\!\!\!\!\!\!n(r)+
n(s),\d}(\s) \hj_{n(r)+n(s),\d}(m\!+\!n\!+\!\srac{n(r)+n(s)}{\r(\s)}) \quad
\quad \nn\\
 &+ (m\!+\!\srac{n(r)}{\r(\s)})\;\d_{m+n+\frac{n(r)+n(s)}{\r(\s)},\,0}\;\sG_{\nrm;-\nrn}(\s) \\\![\hjbb_\nrm(m\!+\!
\srac{n(r)}{\r(\s)}), \hjbb_\nsn(n\!+\!\srac{n(s)}{\r(\s)}) ] \!&\!=\!
i\scf_{\nrm;\nsn}{}^{\!\!\!\!\!\!n(r)+
n(s),\d}(\s) \hjbb_{n(r)+n(s),\d}(m\!+\!n\!+\!\srac{n(r)+n(s)}{\r(\s)})
\quad \quad \nn\\
 &+
(m\!+\!\srac{n(r)}{\r(\s)})\;\d_{m+n+\frac{n(r)+n(s)}{\r(\s)},\,0}\;\sG_{\nrm;-\
nrn}(\s)
\end{align}
\end{subequations}
where $\sharp$ denotes the rectified right-mover modes. The existence of
such a rectification is equivalent to the following conditions
\begin{subequations}
\label{EqB.2}
\begin{gather}
\theta(n(r),\s)_\m{}^\k \theta(n(s),\s)_\n{}^\l \sG_{-n(r),\k ;-n(s),\l}
(\s) = \sG_{\nrm;\nsn} (\s) \label{Eq B.2a} \\
\theta(n(r),\s)_\m{}^\k \theta(n(s),\s)_\n{}^\l \scf_{-n(r),\k
;-n(s),\l}{}^{-n(t),\ep}(\s) =\scf_{\nrm;\nsn}{}^\ntd (\s)
   \theta(n(t),\s)_\de{}^\ep \label{Eq B.2b}
\end{gather}
\end{subequations}
on the twisted metric and twisted structure constants of sector $\s$.

Finding such a rectification $\{ \theta \}$ for a given twisted right-mover
current algebra is non-trivial, but this
rectification problem has in fact been solved on a case-by-case basis

$\bullet$ the WZW permutation orbifolds \cite{Big,Big',Perm}

$\bullet$ the inner-automorphic WZW orbifolds \cite{Big} on simple $\gfrak$

$\bullet$ the outer-automorphic WZW orbifolds \cite{Big',so2n} on simple $\gfrak$

\noindent for all basic orbifold types. Given these results, one might
expect that all twisted right-mover current algebras can be rectified.
We emphasize however that this has not yet been systematically discussed
for more general twisted current algebras, such as the doubly-twisted
current algebras of Refs.~\cite{TVME,Coset} -- which arise from automorphisms
that are compositions of the basic types above.

In the application of the text, we will need the equal-time form of the
rectified current algebra. Towards this, we first use the mode algebra
\eqref{EqB.1} to obtain the equal-time algebra of the {\it unrectified}
orbifold currents on the cylinder \cite{Geom}:
\begin{subequations}
\label{EqB.3}
\begin{align}
& \quad \quad \hj_\nrm (\xi,t) \equiv \sum_{m \in \Zint} \hj_\nrm (\mnrrs)
e^{-i (\mnrrs)(t+\xi)} =
   \hj_{n(r) \pm \r(\s) ,\m} (\xi,t) \quad \\
&\quad \quad \hjb_\nrm (\xi,t) \equiv \sum_{m \in \Zint} \hjb_\nrm (\mnrrs)
e^{i (\mnrrs)(t-\xi)} =
   \hjb_{n(r) \pm \r(\s) ,\m} (\xi,t) \\
&\hj_\nrm (\xi +2\pi ,\s) = e^{-2\pi i\frac{n(r)}{\r(\s)}} \hj_\nrm (\xi
,\s) ,\quad
   \hjb_\nrm (\xi +2\pi ,\s) =e^{-2\pi i\frac{n(r)}{\r(\s)}} \hjb_\nrm (\xi
,\s)  \label{Eq B.3c}
\end{align}
\end{subequations}
\begin{subequations}
\label{EqB.4}
\begin{align}
[ \hj_\nrm (\xi,t,\s) ,\hj_\nsn (\eta,t,\s) ] &= 2\pi i \BIG{(}
\scf_{\nrm;\nsn}{}^{n(r)+n(s),\de} (\s)
\hj_{n(r)+n(s) ,\de} (\eta,t,\s) \quad \nn \\
& +\de_{n(r)+n(s) ,0\, \text{mod } \r(\s)} \sG_{\nrm;-\nrn} (\s) \pl_\xi
\BIG{)} \de_\nrrsf (\xi -\eta) \\
[ \hjb_\nrm (\xi,t,\s) ,\hjb_\nsn (\eta,t,\s) ] &= 2\pi i \BIG{(}
\scf_{\nrm;\nsn}{}^{n(r)+n(s),\de} (\s)
\hjb_{n(r)+n(s) ,\de} (\eta,t,\s) \quad \nn \\
& -\de_{n(r)+n(s) ,0\, \text{mod } \r(\s)} \sG_{\nrm;-\nrn} (\s) \pl_\xi
\BIG{)} \de_\nrrsf (\xi -\eta) \label{Eq B.4b}
\end{align}
\begin{equation}
[ \hj_\nrm (\xi,t,\s) ,\hjb_\nsn (\eta,t,\s)] =0 ,\quad \srange \,.
\label{Eq B.4c}
\end{equation}
\end{subequations}
Except for the range of $\xi$, the twisted left-mover current $\hj$ here is
the same current called $\hj^{(+)}$ in Eq.~\eqref{Eq2.5}, and the
phase-modified Dirac delta function $\de_{n(r)/\r(\s)} (\xi-\eta)$ is that
given in Eq.~\eqref{Eq2.8}.

The rectified right-mover twisted currents are then defined\footnote{The
definition \eqref{Eq B.5a} is the cylinder analogue of the
rectified right-mover twisted currents on the sphere given in Ref.~\cite{Big}.}
in terms of the rectified modes \eqref{Eq B.1a} as follows:
\begin{subequations}
\label{EqB.5}
\begin{gather}
\hjbb_\nrm (\xi,t,\s) \equiv \sum_{m \in \Zint} \hjbb_\nrm (\mnrrs)
e^{-i(\mnrrs) (t-\xi)} = \theta (n(r),\s)_\m{}^\n
   \hjb_\mnrn (\xi,t,\s) \label{Eq B.5a} \\
\hjbb_\nrm (\xi+2\pi ,t,\s)= e^{\tp \nrrs} \hjbb_\nrm (\xi,t,\s) \,.
\end{gather}
\end{subequations}
Then the {\it rectified} equal-time current algebra
\begin{subequations}
\label{EqB.6}
\begin{align}
[ \hj_\nrm (\xi,t,\s) ,\hj_\nsn (\eta,t,\s) ] &= 2\pi i \BIG{(}
\scf_{\nrm;\nsn}{}^{n(r)+n(s),\de} (\s)
\hj_{n(r)+n(s) ,\de} (\eta,t,\s) \quad \nn \\
& +\de_{n(r)+n(s) ,0\, \text{mod } \r(\s)} \sG_{\nrm;-\nrn} (\s) \pl_\xi
\BIG{)} \de_\nrrsf (\xi -\eta) \\
[ \hjbb_\nrm (\xi,t,\s) ,\hjbb_\nsn (\eta,t,\s) ] &= 2\pi i \BIG{(}
\scf_{\nrm;\nsn}{}^{n(r)+n(s),\de} (\s)
\hjbb_{n(r)+n(s) ,\de} (\eta,t,\s) \quad \nn \\
& -\de_{n(r)+n(s) ,0\, \text{mod } \r(\s)} \sG_{\nrm;-\nrn} (\s) \pl_\xi
\BIG{)} \de_{-\nrrsf} (\xi -\eta) \label{Eq B.6b} \\
[ \hj_\nrm (\xi,t,\s) ,\hjbb_\nsn (\eta,t,\s)] &=0 ,\quad \srange
\end{align}
\end{subequations}
is obtained from the unrectified algebra \eqref{EqB.4}. The only difference
between this rectified algebra and Eq.~\eqref{EqB.4} is the phase of the
$\de$-function in Eq.~\eqref{Eq B.6b}.

We will also need to consider the orbifold stress tensors on the cylinder
\begin{subequations}
\label{EqB.7}
\begin{gather}
\hat{T}_\s (\xi,t) \equiv \frac{1}{2\pi}\sum_{m\in \Zint} L_\s (m)
e^{-im(t+\xi)} =\frac{1}{2\pi}{\cL}_{\sgb (\s)}^{\nrm;\mnrn} (\s) :\!
   \hj_\nrm (\xi,t) \hj_\mnrn (\xi,t)\!: \label{Eq B.7a} \\
\hat{\bar{T}}_\s (\xi,t) \equiv \frac{1}{2\pi} \sum_{m\in \Zint} \bar{L}_\s
(m) e^{im(t-\xi)}= \frac{1}{2\pi} {\cL}_{\sgb (\s)}^{\nrm;\mnrn}(\s) :\!
   \hjb_\nrm (\xi,t) \hjb_\mnrn (\xi,t)\!: \label{Eq B.7b} \\
\hat{T}_\s (\xi+2\pi ,t)=\hat{T}_\s (\xi,t) ,\quad \hat{\bar{T}}_\s
(\xi+2\pi ,t)= \hat{\bar{T}}_\s (\xi,t) \label{Eq B.7c}
\end{gather}
\end{subequations}
where the normal ordering $:\, \cdot \,:$ is defined by the mode form of
operator-product normal ordering (see Eqs.~\eqref{Eq 2.1a} and
\eqref{EqA.3}).
Then we find the following equal-time operator algebra
\begin{subequations}
\label{EqB.8}
\begin{align}
&\!\![ \hat{T}_\s (\xi,t) ,\hat{T}_\s (\eta,t) ]=\!i \left( (\hat{T}_\s
(\xi,t) \!+\!\hat{T}_\s (\eta,t))
    -\frac{c_g}{24 \pi} (\pl^2_\xi +1)\right) \pl_\xi \de (\xi-\eta)  \\
&\!\![ \hat{\bar{T}}_\s (\xi,t) ,\hat{\bar{T}}_\s (\eta,t) ]= \!-i \left(
(\hat{\bar{T}}_\s (\xi,t) \!+\!\hat{\bar{T}}_\s
   (\eta,t)) -\frac{c_g}{24 \pi} (\pl^2_\xi +1) \right) \pl_\xi \de
(\xi-\eta) \\
& \bigspc \bigspc \bigspc [\hat{T}_\s (\xi,t), \hat{\bar{T}}_\s (\eta,t) ]=0 \\
& \quad \quad c_g \equiv 2\sG_{\nrm;\nsn}(\s)
{\cL}_{\sgb(\s)}^{\nrm;\nsn}(\s) =2G^{ab} L_g^{ab} =\sum_I \frac{2k_I
\text{dim }\gfrak^I}{2k_I+Q_I}
\end{align}
\end{subequations}
\begin{subequations}
\label{EqB.9}
\begin{gather}
[ \hat{T}_\s (\xi,t) ,\hj_\nrm (\eta,t,\s) ] = -i\pl_\eta \left( \hj_\nrm
(\eta,t,\s) \de(\xi-\eta) \right) \\
[ \hat{\bar{T}}_\s (\xi,t) ,\hjbb_\nrm (\eta,t,\s) ] = i\pl_\eta \left(
\hjbb_\nrm (\eta,t,\s) \de (\xi-\eta) \right) \label{Eq B.9b} \\
[\hat{T}_\s (\xi,t) ,\hjbb_\nrm (\eta,t,\s)] =[\hat{\bar{T}}_\s (\xi,t)
,\hj_\nrm (\eta,t,\s)] =0
\end{gather}
\end{subequations}
which, except for the commutators with the rectified currents $\hjbb$, was
given earlier in Ref.~\cite{Geom}. We also mention that the right-mover stress
tensor
can be rewritten in terms of $\hjbb$:
\begin{gather}
\hat{\bar{T}}_\s (\xi,t) =\frac{1}{2\pi} {\cL}_{\sgb (\s)}^{\nrm;\nsn} (\s)
:\!\hjbb_\nrm (\xi,t,\s) \hjbb_\nsn (\xi,t,\s)\!: \,. \label{EqB.10}
\end{gather}
The argument for \eqref{EqB.10} follows the same line as that given at the
end of App.~A, using now the relations
\begin{subequations}
\label{EqB.11}
\begin{gather}
\sG^{\nrm;\nsn} (\s) \theta (n(r),\s)_\m{}^\k \theta (n(s),\s)_\n{}^\l
=\sG^{-n(r),\k;-n(s),\l}(\s) \bigspc \label{Eq B.11a} \\
\bigspc \Longrightarrow \,\,\,\, {\cL}_{\sgb (\s)}^{\nrm;\nsn}(\s) \theta
(n(r),\s)_\m{}^\k \theta (n(s),\s)_\n{}^\l ={\cL}_{\sgb (\s)}^{-n(r)\k
;-n(s)\l} (\s)
\end{gather}
\end{subequations}
which follow from Eqs.~\eqref{Eq B.2a} and \eqref{EqA.9}.

For {\it classical} closed-string WZW orbifolds, we list the following
rectified relations
\begin{subequations}
\label{EqB.12}
\begin{gather}
\hat{T}_\s (\xi,t) = \srac{1}{4\pi} \sG^{\nrm;\nsn} (\s) \hj_\nrm
(\xi,t,\s) \hj_\nsn (\xi,t,\s) \\
\hat{\bar{T}}_\s (\xi,t) =\srac{1}{4\pi} \sG^{\nrm;\nsn} (\s) \hjbb_\nrm
(\xi,t,\s) \hjbb_\nsn (\xi,t,\s)
\end{gather}
\end{subequations}
\begin{subequations}
\label{EqB.13}
\begin{gather}
\hj_\nrm (\xi) = 2\pi \hei (\xi)_\nrm{}^\nsn \hp_\nsn (\hB ,\xi) +
\frac{1}{2} \pl_\xi \hx^\nsn (\xi)
   \he(\xi)_\nsn{}^\ntd \sG_{\ntd;\nrm} (\s) \\
\hjbb_\nrm (\xi) \!= \!2\pi \heb^{\sharp \,-1} (\xi)_\nrm{}^{\!\nsn}
\hp_\nsn (\hB ,\xi) - \!\frac{1}{2} \pl_\xi
   \hx^\nsn (\xi) \heb^{\sharp} (\xi)_\nsn{}^{\!\ntd} \sG_{\ntd;\nrm} (\s) \\
\heb_\nrm (\st,\xi) = -i\hg(\st,\xi) \hpl_\nrm \hg^{-1} (\st,\xi) =
\heb(\xi)_\nrm{}^\nsn \st_\nsn
   = \heb^{\sharp} (\xi)_\nrm{}^\nsn \st^{\sharp}_\nsn
\end{gather}
\begin{gather}
\{ \hj_\nrm (\xi,t,\s) ,\hx_\s^\nsn (\eta,t) \} = -\tp \hei
(\eta,t,\s)_\nrm{}^\nsn \de_\nrrsf (\xi-\eta) \\
\{ \hjbb_\nrm (\xi,t,\s) ,\hx_\s^\nsn (\eta,t) \} = -\tp \heb^{\sharp \,-1}
(\eta,t,\s)_\nrm{}^\nsn \de_{-\nrrsf} (\xi-\eta)
\end{gather}
\end{subequations}
\begin{subequations}
\label{EqB.14}
\begin{gather}
\{ \hj_\nrm (\xi,t,\s) ,\hg(\st,\eta,t,\s) \} = 2\pi \hg(\st,\eta) \st_\nrm
\de_\nrrsf (\xi-\eta) \quad \\
\quad \quad \{ \hjbb_\nrm (\xi,t,\s) ,\hg(\st,\eta,t,\s) \} = -2\pi
\st_\nrm^\sharp \hg(\st,\eta) \de_{-\nrrsf} (\xi-\eta)
\end{gather}
\end{subequations}
\begin{gather}
\{ \hx_\s^\nrm (\xi,t) ,\hx_\s^\nsn (\eta,t) \} =0  \label{EqB.15}
\end{gather}
which have been selected for comparison with the analogous strip results of
the text. To obtain these results, we used Eq.~(B$.2$), the
corresponding (unrectified) closed-string orbifold relations in
Ref.~\cite{Geom}, and the definitions:
\begin{subequations}
\label{EqB.16}
\begin{gather}
\st_\nrm^{\sharp} \equiv \theta (n(r),\s)_\m{}^\n \st_{\mnrn} \label{Eq
B.16a} \\
[ \st_\nrm^\sharp ,\st_\nsn^\sharp ] = i\scf_{\nrm;\nsn}{}^\ntd (\s)
\st_\ntd^\sharp  \label{Eq B.16b} \\
\heb^{\sharp} (\xi)_\nrm{}^\nsn \equiv \heb (\xi)_\nrm{}^{-n(s),\l}
\theta^{-1} (n(s),\s)_\l{}^\n \label{Eq B.16c} \,.
\end{gather}
\end{subequations}
Here $\hjbb ,\st^{\sharp}$ and $\heb^{\sharp}$ are respectively the
rectified right-mover currents, the rectified twisted representation
matrices and the rectified right-invariant twisted vielbein on the group
orbifold.

Looking back over the rectified results of this appendix, we may finally
consider the following strip$\leftrightarrow$cylinder map:
\begin{subequations}
\label{EqB.17}
\begin{gather}
\hj^{(+)} \leftrightarrow \hj ,\quad \hj^{(-)} \leftrightarrow \hjbb \\
\he \leftrightarrow \he ,\quad \heb \leftrightarrow \heb^{\sharp} ,\quad
\hg \st \leftrightarrow \hg \st ,\quad \st \hg \leftrightarrow \st^{\sharp}
\hg \,.
\end{gather}
\end{subequations}
One finds that under this map, the twisted open-string properties in
Eqs.~\eqref{Eq2.7},\eqref{Eq3.12}, \eqref{Eq3.16} and \eqref{Eq3.38} of the
text are locally twisted WZW, that is, they are isomorphic
in the bulk to the rectified closed-string properties in
Eqs.~\eqref{EqB.6}, \eqref{EqB.13}, \eqref{EqB.14} and \eqref{EqB.15}.

\section{More About the Phase-Modified Delta Functions}

The phase-modified Dirac delta functions $\de_{n(r)/\r(\s)} (\xi \pm
\eta)$ were defined in Eq.~\eqref{Eq2.8}, and the case
$\de_{n(r)/\r(\s)} (\xi - \eta)$ was studied on the cylinder in
Ref.~\cite{Geom}. For the computations of the text,
one needs further information about these delta functions on the strip.

We begin with the strip identities
\begin{subequations}
\label{EqC.1}
\begin{gather}
\hat{A}_\nrm^{(\pm)}{}^\nsn (\xi,t) \de_{\frac{n(t)}{\r(\s)}} (\pm \xi
-\eta) = \hat{A}^{(+)}_\nrm{}^\nsn (\eta,t)
\de_{\frac{n(r)-n(s)+n(t)}{\r(\s)}}
   (\pm \xi -\eta) \\
\hat{A}_\nrm^{(\pm)}{}^\nsn (\xi,t) \de_{\frac{n(t)}{\r(\s)}} (\pm \xi
+\eta) = \hat{A}^{(-)}_\nrm{}^\nsn (\eta,t)
\de_{\frac{n(r)-n(s)+n(t)}{\r(\s)}}
   (\pm \xi +\eta) \\
\text{ for any } \hat{A} \text{ s.t. } \hat{A}^{(\pm )}_\nrm{}^\nsn
(-\xi,t) = \hat{A}^{(\mp)}_\nrm {}^\nsn (\xi,t) \bigspc \nn \\
\quad \quad \quad \text{ and } \hat{A}^{(+)}_\nrm{}^\nsn (\pi,t) = e^{-\tp
\srac{n(r)-n(s)}{\r(\s)}} \hat{A}^{(-)}_\nrm {}^\nsn (\pi,t) \label{Eq
C.1c} \\
0 \leq \xi ,\eta \leq \pi
\end{gather}
\end{subequations}
which apply in particular to the twisted strip currents $\hat{A}^{(\pm)}
=\hj^{(\pm)}$ when $n(s) =0$. The $\de (\xi -\eta)$ identities here
appeared in
Ref.~\cite{Geom} as a consequence of monodromy, which is here replaced by the
strip boundary condition \eqref{Eq C.1c}.

Following Ref.~\cite{Geom}, we note the special case of Eq.~\eqref{EqC.1} with
$n(t)=n(s)-n(r)$:
\begin{subequations}
\label{EqC.2}
\begin{gather}
\hat{A}_\nrm^{(\pm)}{}^\nsn (\xi,t) \de_{\frac{n(s)-n(r)}{\r(\s)}} (\pm \xi
-\eta) = \hat{A}_\nrm^{(+)}{}^\nsn (\eta,t) \de (\pm \xi -\eta) \\
\hat{A}_\nrm^{(\pm)}{}^\nsn (\xi,t) \de_{\frac{n(s)-n(r)}{\r(\s)}} (\pm \xi
+\eta) = \hat{A}_\nrm^{(-)}{}^\nsn (\eta,t) \de (\pm \xi +\eta) \,.
\end{gather}
\end{subequations}
This case leads directly to the following integral identities on the strip
\begin{subequations}
\label{EqC.3}
\begin{equation}
\int_0^\pi \!\!\foot{d\eta \hat{A}^{(+)}_\nrm{}^\nsn (\eta)
\de_{\frac{n(s)-n(r)}{\r(\s)}} (\eta -\xi) } \!=\! \left\{
\begin{array}{ll}
\foot{ \frac{1}{2} \hat{A}^{(+)}_\nrm{}^\nsn (0), }& \foot{ \xi=0 } \\
\foot{ \hat{A}^{(+)}_\nrm{}^\nsn (\xi) ,}& \foot{ 0 <\xi <\pi } \\
\foot{ \frac{1}{2} \hat{A}^{(+)}_\nrm{}^\nsn (\pi), }& \foot{ \xi=\pi }
\end{array} \right.
\end{equation}
\begin{equation}
\int_0^\pi \!\!\foot{d\eta \hat{A}^{(-)}_\nrm{}^{\!\nsn} (\eta)
\de_{\frac{n(s)-n(r)}{\r(\s)}} (-\eta +\xi)} \!=\! \left\{
\begin{array}{ll}
\foot{ \frac{1}{2} \hat{A}^{(-)}_\nrm{}^{\!\nsn} (0), }& \foot{ \xi=0 } \\
\foot{ \hat{A}_\nrm^{(-)}{}^{\!\nsn} (\xi), }& \foot{ 0 <\xi <\pi } \\
\foot{ \frac{1}{2} \hat{A}^{(-)}_\nrm{}^{\!\nsn} (\pi) =\frac{1}{2} e^{\tp
\trac{n(r)-n(s)}{\r(\s)}} \hat{A}^{(+)}_\nrm{}^{\!\nsn} (\pi), } & \foot{
\xi=\pi }
\end{array} \right.
\end{equation}
\begin{equation}
\int_0^\pi \!\!\foot{d\eta \hat{A}^{(+)}_\nrm{}^{\!\nsn} (\eta)
\de_{\frac{n(s)-n(r)}{\r(\s)}} (\eta +\xi)} \!=\! \left\{
\begin{array}{ll}
\foot{ \frac{1}{2} \hat{A}^{(+)}_\nrm{}^{\!\nsn} (0), }& \foot{ \xi=0 }\\
\foot{ 0, }& \foot{ 0 <\xi <\pi } \\
\foot{ \frac{1}{2} \hat{A}^{(-)}_\nrm{}^{\!\nsn} (\pi) = \frac{1}{2} e^{\tp
\trac{n(r)-n(s)}{\r(\s)}} \hat{A}^{(+)}_\nrm{}^{\!\nsn} (\pi), } & \foot{
\xi=\pi }
\end{array} \right.
\end{equation}
\begin{equation}
\int_0^\pi \!\!\foot{d\eta \hat{A}^{(-)}_\nrm{}^{\!\nsn} (\eta)
\de_{\frac{n(s)-n(r)}{\r(\s)}} (-\eta -\xi)} \!=\! \left\{
\begin{array}{ll}
\foot{ \frac{1}{2} \hat{A}^{(+)}_\nrm{}^{\!\nsn} (0), }& \foot{ \xi=0 } \\
\foot{ 0, } & \foot{ 0 <\xi <\pi } \\
\foot{ \frac{1}{2} \hat{A}^{(+)}_\nrm{}^{\!\nsn} (\pi) ,} & \foot{ \xi=\pi }
\end{array} \right.
\end{equation}
\end{subequations}
which generalize the integral identities given in Ref.~\cite{Giusto} for
ordinary delta functions on the strip. Comparing Eqs.~(C$.3$a,b) with
Ref.~\cite{Giusto}, one
confirms that the same results are obtained with or without the phase
modification of $\de (\eta-\xi)$. This is easily understood because the
phase of
$\de_{n(r)/\r(\s)}$ is non-trivial only when the argument of the delta
function is $2\pi$, which is not achieved in the integrals (C$.3$a,b):
\begin{gather}
\de_{\nrrsf} (\xi -\eta) \equiv \de (\xi -\eta) \text{   when } 0\leq
\xi,\eta \leq \pi \,. \label{Eq2.9}
\end{gather}
It is not clear however that the phase-modifications can always be ignored
in spatial derivatives of $\de_{n(r)/\r(\s)}(\xi -\eta)$.

A simple consequence of Eq.~\eqref{EqC.3} is the identity
\begin{equation}
\label{EqC.4}
\int_0^\pi \!\!\!d\eta \left( \hat{A}^{(\pm)}_\nrm{}^{\!\nsn} (\eta)
\de_{\frac{n(s)-n(r)}{\r(\s)}} (\pm \eta \mp \xi) + (\eta \leftrightarrow
-\eta ) \right) \!=\! \hat{A}^{(\pm)}_\nrm{}^{\!\nsn} (\xi)
\end{equation}
and identities similar to Eqs.~\eqref{EqC.1}-\eqref{EqC.4} are easily
derived for twisted fields with any number of
indices. For example, the corresponding identity on $\hat{A}_{\nrm;\nsn}$
\begin{equation}
\label{EqC.5}
\int_0^\pi \!\!\!d\eta \left( \hat{A}^{(\pm)}_{\nrm ;\nsn} (\eta)
\de_{\frac{-n(s)-n(r)}{\r(\s)}} (\pm \eta \mp \xi) + (\eta \leftrightarrow
-\eta ) \right) \!=\! \hat{A}^{(\pm)}_{\nrm ;\nsn} (\xi)
\end{equation}
is obtained from Eq.~\eqref{EqC.4} by replacing $n(s) \rightarrow -n(s)$ in
the phase-modified delta functions.

\section{Examples of the New Non-Commutative Geometries}

In this appendix, we consider some simple examples of the new
non-commutative geometries associated to the $\{ x,x\}$ brackets of the
open WZW strings.

Because no non-abelian examples were worked out in Ref.~\cite{Giusto}, we begin
with two examples in the case of {\it untwisted} open WZW strings
$A_g^{open}$.
For these examples we will need the following explicit forms of the adjoint
action $\Omega (x)$ and the vielbein $e(x)$
\begin{subequations}
\label{EqD.1}
\begin{gather}
\Omega (x) =g^{-1} (T^{adj},x) =e^{-iY(x)} ,\quad e(x) = \smal{ \left(
\frac{e^{iY(x)} -1}{iY(x)} \right)} \\
 Y(x) \equiv x^i e_i^a (0) T_a^{adj} = x^a T^{adj}_a ,\quad a=1,\ldots
,\text{dim } g
\end{gather}
\end{subequations}
which hold for any group manifold. Here $g$ is the group element, $T^{adj}$
is the matrix adjoint rep of Lie $g$ and we have chosen $e(0)=1$ so that
the Einstein-space indices are equivalent to the tangent-space indices $a
\simeq i$.

Example 1: $A^{open}_{\su (2)}$

\noindent For $g=\su(2)$, we choose root length 1 and the standard
Cartesian basis
\begin{subequations}
\label{EqD.2}
\begin{gather}
(T_a^{adj} )_{bc} =i (R_a )_{bc} =-i \ep_{abc} ,\quad \ep_{123} =1 ,\quad
x^a T_a^{adj} = i\vec{x} \cdot \vec{R} ,\quad a,b,c \in \{ 1,2,3 \} \\
(\vec{x} \cdot \vec{R})^{2m+1} = (-|\vec{x}|^2 )^m (\vec{x} \cdot \vec{R}) ,\quad
(\vec{x} \cdot \vec{R})^{2m+2}
   = (-|\vec{x}|^2 )^m (\vec{x} \cdot \vec{R})^2 ,\quad m\geq 0
\end{gather}
\end{subequations}
which allows us to evaluate the following geometric quantities:
\begin{subequations}
\label{EqD.3}
\begin{gather}
G_{ab} =k\de_{ab} ,\quad g(T,\xi) = e^{i \vec{x}(\xi) \cdot T}  \\
\Omega (\xi) = \one +\frac{\sin | \vec{x}(\xi)|}{|\vec{x}(\xi)|}
   \vec{x}(\xi) \cdot \vec{R} + \frac{1- \cos |\vec{x}(\xi)|}{|\vec{x}(\xi)|^2}
(\vec{x}(\xi) \cdot \vec{R})^2 ,\quad e(0)_i{}^a =\de_i{}^a \\
G_{ij} (x(\xi)) = k \left( \de_{ij} + \frac{2\cos |\vec{x}(\xi)|
+|\vec{x}(\xi)|^2 -2}{|\vec{x}(\xi)|^4} ((\vec{x}(\xi)\cdot \vec{R})^2)_{ij}
\right) \,.
\end{gather}
\end{subequations}
Except for the range $0\leq \xi \leq \pi$, these forms hold as well for the
corresponding closed-string WZW model $A_{\su (2)}$. Then the
non-commutative
geometry of the Giusto-Halpern open string $A_{\su (2)}^{open}$
\begin{subequations}
\label{EqD.4}
\begin{gather}
\{ x^i (\xi,t) ,x^j (\eta,t) \} =i\pi \left\{ \begin{array}{cc} -\Psi^{ij}
(0,0) & \text{ if } \xi=\eta=0 \\ \Psi^{ij} (\pi,\pi) & \text{ if }
\xi=\eta =\pi \\
   0 & \text{ otherwise} \end{array} \right. \\
\Psi^{ij} (\xi,\xi) = -\frac{|\vec{x} (\xi)| \sin |\vec{x}(\xi)|}{k(1 -\cos
|\vec{x}(\xi)|)} (\vec{x}(\xi) \cdot \vec{R})_{ij} \label{Eq D.3d}
\end{gather}
\end{subequations}
follows directly from the results in Ref.~\cite{Giusto} or Eq.~\eqref{Eq3.39}.

Example 2: $A_{\su (2) \oplus \su(2)}^{open}$

\noindent We next consider the untwisted open WZW string on $g=\su(2)
\oplus \su(2)$, whose $\Zint_2$ permutation symmetry will be modded out to
obtain
our third (twisted) example below.

The required geometric quantities for this open WZW string
\begin{subequations}
\label{EqD.5}
\begin{gather}
G_{aI;bJ} = k\de_{IJ} \de_{ab},\quad I,J=0,1,\quad a,b,c=1,2,3 \\
g(T,\xi) = exp \left[ i \left( \begin{array}{cc} x^{0a}(\xi) T_a^{(0)} & 0
\\ 0 & x^{1a}(\xi) T_a^{(1)} \end{array} \right)
   \right] ,\quad T^{(0)} \simeq T^{(1)} \simeq T \\
\Omega (\xi)_{aI}{}^{bJ} = \de_I{}^J \left( \de_a{}^b
   +\frac{\sin |\vec{x}^I (\xi)|}{|\vec{x}^I (\xi)|} (\vec{x}^{I} (\xi) \cdot
\vec{R})_{ab} +\frac{1 -\cos |\vec{x}^I (\xi)|}
   {|\vec{x}^I (\xi)|^2} ((\vec{x}^I (\xi) \cdot \vec{R})^2 )_{ab} \right) \\
G_{iI;jJ} (x(\xi)) =k\de_{IJ}\left( \de_{ij} +\frac{2\cos |\vec{x}^I(\xi)|
+|\vec{x}^I(\xi)|^2 -2}{|\vec{x}^I (\xi)|^4} ((\vec{x}^I (\xi)
   \cdot \vec{R})^2)_{ij} \right) \\
x(\xi) =\{ x^{iI}(\xi) \} ,\,\, i=1,2,3,\,\, I=0,1 ,\quad \vec{x}^I
(\xi)\cdot \vec{R} \equiv x^{aI}(\xi) R_a ,\quad e(0)_{iI}{}^{aK} = \de_i{}^a
\de_I{}^K
\end{gather}
\end{subequations}
are easily read off as two copies of those given in the previous example.
Correspondingly, the non-commutative geometry of $A_{\su (2) \oplus
\su(2)}^{open}$
\begin{subequations}
\label{EqD.6}
\begin{gather}
\{ x^{iI} (\xi,t) ,x^{jJ} (\eta,t) \} =i\pi \left\{ \begin{array}{cc}
-\Psi^{iI;jJ} (0,0) & \text{ if } \xi=\eta=0 \\ \Psi^{iI;jJ} (\pi,\pi) &
\text{ if }
   \xi=\eta =\pi \\ 0 & \text{ otherwise} \end{array} \right. \\
\Psi^{iI;jJ} (\xi,\xi) = -\de^{IJ} \frac{|\vec{x}^I (\xi)|\sin |\vec{x}^I
(\xi)|}{k (1 -\cos |\vec{x}^I (\xi)|)} (\vec{x}^I (\xi)\cdot \vec{R})_{ij}
\end{gather}
\end{subequations}
decomposes, as expected, into two non-interacting copies of the geometry of
$A_{\su (2)}^{open}$.

Example 3: $A^{open}_{su(2) \oplus su(2)}(H) /H ,\,\,\, H=\Zint_2$(perm).

\noindent As our last example, we discuss the twisted non-commutative
geometry of the twisted sector of the open-string orbifold $A^{open}_{su(2)
\oplus su(2)} (\Zint_2)/\Zint_2$, where the $\Zint_2$ is the permutation
symmetry which exchanges the two copies of $\su (2)$.

Following the development of this paper, we begin with the data of the
single twisted left-mover sector $(\s =1)$ of the closed-string $\Zint_2$
permutation orbifold $A_{su(2) \oplus su(2)} (\Zint_2)/\Zint_2$. This data
includes for example the simplest orbifold affine algebra \cite{Chr}
\begin{subequations}
\label{EqD.7}
\begin{gather}
[\hj_{\hat{j}a} (m\!+\!\srac{\hat{j}}{2}) ,\hj_{\hat{l}b}
(n\!+\!\srac{\hat{l}}{2})] =i\ep_{abc} \hj_{\hat{j} +\hat{l},c} (m\!+\!n\!+
   \!\srac{\hat{j}+\hat{l}}{2}) +2k\de_{ab} (m\!+\!\srac{\hat{j}}{2})
\de_{m+n+\frac{\hat{j} +\hat{l}}{2} ,0} \\
\bar{n}(r) \rightarrow \bar{\hat{j}} \in \{ 0,1 \} ,\quad \m \rightarrow a
=1,2,3
\end{gather}
\end{subequations}
which provides the current-algebraic input \eqref{Eq 2.1b} for this case.

At the geometric level, the following twisted quantities
\begin{subequations}
\label{EqD.8}
\begin{gather}
\s =1: \quad \sG_{\hat{j}a;\hat{l}b} =2k\de_{ab} \de_{\hat{j}+\hat{l}
,0\,\text{mod }2} ,\quad \st_{\hat{j}a} (T) = T_a \tau_{\hat{j}} \\
\hg (\st(T),\xi) = \exp[ i( \hx^{0a}(\xi) \one_2 + \hx^{1a} (\xi) \tau_1 )
\otimes T_a]
   ,\quad \vec{\tau} =\text{ Pauli matrices} \\
\ho (\hx) = e^{-i\hY(\hx)} ,\quad \hY(\hx(\xi)) \equiv \hx^{\hat{j}a}(\xi)
T_a^{adj} \tau_{\hat{j}} ,\quad \he(\hx) =\smal{ \left(
   \frac{e^{i\hY (\hx)}-1}{i\hY (\hx)} \right)}  \\
\ho (\hx(\xi)) =\frac{1}{2} \left( \begin{array}{cc} \ho(\sxh^+ (\xi))
+\ho(\sxh^- (\xi)) &\ho(\sxh^+ (\xi)) -\ho(\sxh^-(\xi))\\
   \ho(\sxh^+ (\xi)) -\ho(\sxh^- (\xi)) &\ho(\sxh^+ (\xi)) +\ho(\sxh^-
(\xi)) \end{array} \right) \\
\ho(\sxh^\pm (\xi)) \!\equiv \one \!+\!\frac{\sin |\sxh^\pm
(\xi)|}{|\sxh^\pm (\xi)|} \sxh^\pm (\xi) \!\cdot \!\vec{R} + \!\frac{1\!-\!\cos
    |\sxh^\pm (\xi)|}{|\sxh^\pm (\xi)|^2} (\sxh^\pm (\xi)\!\cdot \!\vec{R})^2 \\
\sxh^{\pm,a} (\xi)\equiv \!\hx^{0,a}(\xi) \pm \hx^{1,a}(\xi)
\end{gather}
\begin{gather}
\hat{G}_{\hat{j}a;\hat{l}b} (\hx(\xi)) = k \left( G({\sxh}^+ (\xi)) +
(-1)^{\hat{j}+\hat{l}} G({\sxh}^- (\xi)) \right)_{ab} \\
G({\sxh}^\pm (\xi)) \equiv \one + \frac{({\sxh}^\pm (\xi)\cdot \vec{R})^2}{|
{\sxh}^\pm (\xi)|^4} (2 \cos{| {\sxh}^\pm (\xi)|}
   +|{\sxh}^\pm (\xi)|^2 -2)
\end{gather}
\end{subequations}
follow from the definitions of the text. Except for the range $0\leq \xi
\leq \pi$, these are the same formulae given in Ref.~\cite{Geom} for the
closed-string orbifold $A_{su(2) \oplus su(2)} (\Zint_2)/ \Zint_2$. Then
the twisted non-commutative geometry of sector $\s =1$ of the
open-string orbifold $A^{open}_{su(2) \oplus su(2)} (\Zint_2)/\Zint_2$
\begin{subequations}
\label{EqD.9}
\begin{gather}
\{ \hx^{\hat{j}a} (\xi,t) ,\hx^{\hat{l}b} (\eta,t) \} = i\pi \left\{
\begin{array}{cc} -\hPs^{\hat{j}a;\hat{l}b} (0,0) & \text{ if } \xi =\eta
=0 \\
   \hPs^{\hat{j}a;\hat{l}b} (\pi,\pi) & \text{ if } \xi =\eta =\pi \\ 0 &
\text{ otherwise} \end{array} \right.
\end{gather}
\begin{align}
& \hPs^{\hat{j}a;\hat{l}b} (0,0)= -\frac{1}{4k} \!\BIG{(} \frac{|\sxh^+
(0)| \sin|\sxh^+ (0)|}{1- \cos|\sxh^+ (0)|} (\sxh^+(0)\!\cdot \!\vec{R})_{ab} +
\bigspc \nn \\
& \bigspc \bigspc \bigspc +(-1)^{\hat{j}+\hat{l}} \frac{|\sxh^-(0)|
\sin|\sxh^-(0)|}{1- \cos|\sxh^-(0)|} (\sxh^-(0) \!\cdot \!\vec{R})_{ab} \BIG{)} \\
& \hPs^{\hat{j}a;\hat{l}b} (\pi,\pi)= -\frac{1}{4k} \left[ (-1)^{\hat{j}}
\left( (\hps^+ \hps^-)^t
   \!-(\hps^- \hps^+) \right) \!+(-1)^{\hat{l}} \left( (\hps^- \hps^+)^t
\!-(\hps^+ \hps^-) \right) \right]_{ab} \\
&\hps^\pm \equiv \one + \srac{1}{2} (\sxh^\pm (\pi) \cdot \vec{R}) + \left(
\frac{1}{|\sxh^\pm (\pi)|^2} - \frac{\sin |\sxh^\pm (\pi)|}
   {2 |\sxh^\pm (\pi)|(1-\cos |\sxh^\pm (\pi)|)} \right) (\sxh^\pm (\pi)
\cdot \vec{R})^2
\end{align}
\end{subequations}
follows from Eqs.~\eqref{EqD.8} and \eqref{Eq3.38}. As in the text,
superscript $t$ is matrix transpose. The example above is the simplest
permutation-twisted open WZW string.

\vskip .5cm
\addcontentsline{toc}{section}{References}

\renewcommand{\baselinestretch}{.4}\rm
{\footnotesize

\providecommand{\href}[2]{#2}\begingroup\raggedright\endgroup

\pagebreak

\end{document}